\pdfoutput=1

\documentclass[apj]{emulateapj}

\usepackage{graphicx,psfig,fancyhdr,natbib,subfigure}
\usepackage{epsfig, psfig, epsf}
\usepackage{amsmath, cancel}
\usepackage{amssymb}
\usepackage{lscape}
\usepackage{dcolumn}
\usepackage{bm}
\usepackage{hyperref,ifthen}
\usepackage{verbatim}

 \def\apjl{ApJ~Lett.} 
\def\apjs{ApJS}

 \def\aap{Astron.~\&~Astrophys.}

\def \Omb {\Omega_{\rm b}}

\def \Omlam {\Omega_{\Lambda}}
\def \Omm {\Omega_{\rm m}}

\def \gsim { \lower .75ex \hbox{$\sim$} \llap{\raise .27ex \hbox{$>$}} }
\def \lsim { \lower .75ex \hbox{$\sim$} \llap{\raise .27ex \hbox{$<$}} }

\newcommand{\sqdeg}{deg$^{-2}$}
\newcommand{\lya}{Ly$\alpha$}
\newcommand{\lyaf}{Ly\,$\alpha$\ forest}

\newcommand{\civ}{C\,{\sc iv}\ }

\newcommand{\mgii}{Mg\,{\sc ii}\ }

\begin{document}

\shorttitle{The SDSS-III BOSS: Quasar Target Selection for DR9}
\shortauthors{}

\title{The SDSS-III Baryon Oscillation Spectroscopic Survey: \\
  Quasar Target Selection for Data Release Nine}

\author{
  Nicholas P. Ross\altaffilmark{1,2}, 
  Adam D. Myers\altaffilmark{3,4}, 
  Erin S. Sheldon\altaffilmark{5}, 
  Christophe Y{\`e}che\altaffilmark{6}, 
  Michael A. Strauss\altaffilmark{7},  
  Jo Bovy\altaffilmark{8},  
  Jessica A. Kirkpatrick\altaffilmark{1,9},
  Gordon T. Richards\altaffilmark{10},
  \'Eric Aubourg\altaffilmark{6,11}, 
  Michael R. Blanton\altaffilmark{8}, 
  W. N. Brandt\altaffilmark{2}, 
  William C. Carithers\altaffilmark{1}, 
  Rupert A.C. Croft\altaffilmark{12},  
  Robert da Silva\altaffilmark{13}, 
  Kyle Dawson\altaffilmark{14}, 
  Daniel J. Eisenstein\altaffilmark{15,16}, 
  Joseph F. Hennawi\altaffilmark{17}, 
  Shirley Ho\altaffilmark{1}, 
  David W. Hogg\altaffilmark{8}, 
  Khee-Gan Lee\altaffilmark{7}, 
  Britt Lundgren\altaffilmark{18}, 
  Richard G. McMahon\altaffilmark{19},  
  Jordi Miralda-Escud\'{e}\altaffilmark{20,21}, 
  Nathalie Palanque-Delabrouille\altaffilmark{6}, 
  Isabelle P\^aris\altaffilmark{22},  
  Patrick Petitjean\altaffilmark{22}, 
  Matthew M. Pieri\altaffilmark{23,24},  
  James Rich\altaffilmark{6},  
  Natalie A. Roe\altaffilmark{1},  
  David Schiminovich\altaffilmark{25},
  David J. Schlegel\altaffilmark{1},  
  Donald P. Schneider\altaffilmark{2},  
  An\v{z}e Slosar\altaffilmark{5},  
  Nao Suzuki\altaffilmark{1},  
  Jeremy L. Tinker\altaffilmark{1,8}, 
  David H. Weinberg\altaffilmark{23},  
  Anya Weyant\altaffilmark{26},  
  Martin White\altaffilmark{1,9}, 
  W. Michael Wood-Vasey\altaffilmark{26} 
}
\altaffiltext{1}{Lawrence Berkeley National Laboratory, 1 Cyclotron Road, Berkeley, CA 92420, USA} 
\email{npross@lbl.gov}
\altaffiltext{2}{Department of Astronomy and Astrophysics, The Pennsylvania State University, 525 Davey Laboratory, University Park, PA 16802, USA}
\altaffiltext{3}{Department of Astronomy, MC-221, University of Illinois, 1002 West Green Street, Urbana, IL 61801, USA}
\altaffiltext{4}{Department of Physics and Astronomy, University of Wyoming, Laramie, WY 82071, USA} 
\altaffiltext{5}{Brookhaven National Laboratory, Blgd 510, Upton, NY 11375, USA}
\altaffiltext{6}{CEA, Centre de Saclay, IRFU, 91191 Gif-sur-Yvette, France}
\altaffiltext{7}{Department of Astrophysical Sciences, Princeton University, Princeton, NJ 08544, USA}
\altaffiltext{8}{Center for Cosmology and Particle Physics, New York University, 4 Washington Place, New York, NY 10003, USA}
\altaffiltext{9}{Department of Physics, 366 LeConte Hall, University of California, Berkeley, CA 94720, USA}
\altaffiltext{10}{Department of Physics, Drexel University, 3141 Chestnut Street, Philadelphia, PA 19104, U.S.A}
\altaffiltext{11}{APC, Universit\'{e} Paris Diderot-Paris 7, CNRS/IN2P3, CEA, Observatoire de Paris, 10, rue  A. Domon \& L. Duquet,  Paris, France.}
\altaffiltext{12}{Bruce and Astrid McWilliams Center for Cosmology, Carnegie Mellon University, Pittsburgh, PA 15213, USA}
\altaffiltext{13}{Department of Astronomy \& Astrophysics, University of California, Santa Cruz, Santa Cruz, CA, 95064, USA}
\altaffiltext{14}{Department of Physics and Astronomy, University of Utah, UT, USA}
\altaffiltext{15}{Steward Observatory, 933 North Cherry Avenue, Tucson, AZ 85721, USA}
\altaffiltext{16}{Harvard College Observatory, 60 Garden St., Cambridge, MA 02138, USA}
\altaffiltext{17}{Max-Planck-Institut f\"ur Astronomie, Konigstuhl 17, 69117 Heidelberg, Germany}
\altaffiltext{18}{Department of Physics, Yale University, New Haven, CT 06511, USA}
\altaffiltext{19}{Institute of Astronomy, University of Cambridge, Madingley Road, Cambridge CB3 0HA, UK}
\altaffiltext{20}{Instituci\'{o} Catalana de Recerca i Estudis Avan\c{c}ats, Barcelona, Catalonia, Spain}
\altaffiltext{21}{Institut de Ci\`{e}ncies del Cosmos, Universitat de Barcelona/IEEC, Barcelona 08028, Catalonia, Spain}
\altaffiltext{22}{Universit\'e Paris 6 et CNRS, Institut d'Astrophysique de Paris, 98bis blvd. Arago, 75014 Paris, France}
\altaffiltext{23}{Astronomy Department and Center for Cosmology and AstroParticle Physics, Ohio State University, 140 West 18th Avenue, Columbus, OH 43210, USA}
\altaffiltext{24}{Center for Astrophysics and Space Astronomy, University of Colorado, 389 UCB, Boulder, CO 80309, USA}
\altaffiltext{25}{Department of Astronomy, Columbia University, 550 West 120th Street, New York, NY 10027, USA}
\altaffiltext{26}{Department of Physics and Astronomy, University of Pittsburgh, 100 Allen Hall, 3941 O'Hara St, Pittsburgh PA 15260, USA}

\date{\today}

\begin{abstract}
The SDSS-III Baryon Oscillation Spectroscopic Survey (BOSS), a
five-year spectroscopic survey of 10,000 deg$^{2}$, achieved first
light in late 2009. One of the key goals of BOSS is to measure the
signature of baryon acoustic oscillations in the distribution of
Ly$\alpha$ absorption from the spectra of a sample of $\sim$150,000
$z>2.2$ quasars.  Along with measuring the angular diameter distance
at $z\approx2.5$, BOSS will provide the first direct measurement of
the expansion rate of the Universe at $z > 2$.  One of the biggest
challenges in achieving this goal is an efficient target selection
algorithm for quasars in the redshift range $2.2 < z < 3.5$, where
their colors tend to overlap those of the far more numerous stars.
During the first year of the BOSS survey, quasar target selection
methods were developed and tested to meet the requirement of
delivering at least 15 quasars deg$^{-2}$ in this redshift range, with
a goal of 20, out of 40 targets deg$^{-2}$ allocated to the quasar
survey. To achieve these surface densities, the magnitude limit of the
quasar targets was set at $g\leq22.0$ or $r\leq21.85$.
While detection of the BAO signature in the distribution of Ly$\alpha$
absorption in quasar spectra does not require a uniform target
selection algorithm, many other astrophysical studies do.  We have
therefore defined a uniformly-selected subsample of 20 targets
deg$^{-2}$, for which the selection efficiency is just over 50\%
($\sim$10 $z> 2.20$ quasars deg$^{-2}$).  This ``CORE'' subsample will
be fixed for Years Two through Five of the survey.  For the remaining
20 targets deg$^{-2}$, we will continue to develop improved selection
techniques, including the use of additional data sets beyond the SDSS
imaging data.
In this paper we describe the evolution and implementation of the BOSS
quasar target selection algorithms during the first two years of BOSS
operations (through July 2011), in support of the science
investigations based on these data, and we analyze the spectra
obtained during the first year.  During this year, \hbox{11,263} new
$z>2.20$ quasars were spectroscopically confirmed by the BOSS, roughly
double the number of previously known quasars with $z>2.20$.  Our
current algorithms select an average of 15 $z > 2.20$ quasars
deg$^{-2}$ from 40 targets deg$^{-2}$ using single-epoch SDSS imaging.
Multi-epoch optical data and data at other wavelengths can further
improve the efficiency and completeness of BOSS quasar target
selection.
\end{abstract}


\maketitle

\begin{figure*}
  \begin{center}
    \includegraphics[height=8.0cm,width=18.0cm]
    {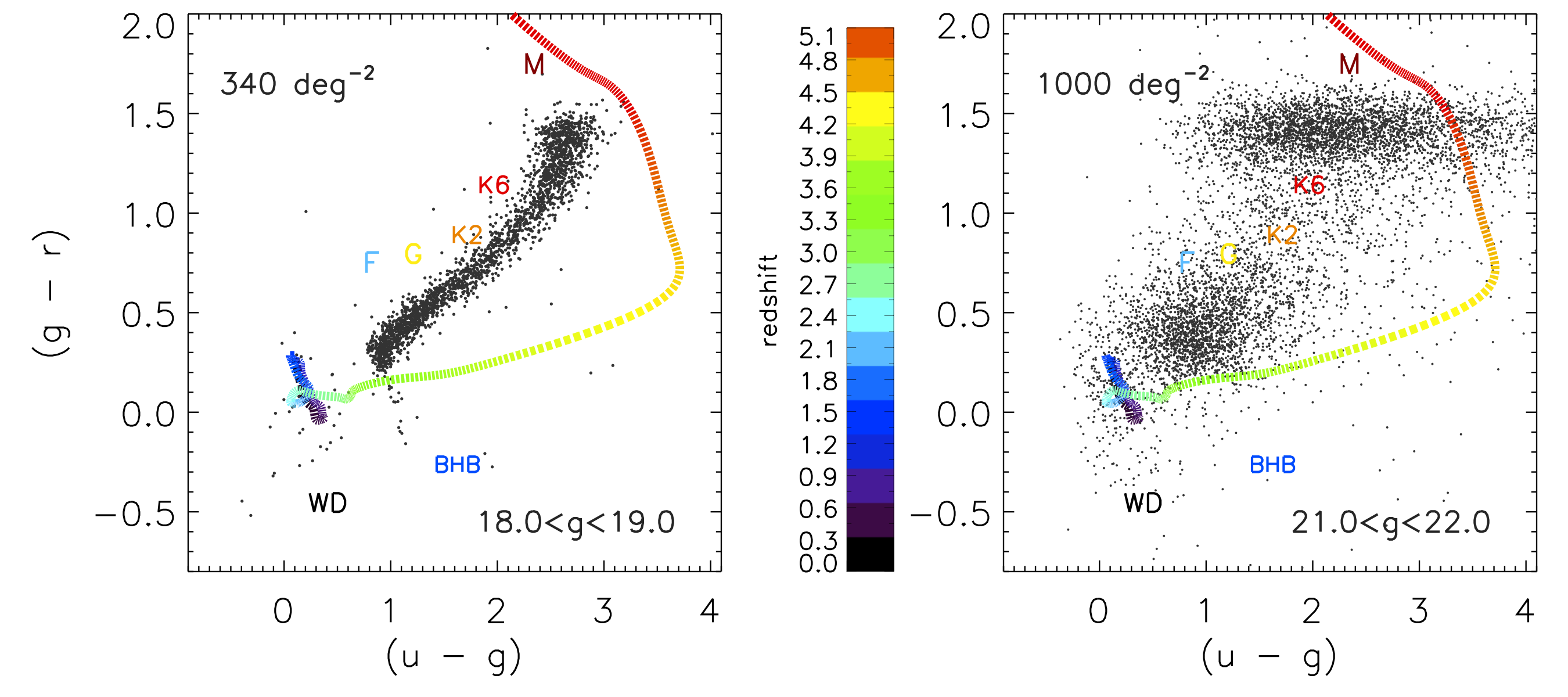}
    \caption{Color-color diagrams of point sources drawn from 7 deg$^{2}$
      (the BOSS spectrograph field of view) in the SDSS photometric
      database.  {\it (Left)} 2,400 objects with $18.0<g<19.0$, and {\it
        (Right)} 7,000 objects with $21.0<g<22.0$.  Most of the objects shown
      are stars; low-redshift ($z < 2.2$) quasars lie preferentially in the
      region $u-g<0.6, g-r>0$ where very few stars are found.  At $z > 2.2$,
      quasars become systematically redder in $u-g$ as the Ly$\alpha$ forest
      moves into the $u$-band and Ly$\alpha$ emission moves into $g$.
      At $z \sim 2.7$, quasars have colors
      similar to those of blue horizontal branch (BHB) stars. The larger
      photometric errors at faint magnitudes broaden the stellar locus
      considerably (especially in the $u$-band for the reddest stars, which
      gives rise to the spread at $g-r \sim 1.5$), illustrating the
      challenges involved in selecting faint objects by their colors. Tracks
      for the quasar locus, as presented in Bovy et al. (2011b, in prep.) are also shown, 
      with the corresponding redshift given by the color-bar legend. Approximate
      surface densities are quoted, and stellar classifications are given as
      a guide.}
    \label{fig:DJS_killerfig}
  \end{center}
\end{figure*}
\section{Introduction}
\label{sec:intro}

\subsection{The Baryon Oscillation Spectroscopic Survey}
The current Cosmic Microwave Background (CMB) data are in excellent
agreement with the theoretical predictions of a flat cosmological
model with cold dark matter which is dominated by dark energy with an
equation of state parameter, $w = -1$ \citep[$\Lambda$CDM;
][]{Komatsu11, Larson11}. Acoustic peaks in the CMB anisotropy power
spectrum are generated by cosmological perturbations exciting sound
waves in the relativistic plasma of the early universe \citep{SZ70,
PeeblesYu70, BondEfstathiou84, BondEfstathiou87, Holtzman89,
Meiksin99}. The scale of these peaks, which is set by the sound
horizon at last scattering \citep{EisensteinHu98, BlakeGlazebrook03,
Seo03}, can be used as a cosmological standard ruler. These baryon
acoustic oscillations (BAO) are present in the distribution of matter
at late times as well, and were first measured in the large-scale
distribution of galaxies by \citet{Eisenstein05} and \citet{Cole05}.

BAO should also be present in the distribution of neutral hydrogen gas
in the intergalactic medium, and thus should be observable in the
Lyman-$\alpha$ forest (Ly$\alpha$F) absorption spectra of distant
quasars \citep{White03, McDonald07, Slosar09, Norman09LyAF,
Barenboim10, White10, McQuinn11}. Measurements of BAO in the
Ly$\alpha$F would provide the first measurements of cosmic expansion
and the angular diameter distance at redshift $z>2$ (other than the
CMB itself), a regime not constrained by current data, thus giving
important constraints on, and tests of, the standard cosmological
model.

The Sloan Digital Sky Survey (\citealt{York00}) is now in its third
phase (SDSS-III; \citealt{Eisenstein11}), and is carrying out a
combination of four interleaved surveys that will continue until the
summer of 2014. One of those surveys, the Baryon Oscillation
Spectroscopic Survey
(BOSS\footnote{\href{http://www.sdss3.org/surveys/boss.php}{\tt
http://www.sdss3.org/surveys/boss.php}}) commenced operations in late
2009, and is using essentially all the dark time for SDSS-III. The key
goal of the BOSS is to measure the absolute cosmic distance scale and
expansion rate to an accuracy of a few percent from the signature of
BAO in the distribution of galaxies and neutral hydrogen
\citep{Schlegel07, Schlegel09}. This will be achieved by measuring
spectroscopic redshifts for $\approx 1.5$ million luminous red
galaxies and, simultaneously, the Ly$\alpha$F towards $\approx$150,000
high-redshift quasars\footnote{In this paper we will use the language
``low'' and ``high'' redshift to indicate objects with $z<2.2$ and
$z>2.2$, respectively. The term ``mid-$z$'' will be used to explicitly
refer to quasars with $2.2<z<3.5$.}. Both samples aim to constrain the
equation of state of dark energy by measuring the angular diameter
distance, $d_{A}$, and the Hubble Parameter, $H(z)$, at $z = 0.3, 0.6$
and $\sim$2.5.  In addition to the cosmology goals, the unprecedented
dataset of $z\sim$2.5 quasars will enable tests of black hole growth,
wind and feedback models and provide insights into the links between
galaxy formation, evolution and luminous AGN activity. Using data from
the original SDSS quasar survey will also allow studies of
spectroscopic variability.  BOSS uses the same 2.5m Sloan Foundation
telescope \citep{Gunn06} that was used in SDSS-I/II, but since BOSS
will observe fainter targets, the fiber-fed spectrographs have been
significantly upgraded. These upgrades include: new CCDs with improved
blue and red response; 1000 $2''$ instead of 640 $3''$ optical
diameter fibers; higher throughput gratings over a spectral range of
3600--10000\AA\ at a resolution of about 2000, and improved optics.

\begin{table}
    \begin{center}
        \begin{tabular}{lrr}
            \hline
             \hline
             Quantity/units                                    &   Year  & Full \\
                                                                                              & One  &Survey$^{*}$ \\
              \hline 
              Area (deg$^{2}$)                                                   &  880  & 10,200\\
              Target density in NGC (deg$^{-2}$)                & 80 &  $\approx$50  \\
              Target density in SGC (deg$^{-2}$)                     & 70  & $\approx$40\\
              Total number of Targets / $1\times10^{3}$          & 133 & $\sim$440 \\
              Efficiency                                                           &  0.26  & $\gtrsim$0.40 \\        
              Number of $z>2.2$ quasars / $1\times10^{3}$     & 13.5 & 175 \\
             \hline 
              \hline
           \end{tabular}  
        \caption{$^{*}$Projection based on observations through April 2011 and DR9 target selection.}
           \label{tab:topline} 
    \end{center}
\end{table}

\subsection{Quasar Target Selection in BOSS}
Quasars have colors distinct from those of the much more numerous
stars in the five-color photometry of the SDSS \citep{Fan99}.
Unobscured quasars have very blue continua, without any breaks redward
of the Ly$\alpha$ emission line, and so can be distinguished from hot
stars which have a strong Balmer break in the $u-g,g-r$ color-color
diagram (Figure~\ref{fig:DJS_killerfig}).  In particular, at $z <
2.2$, quasars have a UV excess (as measured by $u-g$) that
distinguishes them from most stars, and they lie well away from the
stellar locus at most higher redshifts (but see below).  SDSS-I/II
targeted quasars for spectroscopy \citep{Richards02} by selecting
point sources which lie far from the locus of stars in color-color
space (and all extended sources with a strong UV excess), as well as
point sources with radio emission from the Faint Radio Sources at
Twenty cm (FIRST) survey \citep{Becker95}.  The majority of the more
than 100,000 quasars spectroscopically observed by SDSS
\citep{Schneider10} were targeted in this way.

The Ly$\alpha$ forest enters the sensitive range of the BOSS
spectrographs at $z > 2.2$, and the number density of quasars falls
dramatically at $z>3$ \citep{Osmer82, SSG95, Richards06}, so BOSS
quasar target selection is designed to focus on the range $2.2 < z <
3.5$.  However, at $z \sim 2.7$, SDSS quasar colors are very similar
to those of A stars and blue horizontal branch stars \citep{Fan99},
thus the optimal quasars for studying the Ly$\alpha$ forest are the
most difficult ones for BOSS to target.  Indeed, the SDSS-I/II quasar
target selection algorithm deliberately sparse-sampled objects in the
region of color space where $z = 2.7$ quasars should lie
\citep{Richards02}, in an attempt to minimize the contamination by
stars.

The BOSS survey requirements are for spectroscopy of 15 or more $z >
2.2$ quasars deg$^{-2}$ (150,000 quasars over the BOSS footprint of
10,000 deg$^2$; \citealt{Eisenstein11}).  Combining calculations from
\citet{McDonald07} and \citet{McQuinn11} with the luminosity function
given by \citet{Jiang06}, we find that targeting to a magnitude of
$g<22$ {\it with perfect completeness} will provide a surface density
of $20\ z>2.2$ quasars deg$^{-2}$.  This magnitude limit is
approaching the detection limit of SDSS photometry
\citep{Abazajian04}, meaning that photometric errors will
significantly broaden the stellar locus
(Figure~\ref{fig:DJS_killerfig}) and star-galaxy separation will be a
factor.  Contamination at both the bright and the faint end of the
BOSS target range is mainly from metal-poor halo A and F stars, faint
lower redshift ($z\sim0.8$) quasars, and compact galaxies. To put
these requirements into perspective, the final quasar catalog from the
original SDSS-I/II quasar survey \citep{Schneider10} contained
\hbox{17,582} $z>2.2$ objects over 9380 deg$^2$, while the 2dF-SDSS
LRG And QSO (2SLAQ) survey \citep{Croom09a}, which observed to
$g<21.85$ and concentrated on UV-excess objects, contained
\hbox{1,110} such quasars selected over 192 deg$^2$.  The original 2dF
QSO redshift survey (2QZ; \citealt{Croom04}) focused on the redshift
range $z < 2.1$.

These challenges required a new approach to quasar target selection.
The first year of the BOSS survey (``Year One''; 2009 September
through July 2010) was devoted in part to refining our algorithms for
selecting these objects.  The resulting sample of quasars at $z > 2.2$
is comparable in size to the SDSS high-redshift quasar sample, and of
course reaches much fainter magnitudes with much higher surface
density.  Thus the new sample itself represents the best test of our
selection algorithms, and we modified those algorithms multiple times
through the year.  Year One included roughly three months of
commissioning of the upgraded BOSS spectrographs and instrument
control software as well as a steady ramp-up to full efficiency
operations, so it includes well under 20\% of the anticipated final
sample for the five-year BOSS survey.  As of April 2011, BOSS is on
track to complete its intended 10,000 deg$^2$ of spectroscopic survey
area assuming historical weather patterns and continuation of the
current observing efficiency.
 
Motivated by the first science investigations based on Year One data
\citep[e.g., ][]{Slosar11}, this paper presents the methods and
performance of the quasar target selection during this year. In what
follows,``Year Two'' will refer both to the spectroscopic observations
carried out during BOSS' second year, 2010 August to 2011 July, and
the results of the quasar target selection presented in this paper
over the entire 10,000 deg$^2$ BOSS footprint; the distinction should
be clear from context. Data from spectroscopic observations in Years
One and Two will be included in SDSS Data Release Nine
(DR9\footnote{\href{http://www.sdss3.org/surveys/}{http://www.sdss3.org/surveys/}}).
The final SDSS-III quasar target selection algorithm will appear in a
separate paper.

Background quasars have no causal influence on structure in the
Ly$\alpha$F at the BAO scale\footnote{There may however be some
measurement bias at the $0.1 - 1$\% level for the flux power spectrum,
optical depth and the flux probability distribution, due to
gravitational lensing effects, \citep[see e.g., ][]{Loverde10}.}.
Hence the sample of quasars we use for Ly$\alpha$F cosmological
studies may be quite heterogeneous, with the only consequence that the
window function of the survey will depend on the distribution of the
quasars for which we have spectra. Since the precision of the BAO
measurement improves rapidly with the surface density of quasars (at
fixed spectroscopic signal-to-noise ratio (S/N)), we have implemented
a target selection scheme in BOSS that can maximize the number of
quasars found at $z>2.2$ in any area of the sky, taking advantage of
any available information (e.g., auxiliary data). In Year One, we
explored a variety of methods, settling on our final target selection
algorithms late in the year.

At the same time, in order to use the quasars themselves for
statistical studies (such as luminosity functions or clustering
analyses), we must also produce a uniformly selected sample, which we
refer to hereafter as CORE (\S~\ref{sec:core}).  However, we changed
the definition of the CORE sample several times over Year One, as we
tested various algorithms. Therefore, our fully uniform quasar sample
will not include data from this first year of the survey.  However,
statistical studies (luminosity functions, clustering, and so forth)
can utilize all five years of BOSS data by including moderate
incompleteness corrections for Year One selection relative to the
final CORE algorithm (see \S\ref{sec:discussion}).  We describe the
evolution of our algorithms in detail in this paper, concluding with a
description of the method we finally adopted.  We give the target
selection for both Years One and Two, and thus for the DR9, and
analyze our performance from spectra obtained in Year One. By the end
of Year Two, quasar target selection (QTS) had been performed over the
whole 10,000~deg$^{2}$ BOSS footprint. Data from Year One were
gathered over 880 deg$^{2}$; see Table~\ref{tab:topline}.

This paper is organized as follows.  In \S~\ref{sec:input_phot} we
describe the SDSS photometry on which the target selection algorithms
are most heavily based.  Section~\ref{sec:methods} describes our
methods for selecting quasars \citep[][]{Richards09, Yeche10,
Kirkpatrick11, Bovy11}.  These four papers suggest different, but
complementary, methods, and we have used a union of these techniques
in different combinations through the survey. In
Section~\ref{sec:chunky} we describe the implementation of these
targeting methods through the first year.  In
Section~\ref{sec:results}, we report on the global properties of the
resulting sample, including high-$z$ quasar targeting efficiency, from
the data gathered during the first year of the BOSS, and we compare
the effectiveness of the various methods. In
Section~\ref{sec:discussion} we discuss the production of a
statistical quasar sample. We conclude in
Section~\ref{sec:conclusions} and suggest improvements to BOSS quasar
target selection for the remainder (Years Three, Four and Five) of the
survey. Appendix~\ref{sec:appflags} tabulates the logical cuts used on
the input imaging data.  Appendix~\ref{sec:flowchart_YearOne} gives
more detail about Year One target selection, while
Appendix~\ref{sec:MMT} describes a pre-BOSS pilot survey using the
MMT. Appendix~\ref{sec:zwarning} characterizes the redshift
completeness of our spectroscopic data.

We assume a cosmological model throughout with $\Omb=0.046$,
$\Omm=0.228$, $\Omlam=0.725$ \citep{Komatsu11}.  All optical
magnitudes are quoted in, and based upon, the SDSS approximation to
the AB zero-point system \citep{Oke83, Adelman-McCarthy06}, while all
near-infrared (NIR) magnitudes are based on the Vega
system. Throughout the paper, ``magnitude'' refers to SDSS Point
Spread Function (PSF) magnitudes \citep{Stoughton02}.

\begin{figure*}
  \begin{center}
    \includegraphics[height=20.5cm,width=16.0cm]
    {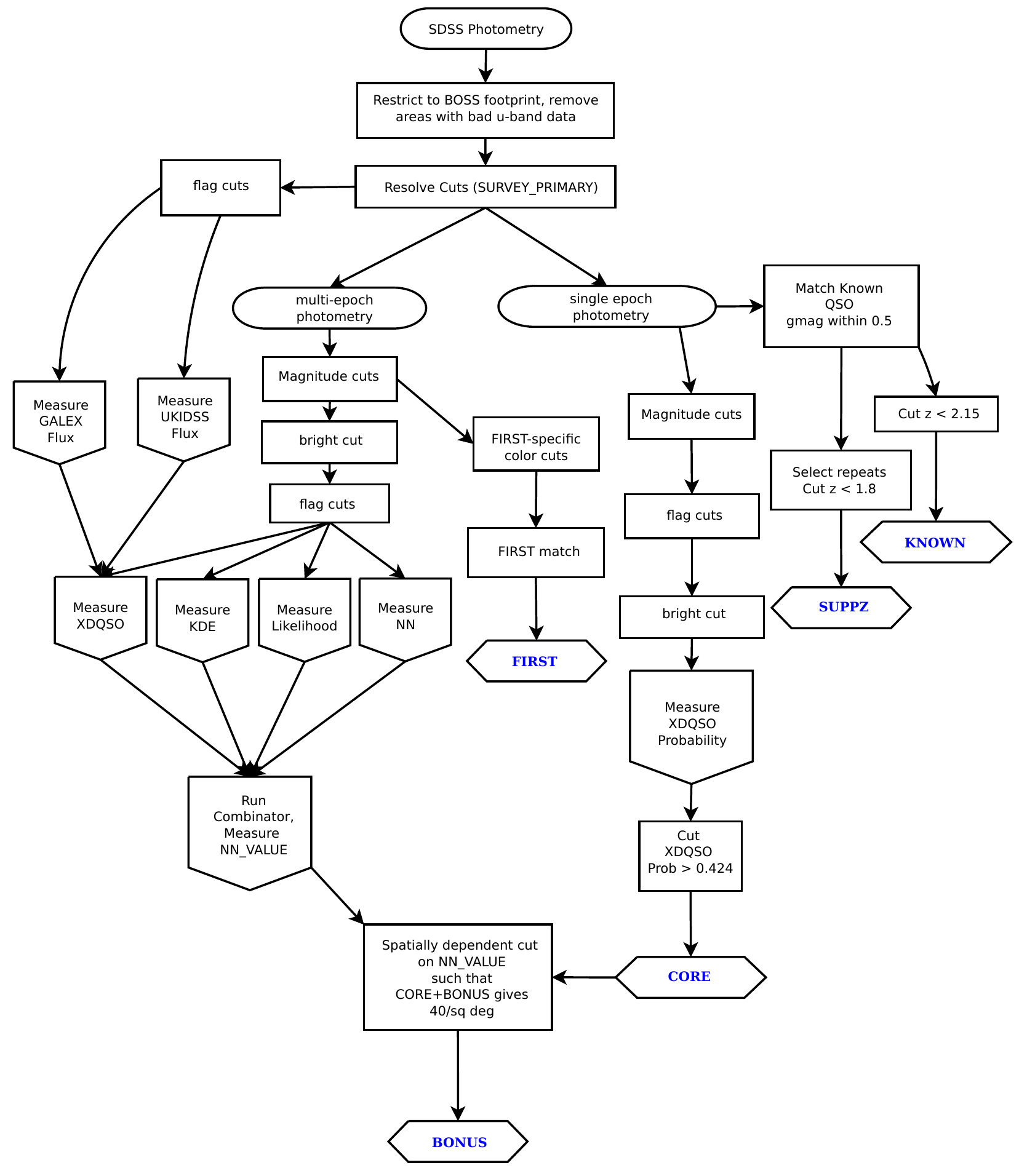}
    \caption{Flowchart for the BOSS quasar target selection, 
      as implemented from the beginning of the second year of BOSS
      observations.  The various broad categories of targets,
      including CORE, BONUS, KNOWN objects, and those detected by the
      FIRST survey, are indicated, and are described in detail in
      Section~\ref{sec:methods}; SUPPZ refers to a small number of
      lower-redshift objects targeted to study the effects of metal
      line absorption (\S~\ref{sec:chunks14to19}). 
      The flowchart for the first year of BOSS
      target selection is given in Appendix~\ref{sec:flowchart_YearOne}.
     The CORE sample is fixed for DR9 and the remainder of the BOSS. 
      Objects which satisfy the XDQSO probability cut of $P(XDQSO)>0.424$
      are selected as CORE, and the QSO\_CORE\_MAIN target flag
      bit is set. 
      CORE selection is based on single-epoch SDSS photometry, but other
      selections use multi-epoch photometry where it is available
      (e.g., in regions where SDSS imaging stripes overlap).}
    \label{fig:flowchart}
  \end{center}
\end{figure*}
\section{SDSS Photometry}
\label{sec:input_phot}

    \subsection{Imaging Data}
    \label{sec:imaging}
    BOSS uses the same imaging data as that of the original SDSS-I/II
    survey, with an extension in the South Galactic Cap (SGC). These data
    were gathered using a dedicated 2.5 m wide-field telescope
    \citep{Gunn06} to collect light for a camera with 30 2k$\times$2k CCDs
    \citep{Gunn98} over five broad bands - {\it ugriz} \citep{Fukugita96};
    this camera has imaged 14,555 unique deg$^{2}$ of the sky, including
    \hbox{7,500 deg$^{2}$} in the North Galactic Cap (NGC) and \hbox{3,100
      deg$^{2}$} in the SGC \citep{Aihara11}. The imaging data were taken on
    dark photometric nights of good seeing \citep{Hogg01}, and objects
    were detected and their properties were measured \citep{Lupton01,
      Stoughton02} and calibrated photometrically \citep{Smith02, Ivezic04,
      Tucker06, Padmanabhan08a}, and astrometrically \citep{Pier03}.
    
    \citet{Padmanabhan08a} present an algorithm which uses overlaps
    between SDSS imaging scans to photometrically calibrate the SDSS
    imaging data.  BOSS target selection uses data calibrated using this
    algorithm from the SDSS Data Release Eight (DR8) database
    \citep[Sec. 3.3; ][]{Aihara11}.  The $2.5^\circ$-wide stripe along the
    celestial equator in the Southern Galactic Cap, commonly referred to
    as ``Stripe 82'' was imaged multiple times, with up to 80 epochs at
    each point along the stripe spanning a 10-year baseline
    \citep{Abazajian09}.  In Section~\ref{sec:chunky} we will discuss how
    the commissioning phase of BOSS used coadded catalogs in SDSS Stripe
    82, generated by averaging the photometric measurements from $\sim 20$
    individual repeat scans; the details are discussed in
    Appendix~\ref{sec:VarCats} and in \citet{Kirkpatrick11}.
    
    Roughly 50\% of the SDSS footprint has been imaged more than once
    \citep{Aihara11}; combining the photometric measurements in these
    overlap regions reduces the flux errors.
    
    Using the imaging data, BOSS quasar target candidates are selected for
    spectroscopic observation based on their PSF fluxes and colors in SDSS
    bands. Fluxes that are used for quasar target selection are corrected
    for Galactic dust extinction according to the maps of \citet{SFD}. All
    objects classified as point-like ({OBJC\_TYPE $=$ 6}) and are brighter
    than $g = 22$ {\it or} $r = 21.85$ are passed to the various quasar
    target selection algorithms. The joint magnitude limit was imposed due
    to concerns of the Ly$\alpha$F moving into the $g$-band at
    $z\approx2.3$ resulting in suppressed flux at redshifts greater than
    this. In practice, almost all our targets satisfy both these
    conditions.  Throughout this paper, magnitudes use the asinh scale at
    low flux levels, as described by \citet{Lupton99}.
    
    \subsection{Photometric Pipeline Flag and Logic Cuts}
    \label{sec:phot_flags}
    During processing of the imaging data by the SDSS photometric
    pipeline, a number of photometric flags are set for each detected
    object \citep{Stoughton02}. These are generated by the SDSS
    photometric pipeline \citep[][]{Lupton01}, the {\it Resolve} algorithm
    \citep{Aihara11}, and by photometric calibration
    \citep{Padmanabhan08a}. Some of these flags indicate problems with the
    de-blending of close pairs of objects.  Other flags are set due to
    poor or unreliable photometry, e.g., if an object was saturated due to
    a bright star's diffraction spike or an object was too close to the
    edge of a frame.  If these flags are ignored, they can lead to
    artifacts in the imaging data being selected as quasar
    targets. Details of these flags are given in \citet{Stoughton02} and
    have been updated in DR8 \citep{Aihara11}.
    
    There are four distinct sets of quasars targeted by BOSS: targets
    selected by a uniform method, targets selected in a non-uniform way,
    matches to previously known $z > 2.2$ quasars, and matches to objects
    in the FIRST survey. We refer to these subsets of targets in this
    paper as CORE, BONUS, KNOWN and FIRST, respectively.
    
    Each of our targeting algorithms has different imaging flag cuts,
    as well as different flux limits imposed. We refer to these criteria
    collectively as ``logic cuts.'' All such cuts are applied using
    single-epoch data with one exception: color cuts made on FIRST targets
    use coadded, multi-epoch data wherever these are available.  FIRST
    objects are thus not considered to be part of the CORE statistical
    sample, unless they independently meet the CORE selection criteria.
    The logic cuts are described in detail in Appendix~\ref{sec:appflags}.

\section{Methods for BOSS Quasar Target Selection}
\label{sec:methods}

    \subsection{Philosophy of CORE and BONUS}  
    \label{sec:core}
    The methods, data and logic flag cuts for BOSS Quasar Target
    Selection (QTS) are summarized in Fig.~\ref{fig:flowchart}.  During
    Year One, we carried out QTS and designed spectroscopic plates on
    areas of $\sim$100-300 deg$^{2}$ at a time.  We refer to these areas,
    within which all the algorithms used in QTS are uniform, as
    ``chunks''. Once QTS was more settled in Year Two, the areas of chunks
    could be, and sometimes were, more than $1000$ deg$^{2}$. For guidance
    in the following discussion, Chunks 1 through 9, inclusive, constitute
    Year One, and Chunks 10 through 18, Year Two. Stripe 82 was targeted
    twice with different targeting algorithms: once in Year One (Chunk 1)
    and once in Year Two (Chunk 11).
    
    If an object satisfies the selection criteria of one or more of
    our methods outlined below, bits in the {BOSS\_TARGET1} target flag
    are set. Table~\ref{tab:BOSS_TARGET_FLAGS} gives the flag name, the
    bit value and the short description of the different target selection
    flags.
    \begin{table*}
      \begin{center}
        \begin{tabular}{lclc}
          \hline
          \hline
          {BOSS\_TARGET1} flag   &   bit  &  Description & Used in
	  Year Two?  \\
          \hline
          QSO\_CORE$^{a}$                   &   10  &  Restrictive quasar selection & No \\
          QSO\_BONUS$^{a}$                 &   11  &  Permissive quasar selection & No\\
          QSO\_KNOWN\_MIDZ             &   12  &  Known quasar with $z>2.15$ & Yes\\
          QSO\_KNOWN\_LOHIZ$^{b}$  &    13  &  Known quasar with $z<2.15$ & Yes\\
          QSO\_NN$^{c}$                       &   14  & Neural Net & Yes \\
          QSO\_UKIDSS$^{d}$                &  15  &  K-excess targets & No \\
          QSO\_KDE\_COADD               &  16 &  KDE targets from the Stripe82 coadd & No\\
          QSO\_LIKE                              &  17 &  Likelihood method & Yes\\
          QSO\_FIRST\_BOSS                 & 18  & FIRST radio match  & Yes \\
          QSO\_KDE                              & 19  & Selected by KDE$+\chi^{2}$ & Yes\\
          \hline
          {\bf QSO\_CORE\_MAIN}$^{e}$       & {\bf 40} & {\bf Main survey CORE sample } & {\bf Yes}\\
          {\bf QSO\_BONUS\_MAIN}$^{e,f}$   & {\bf 41}  & {\bf Main survey BONUS sample } & {\bf Yes}\\
          \hline
          QSO\_CORE\_ED           & 42  &  Extreme Deconvolution in CORE & Yes\\
          QSO\_CORE\_LIKE         & 43  &  Likelihood objects that make it into CORE  & Yes\\
          QSO\_KNOWN\_SUPPZ  & 44 & Known quasars with $1.80<z<2.15$ & Yes \\
          \hline
          \hline
        \end{tabular}  
      \end{center}
      \caption[The flag name, bitwise value and the short description of the 
      different target selection flags.]
      {The flag name, bit value and the short description of the  different target selection flags. \\
        $^{a}$QSO\_CORE and QSO\_BONUS were set only for Chunks 1
        and 2, after which the definition
        of CORE and BONUS changed. \\
        $^{b}$These objects are not targeted. \\
        $^{c}$Set if an object is selected by the first stage neural
	network (\S~\ref{sec:nn}).\\
        $^{d}$These objects were only targeted on Chunk 1. \\
        $^{e}$ QSO\_CORE\_MAIN and QSO\_BONUS\_MAIN were
        introduced with Chunk 3, and identify the CORE and BONUS
        samples.  They appear in tandem with another flag
	indicating the specific method that selected each object. \\
       $^{f}$Set if an object is selected by the NN-Combinator.\\}
      \label{tab:BOSS_TARGET_FLAGS}
    \end{table*}
    
    As discussed in the introduction, we wish to define a CORE sample that
    is uniformly selected over the BOSS footprint, for statistical studies
    of quasars, such as measurements of the luminosity function and the
    clustering of quasars.  While these goals do not drive our technical
    requirements, the survey we have designed to measure the BAO signal
    will also provide an unprecedented spectroscopic dataset for studies
    of quasars themselves. Thus, design choices that are roughly neutral
    with regard to cost and impact on the cosmology goals are guided
    by these additional science considerations.
    
    This is the motivation for dividing our quasar targets into two
    broad classes. Since the one (imaging) dataset that we have over the
    entire BOSS footprint is the SDSS single-epoch photometry
    (including the new coverage in the SGC; \citealt{Aihara11}), we
    define quasar CORE targets as a sample of 20 targets \sqdeg,
    which are selected {\it only from this single-epoch imaging data,
      using a uniform algorithm}.  As we shall see, the efficiency of the
    CORE sample is near our goal of 50\%
    (i.e. $\sim$10 out of 20 CORE targets deg$^{-2}$ are $z>2.2$
    quasars). The CORE sample is designed to have a well understood,
    uniform, and reproducible selection function.
    
    In contrast, the ``BONUS'' sample is selected using as many
    methods and additional data as deemed necessary to achieve our
    desired quasar density. The BONUS sample has a target density of 20
    deg$^{-2}$.   The number of BONUS targets 
    added in each region of sky is adjusted to assure that 
    the total density of targets, CORE + BONUS, is uniform across the 
    sky, as we will show in \S~\ref{sec:skydist} below.   However, as
    we detail below, the number of BONUS targets was extended up to 
    60 targets deg$^{-2}$ (and then 40 targets deg$^{-2}$), during the
    BOSS Commissioning and early science phases, for a total (CORE+BONUS)
    of 80 (and then 60) targets deg$^{-2}$. 
    The efficiency of BONUS selection is generally lower than that
    of CORE, despite the use of multiple algorithms and auxiliary data,
    simply because the relatively ``easy'' targets have already
    been picked by CORE and are therefore are not included in BONUS.
    
    Prior to BOSS, there was no extant survey that successfully
    targeted $z > 2.2$ quasars to the depth and surface density and with
    the efficiency we needed.  The first year of BOSS spectroscopy was
    therefore largely a commissioning year for quasar target selection,
    during which we gathered the quasar sample needed to test our various
    algorithms. In particular, it was only at the end of the year that we
    settled on the final CORE and BONUS algorithms.  Thus, the nominal
    CORE-selected objects from the first year are {\em not} a uniformly
    selected sample.  Sec.~\ref{sec:discussion} describes the completeness
    of the final CORE sample in Year One spectroscopy.
    
    Through this first year, we worked on and refined a variety of
    algorithms for BOSS target selection, as it was not clear from the
    outset that any single method could meet our scientific goals. These
    methods include:
    \begin{itemize} 
    \item The Non-parametric Bayesian Classification and Kernel
      Density Estimator \citep[KDE; ][]{Richards04, Richards09}, which
      measures the densities of quasars and stars in color-color space from
      training sets.  \citet{Richards09} showed that this was able to
      identify quasars at $2.2<z<3.5$ from SDSS photometry with an
      efficiency of $46.4\pm5.8$\%, down to a magnitude limit of $i=21.3$,
      approximately $\sim0.5$ magnitudes brighter than the BOSS limit.
    \item A likelihood approach \citep{Kirkpatrick11}, which
      determines the likelihood that each object is a quasar, given its
      photometry and models for the stellar and quasar loci.
    \item A Neural Network (NN) approach from \citet{Yeche10}, which
      takes as input the SDSS photometry and errors.
    \item A variant of the likelihood approach, which accounts for the
      observational errors more properly when determining the stellar locus,
      called ``Extreme Deconvolution'' \citep[XD; ][]{Bovy09}.
      \citet[][]{Bovy11} present full details on how the XD method can be
      used to describe a probabilistic quasar target selection technique,
      called ``XDQSO'', that uses density estimation in flux space to assign
      quasar probabilities to all SDSS point sources. XDQSO was not used in
      Year One target selection, but it did become the CORE method in Year
      Two.
    \end{itemize}
    
    Each of the methods described above has one, or more, key
    parameters; these are summarized in Table~\ref{tab:key_params}, and
    Table~\ref{tab:BOSS_TARGET_FLAGS} gives the associated bitwise target
    flags. We now describe each of these methods in turn, leaving the
    details for the cited papers.  We also introduce a variant of the NN,
    the ``Combined Neural Network'' (a.k.a. the NN-Combinator), which
    incorporates information from all the methods and produces the BONUS
    sample.  We also describe several ancillary methods of selection,
    including objects associated with FIRST radio sources
    (\S~\ref{sec:Radio_Sec2}) and repeat observations of previously known
    $z>2.2$ quasars (\S~\ref{sec:prev_known}).
   
    \begin{table*}
      \begin{center}
        \begin{tabular}{llll}
          \hline
          \hline
          Method             & Key                        & Variable name  & References \\
                                   & Parameter(s)         &  in target files & \\
          \hline
          Likelihood         & $\mathcal{P}$               & {\tt LIKE\_RATIO}      & \citet{Kirkpatrick11}\\
          KDE                   &  KDE$_{prob}$                 & {\tt KDE\_Prob}          & \citet{Richards09} \\
                                    &  {$\chi_{\rm star}^{2}$}    &  {\tt chi2\_star}                 & \citet{Hennawi10} \\
          Neural Network & $y_{NN}$                       & {\tt NN\_XNN} & \citet{Yeche10} \\
                                    & $z_{p {\rm NN}}$             & {\tt NN\_ZNN\_phot} & \citet{Yeche10} \\
          XDQSO              &   P(XDQSO)                &  {\tt QSOED\_PROB}  & \citet{Bovy11} \\
          Combined-NN  &  NN Value                  & {\tt NN\_VALUE}         & this paper \\
          \hline
          \hline
        \end{tabular}  
      \end{center}
      \caption{Key parameters for the various methods and the  
        variable name in the output target files.}
      \label{tab:key_params} 
    \end{table*}
    
    \subsection{Kernel Density Estimation and $\chi^{2}$ cuts}
    \label{sec:KDE}
    \citet{Gray03}, \citet{Gray06}, and \citet{Riegel08} describe the
    KDE classification scheme.  \citet{Richards04} and
    \citet[][]{Richards09} have applied it to the SDSS imaging data to
    produce photometric quasar catalogs with $\approx 10^6$ quasars. The
    principles of the KDE are as follows. A sample of objects of known
    classification (stars and quasars) serves as a training set, from
    which the smoothed distributions of quasar and star probability as a
    function of color are constructed. This allows one to compute the
    probability that any object of interest from the test set is a star,
    ``$\rm KDE\,star\, density$'', or quasar, ``$\rm KDE\, quasar\,
    density$'' \citep[e.g. Fig.~8 in][]{Richards09}.  Based on these
    probabilities, we define the ``KDE probability'' (see
    Fig.~\ref{fig:flowchart} and Table~\ref{tab:key_params}) as:
    \begin{equation} 
      {\rm KDE}_{Prob} = \frac{ {\rm KDE\, quasar\, density}}
      {{\rm KDE\, quasar\, density + KDE\, star\, density}}, 
      \label{eq:KDE_PROB}
    \end{equation} 
    which can be used to decide whether a given object should be
    targeted as a quasar.  As described in Section 3.5 of
    \citet{Richards09}, for our purposes, we define the quasar density
    just for those objects with $2.2 < z < 3.5$; all other quasars are
    put into the ``star'' category.  
    
    \citet{Richards04,Richards09} actually define two
    KDEs, split at $g=21$, with separate color loci (different
    ``trainings'') for the bright and faint estimations.  This
    approach crudely accounts for the very different photometric
    errors of the two sets, given that the KDE method, as implemented,
    does not take errors explicitly into account.   
   
    Roughly 45\% of objects in the KDE catalog of \citet{Richards09}
    in the ``mid-$z$'' range (i.e. the redshift range of interest to BOSS)
    are not stars \citep[Table 4, ][]{Richards09}, based on an analysis of
    the classification efficiency using clustering \citep[e.g.,
    ][]{Myers06}. In the absence of significant contamination by galaxies
    at the faint end of the KDE catalog, the KDE algorithm is thus about
    45\% efficient at the \citet{Richards09} target density of 18.6
    mid-$z$ quasars~$\mathrm{deg^{-2}}$.
    
    We need a higher efficiency for BOSS, so we have applied an
    additional cut beyond that of the Richards et al.\ papers to improve
    the efficiency of the KDE method. This cut is based on the
    $\chi^{2}_{\rm star}$ statistic introduced by \citet{Hennawi10}, which
    quantifies how far a given object is from the stellar locus:
    \begin{equation}
      \chi^{2}_{star} = \sum_{m=ugriz} 
      \frac{ [ f^{m}_{\rm data}  - A f^{m}_{\rm model} ]^{2}}   
      { [\sigma^{m}_{\rm data}]^{2} + A^{2}[\sigma^{m}_{\rm model}]^{2}     }
    \end{equation}
    where $f$ is the flux in each of the five SDSS bands ($m=ugriz$)
    for the data and for the model, $\sigma^m_{\rm data}$ is the flux
    error in each band, $\sigma^m_{\rm model}$ is the model uncertainty in
    each band, and $A$ is a normalization. Following \citet{Hennawi10},
    the stellar locus is defined by a set of $\approx$14,000 stars with
    accurate photometry from SDSS spectroscopic plates, on which all point
    sources were targeted above a flux limit of $i < 19.1$ regardless of
    color \citep{Adelman-McCarthy06}. The minimum distance to the stellar
    locus, $\chi^{2}_{star}$, can the be computed by minimizing the value
    $\chi^{2} (A, g-i)$, where $A$ is the normalization constant relating
    the data to a model, $f^{m}_{\rm data} = A f^{m}_{\rm model}$, and
    $g-i$ is the color chosen as a proxy for stellar temperature. The
    distribution of the minimum distance to the stellar locus, i.e. range
    of $\chi^{2}_{star}$, is shown in Fig.~3 of \citet{Hennawi10}.  The
    crucial strength that the $\chi^2_{star}$ cut adds to our KDE
    selection is the rejection of objects that have colors consistent with
    those of quasars, but have flux errors that make them consistent with
    the stellar locus as well.
    
    The key parameters (Fig.~\ref{fig:flowchart}) for the KDE method
    are the minimum thresholds for selection in both KDE$_{prob}$ and
    $\chi^{2}_{\rm star}$.  Early in Year One, CORE objects were selected
    solely by the KDE algorithm (Section~\ref{sec:chunky}); at that time,
    we applied a limit $\chi^{2}_{\rm star} \ge 7$.  Later, when KDE was
    no longer the CORE algorithm, we relaxed this criterion to
    $\chi^{2}_{\rm star} \ge 3$. Objects selected by the KDE method have
    the QSO\_KDE target flag set.
        
    \subsection{Likelihood Method}
    \label{sec:Kirkpatrick}
    Full details of the Likelihood method, including an in-depth
    analysis of its performance, 
    are presented in \citet{Kirkpatrick11}.  We summarize it briefly
    here. 
    
    Like KDE, the Likelihood method starts with a sample of known
    quasars, and a sample of ``Everything Else'' (EE in what follows),
    i.e., stars and galaxies, with $ugriz$ photometry and errors.  One
    defines likelihoods that a given object with fluxes $f^m$ and
    errors $\sigma^m$ ($m = ugriz$) is drawn from the quasar or
    EE catalog by summing a $\chi^2$-like statistic over the full
    training set: 
    \begin{equation} 
      {\cal L}_{\rm quasar}  = \sum_i \prod_m \sqrt{1 \over 2\,\pi (\sigma^m_i)^2}
      \exp\left( - {\left[f^m - {\rm quasar}^m_i\right]^2 \over
          2\,(\sigma^m)^2}\right)
    \end{equation}
    \begin{equation} 
      {\cal L}_{EE}  = \sum_i \prod_m \sqrt{1 \over 2\,\pi (\sigma^m_i)^2}
      \exp\left( - {\left[f^m - EE^m_i\right]^2 \over
          2\,(\sigma^m)^2}\right).
    \end{equation}
    The sums are over all objects $i$ in the training set.  By restricting
    the sum to those training-set quasars in a specific redshift range,
    one can define an equivalent likelihood that the object in question is
    in this redshift range; in Year One, this was done by summing over
    those quasars with $z > 2.2$.  Given these likelihoods, one defines a
    probability that the object is a quasar to be targeted (compare with
    equation~\ref{eq:KDE_PROB}): 
    \begin{equation} 
      \mathcal{P} =\frac{ {\cal L}_{\rm quasar}(z>2.2) / A_{\rm quasar}} 
      { {\cal L}_{EE}/ A_{EE} + {\cal
          L}_{\rm quasar}({\rm all}\ z)/ A_{\rm quasar}}, 
      \label{eq:likelihood_ratio}
    \end{equation} 
    where the $A$s normalize for the possibly different effective
    solid angles of the quasar and $EE$ training sets.  In the
    denominator, the likelihood sum is over quasars at all redshifts, not
    just those at $z > 2.2$.
    
    Like the KDE method above, this method makes use of the varying
    densities of objects in color space, and includes a $\chi^{2}$
    selection.  Note that it correctly utilizes the flux errors in
    determining whether a given object belongs to the quasar or EE class.
    Potential quasar targets can be ranked by their probability
    $\mathcal{P}$. We define a threshold ($\mathcal{P} \geq 0.234$); for
    $\mathcal{P}$ above this value, we target all objects as quasars.  The
    Likelihood method was chosen as the CORE algorithm near the end of
    Year One (section~\ref{sec:chunk7-11}).  Objects selected by the
    Likelihood method have the QSO\_LIKE target flag set.

    \subsection{Artificial Neural Network} 
    \label{sec:nn}
    We use an Artificial Neural Network (NN) at two stages of the
    selection process.  Full details of this algorithm may be found in
    \citet{Yeche10}.  As in the previous methods, we define training
    sets of known quasars, and objects that are not quasars.  
    
    For the first stage, we use the NN with 10 inputs for each object (the SDSS
    $g$-band magnitude, the five SDSS magnitude errors and the four SDSS
    colors).  The training set for non-quasars is a
    set of $\sim 30,000$ SDSS point sources from SDSS DR7
    \citep{Abazajian09}, selected over the magnitude range
    $18.0<g<22.0$ and with Galactic latitude $b \approx 45^\circ$ to
    average the effects of Galactic extinction. The 
    training set for quasars consisted of spectroscopically confirmed
    quasars from the 2QZ \citep{Croom04}, 2SLAQ 
    \citep{Croom09a}, and the SDSS \citep{Schneider10} quasar catalogs.
    
    The NN developed for target selection has four layers of
    ``neurons'' \citep[see Fig. 3 of ][]{Yeche10}.  The fourth layer
    only has one neuron, providing a single output parameter, $y_{NN}$.
    The quantity $y_{NN}$ quantifies the probability that an input
    object is a quasar, although since $y_{NN}$ can be greater than
    1, it is not a a probability in the formal sense. A photometric
    redshift estimate, $z_{p {\rm NN}}$, is also generated \citep[see Section 5 of
    ][]{Yeche10}, with a cut placed on this photometric redshift estimate,
    $z_{p \rm NN}>2.1$. Objects selected by the NN method have the 
    QSO\_NN target flag set.

    \subsection{Extreme Deconvolution} 
    \label{sec:exd} 
    Extreme deconvolution \citep[XD; ][]{Bovy09} is a method to
    describe the underlying distribution function of a series of points in
    parameter space (e.g., quasars in color space), by modeling that
    distribution as a sum of Gaussians convolved with measurement errors.
    \citet{Bovy11} apply XD to the problem of quasar target selection,
    using flux data from the SDSS DR8.  The so-called ``XDQSO'' method is
    conceptually similar to the Likelihood method, but explicitly models
    the non-uniform errors of the training set from which the quasars and
    stellar/EE loci are derived.  Indeed, the Likelihood method
    effectively double-counts the errors of the training set, since the
    {\it observed} distribution of fluxes from which the Likelihood
    training set is built is the true {\it underlying} distribution
    convolved with the uncertainty distribution.  XD avoids this
    double-counting by deconvolving the underlying distribution of the
    training set.
   
    XDQSO constructs a model of the distribution of the fluxes of
    stars and quasars in different redshift ranges based on training
    samples of known stars and quasars. XDQSO then builds a model of the
    relative-flux distribution as a mixture of 20 Gaussian components and
    fits this model to the training data, taking the heteroscedastic
    nature of the SDSS flux uncertainties fully into account. The XD model
    for the relative-flux distribution is fit in narrow bins in $i$-band
    magnitude and combined with an apparent-magnitude dependent prior
    based on star counts in Stripe 82 and the \citet{HRH07} quasar
    luminosity function. The probability for an object to be a
    mid-redshift quasar ($2.2 < z < 3.5$) is given by the ratio between
    the number density of mid-redshift quasars and that of stars plus all
    quasars at the object's fluxes (in the spirit of
    equation~\ref{eq:likelihood_ratio}) . The probability that a given
    object is a mid-$z$ quasar is then:
    \begin{multline}
      P(\mathrm{QSO}_{\mathrm{midz}} | \{f^m\}) \propto \\
      P(\{f^m/f^i\} | \mathrm{QSO}_{\mathrm{midz}})\,
      P(f^i | \mathrm{QSO}_{\mathrm{midz}})\,P(\mathrm{QSO}_{\mathrm{midz}})\,,
    \end{multline}
    where $m$ indexes the fluxes and $f^i$ is the SDSS $i$-band
    flux. The first factor on the right is given by the XD model for the
    relative-flux (i.e., color) distribution of quasars, while the second
    and third factors are obtained from the quasar luminosity
    function. The underlying relative-flux distribution is convolved with
    the object's flux uncertainties before evaluation. The expressions for
    stars and high/low redshift quasars are similar. Probabilities are
    normalized assuming that these classes exhaust the possibilities
    ($P(\mathrm{QSO}_{\mathrm{midz}}) + P( \mathrm{QSO}_{\mathrm{hiloz}})
    + P(\mathrm{star}) = 1$). Objects are ranked on their mid-redshift
    quasar probability for targeting.
    
    Since XDQSO target selection properly takes the flux uncertainties
    into account both in the training and the evaluation stage, it can be
    trained and evaluated on data of low signal-to-noise ratio.  It can
    also incorporate data from surveys other than SDSS in a
    straightforward way, as we describe for near-infrared and ultraviolet
    surveys below.  The performance of XDQSO, using Stripe 82 data is
    given in \citet{Bovy11} and its performance in Year Two will be
    described in a future paper.  The catalog of SDSS objects selected by
    XDQSO is available through the SDSS-III DR8 Science Archive
    Server\footnote{\href{http://data.sdss3.org/sas/dr8/groups/boss/photoObj/xdqso/xdcore/}
      {http://data.sdss3.org/sas/dr8/groups/boss/photoObj/xdqso/xdcore/}}.
    
    The XDQSO method was not used during Year One, but we then set,
    {\bf and fixed}, XDQSO as CORE for Year Two and the remainder of the
    BOSS. In Section~\ref{sec:discussion} we detail how to replicate the
    CORE selection using XDQSO for the BOSS quasars. Objects selected by
    the XDQSO method have the QSO\_CORE\_MAIN, and sometimes the
    QSO\_CORE\_ED, target flag set (see Section~\ref{sec:discussion}).
    
    \subsubsection{The UKIRT Infrared Deep Sky Survey}
    \label{sec:UKIDSS}
    \citet{Lawrence07} presents an overview of the United Kingdom
    Infrared Telescope (UKIRT) Infrared Deep Sky Survey (UKIDSS). The
    UKIDSS is a collection of five surveys of different coverage and depth
    using the Wide-Field Camera (WFCAM, \citealt{Casali07}) on UKIRT.
    WFCAM has an instantaneous field of view of 0.21 deg$^{2}$, and the
    various surveys employ up to five filters, {\it ZYJHK}, covering the
    wavelength range 0.83-2.37$\mu$m.  The photometric system and
    calibration are described in \citet{Hewett06} and \citet{Hodgkin09},
    respectively. The pipeline processing is described in Irwin et
    al. (2011, in prep.) and the WFCAM Science Archive (WSA) by
    \citet{Hambly08}.  The astrometry is accurate to 0.1\arcsec.
    
    The UKIDSS Large Area Survey (ULAS) aims to map $\sim 4,000$
    deg$^{2}$ of the Northern Sky, which, when combined with the SDSS,
    produces an atlas covering almost an octave in wavelength. The target
    point-source depths of the survey are $Y=20.3, J=19.5, H=18.6, K =
    18.2$ (Vega); the ULAS does not image in the WFCAM $Z$-band.  Unlike
    the SDSS, the ULAS multiband photometry is not taken simultaneously
    \citep[e.g. Sec. 5.2 of ][ Sec. 4.2]{Dye06, Lawrence07}, so the four
    bands have different coverage maps, with the {\it H} and {\it K} bands
    obtained together, and {\it Y} and {\it J} obtained separately. For
    example, the ULAS
    ``DR8Plus''\footnote{\href{http://surveys.roe.ac.uk/wsa/dr8\_las.html}
      {http://surveys.roe.ac.uk/wsa/dr8\_las.html}} coverage is 2,670
    deg$^{2}$, 2,685 deg$^{2}$, 2,795 deg$^{2}$ and 2,810 deg$^{2}$, in
    {\it Y}, {\it J}, {\it H} and {\it K} respectively.
    
    We use the UKIDSS NIR photometry to improve target selection in
    two complementary techniques.  The first is to classify quasars by
    their ``K-excess'' \citep[``KX''; e.g., ][]{Warren00, Croom01KX,
      Sharp02, Chiu07, Maddox08, Smail08, Wu10, Peth11}. The power-law
    quasar SED has an excess in the {\it K}-band over a blackbody stellar
    SED, allowing quasars to be identified (and stars rejected) that would
    be normally excluded from an optical color-only quasar selection
    algorithm - even for dust reddened quasars. \citet{Peth11}
    investigated the KX method and provided an SDSS-UKIDSS matched quasar
    catalog. For BOSS, KX-selected objects were selected early in
    commissioning and had the QSO\_UKIDSS target flag set.  However, the
    very low yield (from admittedly a small target sample) caused us to
    drop this method.
    
    The second method of inclusion of NIR photometry is to improve
    quasar classification, and of particular importance for BOSS,
    photometric redshift estimation, in the XDQSO method.  Including the
    NIR flux information removes many of the optically-based redshift
    degeneracies known for quasars (see Bovy et al. 2011b, in prep.).
    Models were trained for SDSS-only fluxes and various combinations of
    SDSS+UKIDSS data.  $z\sim2.5$ quasars have $(i-K)\sim2.1$ \citep[e.g.,
    ][]{Peth11}; thus given the BOSS quasar survey magnitude limit of
    $i\sim21.8$, the ULAS catalog is too shallow to guarantee 5$\sigma$
    detections of all sources.  We therefore measure aperture magnitudes
    in the UKIDSS images at the positions of SDSS object counterparts;
    even low-significance detections can be used by XDQSO.  Bovy et
    al.~(2011b, in prep) give technical details.  The SDSS (optical) only
    model is used by XDQSO to generate targets for CORE, where the upper
    limit of the mid-$z$ bin is $z=3.5$.  For BONUS, the SDSS+UKIDSS model
    is used to generate targets as an input to the NN-Combinator with an
    upper limit of the mid-$z$ bin extended to $z=4.0$.  This was
    implemented in BONUS from the middle of Year Two (Chunk 16) onwards,
    with significant gains in the yield of $z > 2.2$ quasars.

    \subsubsection{GALEX: The Far and Near UV}
    \label{sec:GALEX}
    The Far (1350 - 1750\AA) and Near (1750 - 2750\AA) ultraviolet
    (FUV and NUV respectively) photometry from the GALEX Small Explorer
    mission \citep{Martin05} also provide information
    that could help discriminate between hot stars and $z\sim0.8$ quasars,
    both of which should have considerably more flux in the UV than a
    $z>2$ quasar because of Ly$\alpha$ absorption along the line of
    sight in the latter.
    
    We have trained the XDQSO technique on SDSS, UKIDSS and GALEX
    input data. Thus we can now perform 11-dimensional quasar target
    selection using the {\it FUV/NUV}$ugrizYJHK$ bands.  The relevant
    GALEX surveys are relatively shallow, e.g. $m_{\rm UV} 
    \approx 20.5$ AB, so most potential BOSS quasar targets are not
    detected at high significance.  Despite this, our tests
    (detailed in Section~\ref{sec:results}) confirmed that GALEX measurements---even at
    low significance---do help with target selection.
    
    We had access to medium-deep GALEX data
    on Stripe 82 at the start of Year Two, when we targeted the Stripe
    for the second time (Chunk 11; \S~\ref{sec:chunk7-11}).  We therefore incorporated 
    the GALEX FUV and NUV fluxes in the XDQSO probabilities. 

    \begin{figure}
      \centering
      \includegraphics[height=6.0cm,width=8.5cm]
      {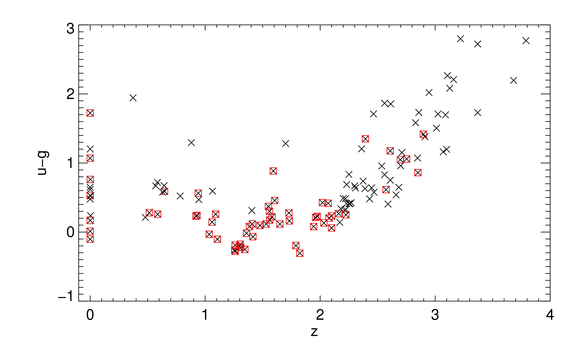}
      \caption[Redshift versus $(u-g)$ color for FIRST-only targets.]
      {Redshift versus $(u-g)$ color for BOSS FIRST quasar targets.
        Objects from the BOSS commissioning were either targeted
        by FIRST, and also a optical selection, (black) crosses, or, they
        were targeted {\it only} as FIRST sources, (red) squares. 
        These early findings inspired our $(u-g)>0.4$ cut to minimize
	contamination from $z < 2.2$ quasars.}
      \label{fig:red_FIRST} 
    \end{figure}
    \subsection{Radio Selection}~\label{sec:Radio_Sec2}
    As in the SDSS-I/II quasar survey, objects that are detected in
    the FIRST radio survey \citep{Becker95} are also incorporated in
    target selection.  Radio stars are rare, thus most radio sources with
    faint, unresolved optical counterparts are quasars. Optical stellar
    objects with $g \leq 22.00$ or $r \leq 21.85$ which have FIRST
    counterparts within 1\arcsec\ are considered as potential quasar
    targets, irrespective of the radio morphology.
    
    In the early BOSS commissioning data (\S~\ref{sec:chunky}), we
    simply selected all such radio matches.  This approach targeted a
    substantial number of quasars with $z < 2.2$, and thus we placed an
    additional color cut, $(u-g)>0.4$, to exclude UV--excess sources at
    lower redshift (Fig.~\ref{fig:red_FIRST}). Thus the QSO\_FIRST flag
    designates objects with $(u-g)>0.4$ that matched a FIRST source. Bluer
    FIRST sources are not rejected outright, but are required to pass one
    of the regular optical color selections to be
    selected. Section~\ref{sec:chunky} describes when in Year One this
    $(u-g)>0.4$ cut was implemented.

    \subsection{Previously Known Objects}	
    \label{sec:prev_known}
    The density of $z>2.2$ quasars known before BOSS started was $\sim
    2$ objects deg$^{-2}$.  Given the superior throughput of the BOSS
    spectrographs over those of SDSS-I/II, we decided to re-observe these
    objects for improved Ly$\alpha$ forest clustering signal. Moreover,
    this allows vital checks of survey quality and uniformity, and the
    data can be used to study the spectroscopic variability of quasars. We
    thus target previously known spectroscopically confirmed $z>2.15$
    quasars from the literature.  We include such objects as targets if
    they match a point source in the target imaging to within 1.5\arcsec,
    or if they match a point source in the target imaging to within
    2\arcsec\ and match the magnitude of that object to within 0.5.
    
    The catalogs of previously known quasars we use include the SDSS DR7
    quasar catalog \citep{Schneider10}, the 2SLAQ quasar catalog
    \citep{Croom09a}, the 2QZ survey \citep{Croom04}, the AAT-UKIDSS-SDSS
    (AUS) survey (Croom et al., in prep), and the MMT-BOSS pilot survey
    (Appendix~\ref{sec:MMT}).
    
    To compare and check our moderate resolution spectra of generally
    fainter quasars to those taken by 10m class telescopes using
    high-resolution spectrographs (e.g. KECK-HIRES and VLT-UVES), we also
    mined the data archives (the
    NED\footnote{http://nedwww.ipac.caltech.edu/}, the Keck Observatory
    Archive\footnote{http://www2.keck.hawaii.edu/koa/public/koa.php} and
    the ESO Science Archive Facility\footnote{http://archive.eso.org/})
    and added those quasars with $z>2.15$ that were not included from the
    above catalogs.
    
    The full sample of known quasars contains $\sim 18,000$ $z>2.15$
    objects.  We assign those objects in the BOSS footprint the
    QSO\_KNOWN\_MIDZ flag and give them highest targeting
    priority in tiling \citep{Blanton03}.  

    We also veto previously known low ($z<2.15$) redshift quasars
    identified from the SDSS-I/II, 2QZ, 2SLAQ and MMT surveys,
    labeling them with the QSO\_KNOWN\_LOHIZ target flag and
    never assigning them
    spectroscopic fibers\footnote{The name for this flag, QSO\_KNOWN\_LOHIZ, 
      is misleading, in that it does not explicitly flag high-$z$ quasars.}. 
    We are confident that we are not inadvertently rejecting any real 
    $z>2.2$ quasars, since the vast majority of these objects were visually
    inspected and identified in the SDSS, 2QZ and MMT surveys
    \citep[][]{Schneider10, Croom05}.  A veto of objects with known
    stellar spectra, again from the SDSS-I/II, 2QZ, 2SLAQ and MMT surveys,
    was not implemented until Chunk 5, because we were not initially
    confident that shallower surveys, at their faint end, would have
    sufficient S/N to correctly identify stars, and that our initial
    matching procedures were not discarding some quasars of utility to
    BOSS. 

    \begin{figure*}
      \centering
      \includegraphics[height=8.0cm,width=16.0cm]
      {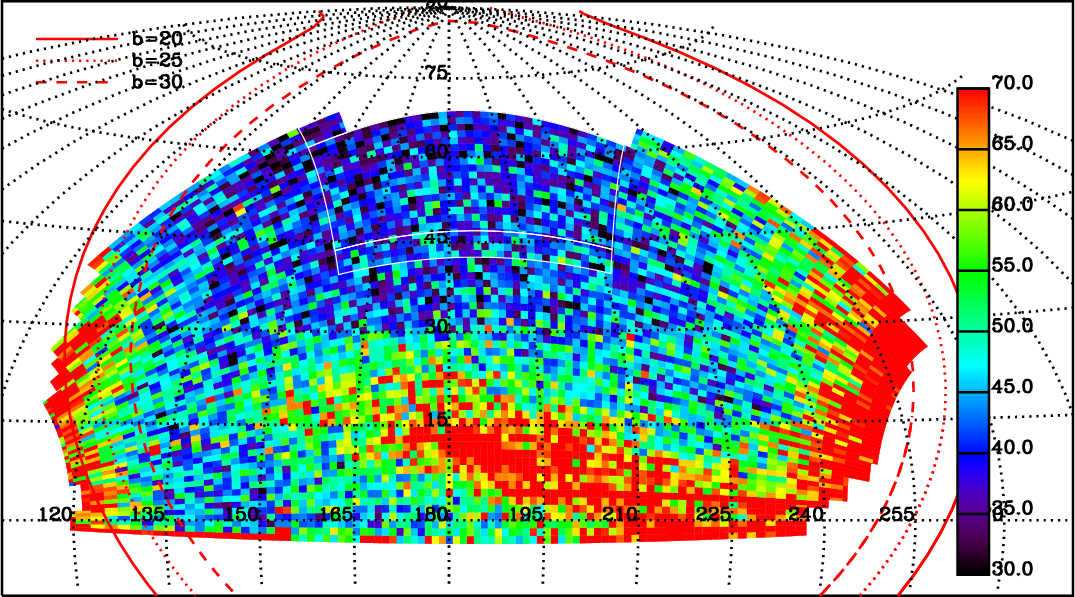}
      \caption[The full-blown NGC ``HeatMap''.]
      {The BOSS quasar target surface density in
	Equatorial coordinates in the NGC, from
        a run of the BOSS QTS with a selection made by combining the three
        Year One methods, KDE, Likelihood and NN, in such a way that the
        average target density over the full given NGC area was $\sim60$
        quasar targets deg$^{-2}$. The color indicates the local number
	density of targets per square degree.  The tidal stream of the
	Sagittarius dwarf spheroidal galaxy is prominent in the region
	$180^\circ<\alpha<240^\circ$, and $0^\circ< \delta <+15^\circ$.
        The white lines show the ``Blind Test Area'', described in \S~\ref{sec:blind_test}.}
      \label{fig:NGC_heatmap}
    \end{figure*}
    \subsection{Combinations of Methods}
    \label{sec:combine}
    Combining results from several of the methods described above in
    target selection requires a method to merge the (overlapping) ranked
    lists from these methods into a single ranked catalog.  The challenge
    is shown in Fig.~\ref{fig:NGC_heatmap}, which shows the surface
    density of the union of those objects selected by the KDE, Likelihood,
    and NN methods with no further refinement, to yield an average target
    density of $\sim60$ targets deg$^{-2}$.  The tidal stream of the
    Sagittarius dwarf spheroidal galaxy \citep{Ibata95,Belokurov06} is
    quite striking in this figure, spanning $180^\circ<\alpha<240^\circ$
    and $0^\circ< \delta <+15^\circ$.  The target density in
    Figure~\ref{fig:NGC_heatmap} varies from 35 to 70 deg$^{-2}$.
   
        \subsubsection{Tuning and Ranking}
        \label{sec:ranking_method}
        In the early stages of commissioning, the target density was
        tuned to 80 deg$^{-2}$ using the KDE method (and its $\chi^{2}_{\rm
          star}$ parameter).  The three main Year One algorithms (KDE,
        Likelihood and NN) were then trained on regions where very early BOSS
        spectroscopy was obtained. This was mainly in Stripe 82 (observed in
        Chunk 1; see \S~\ref{sec:chunk1}), but also some of Chunk 2, yielding
        $\sim$650 $z>2.2$ quasars from $\sim$2000 targets.  For these initial
        tests, the limiting parameters of the KDE, Likelihood and NN methods
        were chosen to give target densities of 80 deg$^{-2}$ each, and each
        produced a ranked list (based on the value of their respective output
        probability parameter) of targets. These three lists were then
        combined to generate the list of the 60 targets deg$^{-2}$ most likely
        to be high-$z$ quasars, finding the interleaving (without repeating
        objects selected by more than one algorithm) of the combined list of
        objects that led to the highest yield of $z > 2.2$ quasars.  That is,
        we first took the first-ranked object from each of the three methods,
        then the second-ranked object, and so on, of course not
        double-counting objects which were selected by more than one method.
        Each of these objects is associated with a ranking parameter (as
        listed in Table~\ref{tab:key_params}), giving us a relative ranking of
        the three methods which we can use for combining other data in which
        one didn't know a priori which objects were actually $z > 2.2$
        quasars.  This technique was tested by splitting the initial data in
        half and running the ranking algorithm to find the thresholds required
        for each of the three methods. Observed targets from the second half
        of the data were also chosen using these calculated thresholds, and
        the yield of $z > 2.2$ quasars was consistent. The result of the
        combined rankings was to allocate targets to the three methods in
        approximately equal quantity and priority.

        \subsubsection{NN-Combinator}
        \label{sec:combinator}
        We found that the outputs of the three methods could be used
        as inputs into a neural net to improve the yield of $z > 2.2$ quasars.
        We refer to this approach in what follows as the NN-Combinator. This
        approach can easily be expanded to allow input from additional
        selection techniques.
        
        The key output parameter of the NN-Combinator is designated as
        the ``NN value'', which is, by design, allowed to change from chunk to
        chunk.  The NN-Combinator used the data from Stripe 82 obtained by
        BOSS (Chunk 1, see Section~\ref{sec:chunk1} below) as an input
        training set. The NN-Combinator was the selection method for BONUS
        from Chunk 7 onwards in the survey, drawing on the inputs of KDE,
        Likelihood, and NN. This replaced the interleaving method described in
        \S\ref{sec:ranking_method}.
        
        In Year Two, with the advent of the XDQSO method, we added the
        results of this method to the NN Combinator.  In particular, near the
        end of Year Two, we used a version of XDQSO that included data from
        UKIDSS (\S~\ref{sec:UKIDSS}) which selected targets to $z = 4$; the
        version of XDQSO used for CORE used SDSS single-epoch photometry only
        and did not incorporate UKIDSS data.

    \subsection{Rationale and Summary}
    As the above makes clear, BOSS quasar selection has been through a
    complex series of changes during its first two years.  Here we recall
    the reasons for this complexity and summarize the main points of this
    history.
    
    BOSS quasar target selection is complex because
    \begin{itemize}
    \item{for the survey's defining science goal, measurement of BAO 
        in the Ly$\alpha$ forest, the primary requirement is a high surface density
        of quasars in the relevant redshift range, not simplicity or homogeneity
        of selection,}
    \item{selection of quasars in the desired redshift range from single-epoch
        SDSS imaging is difficult because of proximity to the stellar locus
        and substantial photometric errors near the magnitude limit for 
        BOSS selection,}
    \item{pre-BOSS quasar samples provided inadequate training sets
        in our desired magnitude and redshift range, so the quasars we
        discovered in this first year allowed us to refine our algorithms as
        the year proceeded. }
    \end{itemize}
    Roughly speaking, the effective survey volume for measurement of
    Ly$\alpha$ forest clustering
    is quadratic in the number of quasars, so even modest gains in efficiency
    have a significant science impact.
    
    As discussed in \S\ref{sec:core}, the goal of CORE selection {\it is}
    to provide a homogeneously selected sample suitable for quasar science.
    Ideally, we would have frozen the CORE algorithm at the very beginning
    of BOSS, but the higher imperative of maximizing efficiency has led
    us to alter CORE as our algorithms improved.  We started by using
    KDE+$\chi^2$ as the CORE algorithm but switched to Likelihood based on its 
    greater flexibility and simplicity.  Finally, we switched from
    Likelihood to XDQSO based on its better performance (at a level of $\sim$one
    additional high-$z$ quasar deg$^{-2}$).  The chunk-by-chunk history
    of these changes is given in \S\ref{sec:chunky} below.
    We intend to maintain a fixed CORE algorithm for Years $2-5$ of the
    survey, and for many purposes we anticipate that completeness corrections
    will allow use of Year One data in statistical studies of the 
    quasar population (see \S\ref{sec:discussion}).
    
    Beyond CORE, we use whatever combinations of data and methods can 
    maximize our targeting efficiency, including known quasars, FIRST
    candidates, and the BONUS sample.  Because the methods described in
    \S\S\ref{sec:KDE}-\ref{sec:exd} have complementary strengths, we draw
    on all of them in creating the BONUS sample.  We have tried different
    methods of forming a combined BONUS list during the first year,
    and we have now settled on the NN-combinator (\S\ref{sec:combinator})
    as our primary tool for doing so.  The individual methods feeding
    into the NN-combinator use co-added SDSS photometry where it is
    available in overlap regions, in contrast to CORE, which relies
    on single-epoch photometry to ensure uniformity.  Auxiliary data
    such as UKIDSS and GALEX photometry are fed into the XDQSO
    selection, which in turn is fed into the NN-combinator.

    \begin{figure*}
  \centering
  \includegraphics[width=14.0cm]
  {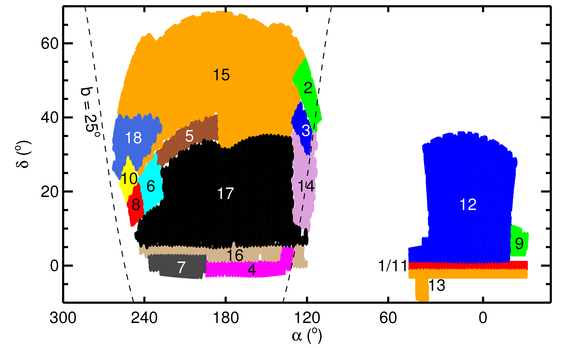}
  \caption[]
  {The targeting footprint for the SDSS-III:BOSS Ly$\alpha$ forest/Quasar
    Survey.  The various chunks are indicated by different colors. 
    Chunks 16, 17 and 18 lie within the footprint of Chunk 15. 
    The full targeting footprint is 10,200 deg$^{2}$, with a
    total of $\approx$430,000 tiled targets.  Roughly $\sim$150,000
    of these targets will have spectra by the end of Year Two observations. 
    The
    global Year One quasar target density is 60.4 targets deg$^{-2}$, and
    the mean target density over all chunks shown is 47.9 targets
    deg$^{-2}$.  The dashed line is at Galactic latitude $b
    =25^{\circ}$.}
  \label{fig:SLC2010_ADM_targets}
\end{figure*}
   
\begin{table*}
  \begin{center}
    \begin{tabular}{lcc rrc ccc}
      \hline
      \hline
      Area    & RA (2000)     & Dec (2000)     & Area          & Total \#                                      &   Galactic        & quasar target density & Method for  \\
      Name  & Range & Range & deg$^2$     & targets (tiled)                            &   latitude cut?   & deg$^{-2}$ (tiled)     & CORE             \\
      \hline
      Chunk 1   & 317.0 -   45.0$^{a}$  &  $-1.25$ - $+1.25$   & 219.93 & 19,205 (18,657) & no  & 87.3 (84.8)   & KDE$^{b}$    \\ 
      Chunk 2   & 108.9 - 131.0  &  35.6 - 56.2   & 143.66 & 11,337 (11,024)         &  no   & 78.9 (76.7)    & KDE   \\ 
      Chunk 3   & 115.7 - 132.8  &  28.8 - 44.4   & 107.34 &   9,476   ( 6,949)       &  $b>25^{\circ}$ & 88.3 (64.7)    & ---$^{d}$  \\ 
      Chunk 4   & 128.7 - 195.0  &  -3.3 -   5.0    & 306.50 & 32,750 (20,679)      &  $b>25^{\circ}$ & 106.9$^{c}$ (67.5)   & ---$^{d}$ \\
      Chunk 5   & 185.0 - 232.2  &  26.2 - 40.7   & 245.82  & 18,533 (13,418)     & no & 75.4  (54.6)  &  ---$^{d}$   \\ 
      Chunk 6   & 225.4 - 244.9  &  13.5 - 30.53 & 186.13 &  19,304 (13,130)     & no & 103.7 (70.5)    &  ---$^{d}$ \\  
      Chunk 7   & 194.0 - 237.9  & -3.6 -    3.2   &  257.01 & 10,783 ( 9,596)      & no & 42.0  (37.3)       &  Likelihood \\ 
      Chunk 8   & 240.2 -  253.1  &  10.5 -  22.9 &   97.82   &  4,004 ( 3,500)      & no  & 40.9 (35.8)        & Likelihood  \\ 
      Chunk 9   & 316.3 - 330.0  &    2.5 -  11.1 &   97.54   &  3,870 ( 3,360)       & $b<-25^{\circ}$ & 39.7 (34.4)    & Likelihood \\ 
      \hline
      {\bf Year One}  &  &              & {\bf 1661.75}  & {\bf 132,923 (100,313)}  &              &     &        {\bf 80.0 (60.4)}  &  \\
      \hline
      Chunk 10 & 245.0 - 258.6  &  17.1 - 30.0  &   91.14 &   3,661 (  3,325)           & no           & 40.2 (36.5)    &  Likelihood     \\
      Chunk 11 & 317.0 -   45.0  &  $|$1.25$|$   & (219.84) &  8,820 ( 8,432)           & no      & 40.1 (38.4)   & variability$^{e}$    \\ 
      Chunk 12 & 324.6 -   45.1  &  0.55 - 36.2  & 2075.9 &  84,038 ( 77,447)         & no & 40.5 (37.3)    &  Likelihood/XDQSO     \\
      Chunk 13 & 317.0 -   45.0  &  -9.9 - -0.8  &  281.7 &  11,051 ( 10,072)           & no & 39.2 (35.8)    &  Likelihood/XDQSO     \\
      Chunk 14 & 111.8 - 131.5  &   9.0  - 36.3  &  347.43 & 14,165  (13,479)          & no & 40.8 (38.8)    &  XDQSO     \\
      Chunk 15 & 118.9 - 263.9  &  -0.8 - 68.7  &  5743.5 & 233,530 (220,029        & no & 40.7 (38.3)    &  XDQSO     \\
      Chunk 16 & 118.9 - 247.3  &  -0.8 - 35.6  &  (3108.3) & [128,250 (120,905)]   & no & 41.3 (38.9)    &  XDQSO     \\
      Chunk 17 & 118.9 - 247.3  &    4.4 - 35.6  &  (2742.4) & [116,471 (107,562)]  & no & 42.5 (39.2)    &  XDQSO     \\
      Chunk 18 & 226.9 - 263.9  &  23.1 - 41.1  &   (337.20) &  [13,372 ( 12,699)]   & no & 39.7 (37.7)    &  XDQSO     \\
      \hline
      {\bf Year Two}     &  &      & {\bf 8539.65}  & {\bf 355,265 (332,784)}  &     &     &     {\bf 41.6 (39.0)}  &  \\
     \hline  
     \hline
      {\bf Total }$^{*}$  &  &  & {\bf 10,201.4}  & {\bf 488,188 (433,097)}  &      &    &      {\bf 47.9 (42.5)}  & \\
      \hline
      \hline
    \end{tabular}  
  \end{center}
  \caption[Details of the DR9 chunks.]
  {Details of the 18 chunks targeted for the first two years
    of BOSS observations. Spectra from each of the 18 chunks will 
    be taken during the first two years, but only an area of
    $\sim$3000 deg$^2$
    will be covered for spectroscopy.  However, we plan to observe {\it all} 
    of the Year One chunks by the end of the Year Two observations. 
    $^a$ The RA and
    Dec ranges give the extremities of each chunk area, and thus do not
    indicate the coordinates of the corners of the chunk footprints.  Chunks 16, 17
    and 18 lie within the area of Chunk 15, hence their areas and targets
    are not counted towards the total. \\ $^{b}$From Single-epoch data. \\
    $^{c}$Chunk 4 uses imaging data in which problems with the
    $u$-band data lead to an excess target density ($>106$ targets deg$^{-2}$). \\ $^{d}$ A ranking
    scheme was used; for Chunks 3-6, CORE included a combination of
    NN, Likelihood, and KDE targets (\S~\ref{sec:chunk3-6}).}
  \label{tab:chunk_targets}
\end{table*}
\section{BOSS Quasar Target Selection for Years One and Two, Chunk by Chunk}
\label{sec:chunky} 
BOSS is a five year project running from 2009 August to the end of
June 2014. Starting in 2009 September, target selection commissioning
(both for the galaxies and quasars) ran alongside commissioning of the
new hardware and reduction software. The hardware commissioning was
essentially complete by 2009 December (data taken earlier were
therefore not of survey quality), but QTS commissioning continued
through 2010 April; during this period the quasar target density was
set appreciably higher (60 or 80 deg$^{-2}$), than for the nominal
survey (40 deg$^{-2}$).  The bulk of the Year One observations from
MJD=55176 (2009 December 11) to MJD=55383 (2010 July 6) were thus QTS
commissioning data.

The targeting chunks into which the Year One and Year Two data were
divided are detailed in Table~\ref{tab:chunk_targets} and
Figure~\ref{fig:SLC2010_ADM_targets}. By the end of Year Two, we had
run target selection over the whole 10,000 deg$^{2}$ imaging
footprint, resulting in $\approx 430,000$ tiled targets.  This target
list is not necessarily final -- if we obtain data that could improve
our target selection efficiency in later years of BOSS, we will rerun
target selection for areas that have not yet been observed.  Spectra
collected during Years One and Two will constitute the DR9, and will
include 150,000 quasar targets, a third to half of which will be
$z>2.2$ quasars.  By the end of Year Two, we will have observed all of
the Year One chunks.  The BOSS quasar target selection changed from
chunk to chunk during the first year, as we gathered data and refined
our algorithms.  These changes in the algorithms are detailed in the
following subsections.

    \subsection{Chunk 1}
    \label{sec:chunk1}
    The first area that we targeted and observed for BOSS was SDSS
    Stripe 82, along the celestial equator in the Southern Galactic Cap.
    The target field covered $317.0^\circ \leq \alpha_{\rm J2000} \leq
    45.0^\circ$, $-1.25^\circ \leq \delta_{\rm J2000} \leq 1.25^\circ$,
    for a total area of 220 deg$^{2}$ (smaller than the $\sim$300
    deg$^{2}$ imaging coverage on the Stripe).
    
    The KDE method, based on single-run data and with a cut at
    $\chi^{2}_{\rm star} \geq 7.0$, was used as the CORE (QSO\_CORE)
    selection for Chunk 1. The KDE method was one of the techniques used
    for BONUS, (QSO\_BONUS) with targets chosen using the coadded data
    described by Section~\ref{sec:input_phot}, and given the flag
    QSO\_KDE\_COADD.  Coadded data were not used in later chunks, thus the
    QSO\_KDE\_COADD flag was used only for Chunk 1.  In Chunk 1, with the
    benefit of coadded data, the quasar and stellar loci were better
    defined than in the standard one-epoch SDSS data. Hence there was far
    more overlap between the samples of sources targeted by all of the
    methods, freeing fibers to be placed on lower-priority KDE targets.
    As most of these lower-priority targets proved to be stars, the
    overall efficiency of selection of the KDE method is thus quite low in
    Chunk~1.
    
    Likelihood targets were selected at a target density of $\sim$35
    targets deg$^{-2}$ using a threshold ${\mathcal P} = 0.10$, Neural
    Network targets at $\sim$20 deg$^{-2}$ with a threshold $y_{NN} =
    0.65$, and KDE targets using the coadded data at $\sim$50 deg$^{-2}$.
    The density of the coadded KDE targets was tuned on a {\it second}
    $\chi^{2}_{\rm star}$ value calculated from coadded data, to obtain
    the required total of 80 targets deg$^{-2}$ total across all
    methods. This second $\chi^{2}_{\rm star}$ parameter is dependent on
    right ascension, but is always $>4.0$.
    
    The final Chunk 1 target densities were approximately 7, 2, 20,
    and 60 targets deg$^{-2}$ for the Known, KX-selected
    (\S~\ref{sec:UKIDSS}) CORE, and BONUS, respectively (with overlap
    between these categories).  At tiling, {\it all} quasar targets were
    given priority over all other BOSS targets (such as galaxies) for
    Chunks 1 and 2. The tiling priorities for all the chunks are given in
    Appendix~\ref{sec:flowchart_YearOne} (Table~\ref{tab:chunk_targets2}).
    
    About 100 deg$^2$ of Chunk 1 was observed following hardware
    commissioning, i.e., after MJD 55176.  Stripe 82 was re-observed in
    2010 Fall as part of Chunk 11 (\S~\ref{sec:chunk7-11}).
    
    \subsection{Chunk 2}
    For this chunk in the NGC (Figure~\ref{fig:SLC2010_ADM_targets}),
    the targeting algorithms were similar, but not identical, to Chunk 1,
    as coadded photometry was not available. The surface density of known
    quasars was lower than in Chunk 1 (since Stripe 82 contains more
    extensive spectroscopy from prior surveys), and with no UKIDSS
    coverage, there were no KX-selected targets. Unresolved optical
    objects that had a match to {\it any} FIRST source
    (\S~\ref{sec:Radio_Sec2}) were included, and given the target flag
    QSO\_FIRST (bit 18).  The CORE method remained the KDE. The Chunk 2
    target densities were approximately 2, 2, and 20 deg$^{-2}$ for the
    Known, radio-selected and CORE objects, respectively.
    
    Objects from the Likelihood, NN and KDE methods were targeted for
    BONUS, using single-epoch data to achieve $\sim$35, 20 and 25 targets
    deg$^{-2}$, respectively. As in Chunk 1, the KDE was tuned on the
    $\chi^{2}_{\rm star}$ parameter to obtain a total of 80 targets
    deg$^{-2}$ over all methods. In Chunk 2, flux errors are larger than
    in Chunk 1, due to the use of single-epoch data. Thus, the stellar
    locus is expanded and there is far less overlap between the targets
    chosen by various methods. The target density of QSO\_BONUS sources is
    thus approximately halved in Chunk 2.
    
    As this chunk used single-epoch data with its larger photometric
    errors, the thresholds for the target selection algorithms were
    modified as follows, giving the target densities above:
    \begin{itemize}
    \item The Likelihood Probability threshold, $\mathcal{P}$, was
      changed from 0.10 to 0.24;
    \item The NN probability parameter, $y_{NN}$, was changed from
      0.65 to 0.70;
    \item The KDE algorithm was retrained, using all available quasar
      spectroscopy to date.
    \end{itemize} At this stage, the list of quasars with
    high-resolution spectroscopy (Section~\ref{sec:prev_known}) were added
    to the database of known quasars, although few lie within the
    boundaries of Chunk 2.
    
    \subsection{Chunks 3, 4, 5 and 6}
    \label{sec:chunk3-6}
    We already had our initial spectroscopic results in hand from
    $\sim$20 plates from Chunks 1 and 2 when we identified targets in
    Chunk 3, and we used these results to refine our algorithms.  In
    particular, we rejected FIRST sources with $(u-g)<0.4$, greatly
    decreasing contamination from $z<2.2$ quasars, but decreasing the
    number of FIRST $z>2.2$ objects by only 10\%.  The resulting FIRST
    target density drops to $\sim$1-2 deg$^{-2}$, 40\% of which turn out
    to be $2.2 < z < 3.5$ quasars (see Fig.~\ref{fig:red_FIRST}).
    
    In the first two chunks, we found that only 1 {\it new} bright
    ($i\leq17.7$) $z > 2.2$ quasar had been discovered from 486 bright
    targets. Thus, a bright limit of $i>17.8$ was set to reduce stellar
    contamination at the bright end. Due to the proximity of Chunk 3 to
    the Milky Way, we also imposed a Galactic latitude cut of
    $b>25^{\circ}$.
    
    There was a change in the target density and methodology from
    those in Chunks 1 and 2. For Chunks 3, 4, 5 and 6, the ranking method
    described in Section~\ref{sec:ranking_method} was adopted, allowing us
    to combine Likelihood, KDE, and NN for CORE at 20 targets deg$^{-2}$.
    All remaining targets, to a total density of 60 deg$^{-2}$, were
    designated as BONUS.  To monitor the CORE and BONUS changes, two new
    target flags, QSO\_CORE\_MAIN (flag bit 40) and QSO\_BONUS\_MAIN (flag
    bit 41)\footnote{Now requiring Long64, or ``LL'' integer type.} were
    introduced.
    
    The final target densities for Chunks 3, 4, 5 and 6 were 2 and 1
    targets deg$^{-2}$ for Known quasar and FIRST targets,
    respectively. The CORE target density was $\approx$ 19, 19, 16, 17
    deg$^{-2}$ in the four chunks respectively, and the BONUS target
    density was roughly 40 deg$^{-2}$.  To provide a more uniform galaxy
    sample, galaxy targets were given precedence over quasar targets in
    tiling (see Appendix~\ref{sec:flowchart_YearOne} and
    Table~\ref{tab:chunk_targets2}).
   
    \subsection{Chunks 7 - 11}
    \label{sec:chunk7-11}
    Chunks 7, 8 and 9 were the first chunks which were targeted at the
    nominal survey target density of 40 quasar targets deg$^{-2}$.  The
    area covered by Chunk 9 in the SGC was not in the original SDSS
    survey, and target selection was done from the DR8 imaging
    \citep{Aihara11}, a region of sky where there were no previously known
    $z > 2.2$ quasars in our catalog (\S~\ref{sec:prev_known}). This
    change led to a lower efficiency (see Section~\ref{sec:results}).
    
    For Chunks 7, 8 and 9, based on the tests described in
    Section~\ref{sec:blind_test}, we set the CORE method to Likelihood,
    while BONUS targets were selected using the NN-Combinator
    (\S~\ref{sec:combinator}). In addition, previously known stars from
    SDSS or 2dF spectroscopy were now vetoed.  The NN photometric redshift
    threshold was relaxed slightly, from $z_{p \rm NN}>2.1$ to 2.0.
    
    Chunk 9 was the last chunk to be observed in Year One, and thus
    the last data included in the spectroscopic sample presented in this
    paper. Target selection for Chunk 10 was performed in the first year
    of BOSS, but the Chunk 10 plates were not observed until the second
    year, after the Summer 2010 shutdown.  Chunk 11, the re-observation of
    Stripe 82, was also observed at the start of the second year of BOSS
    observations. As described in detail by
    \citet{Palanque-Delabrouille11}, a variability-based quasar selection
    was performed for Chunk 11. This led to a significantly higher
    high-$z$ quasar density than elsewhere in the survey, 24 $z>2.15$
    quasars deg$^{-2}$ \citep{Palanque-Delabrouille11}, as we describe
    further in Section~\ref{sec:results}.

    \subsection{Chunks 12 and 13}
    \label{sec:chunks12and13}
    The XDQSO method was introduced in Chunk 12 to test its
    efficiency.  There is substantial overlap between the target list
    of XDQSO and Likelihood, and including the highest ranked 20
    targets deg$^{-2}$ from each yielded a total of 25 targets
    deg$^{-2}$.  We thus defined CORE to be the union of all these targets. 
    The NN-Combinator was retained as the method for BONUS.

    \subsection{Chunks 14-18}
    \label{sec:chunks14to19}
    The Chunk 12 and 13 spectroscopic results demonstrated the
    superiority of XDQSO for the core algorithm.  Therefore, from Chunk 14
    onwards, and for the rest of the BOSS, the ``Extreme Deconvolution''
    algorithm (XDQSO), and that alone, was set to be CORE.  This led to a
    gain of $\sim1$ high-$z$ quasar deg$^{-2}$ in the CORE.
    
    Various further improvements were implemented in BONUS starting
    with Chunk 14. For Chunk 14, a change in fiber collision
    prioritization (see Appendix~\ref{sec:flowchart_YearOne}) led to a
    gain of $\sim1$ quasar deg$^{-2}$.  In Chunk 15 we began a policy of
    re-observing previously known quasars in plate overlap regions,
    leading to a spectroscopic signal-to-noise ratio gain of $\sim15$\%
    per quasar.  In Chunk 16, we incorporated UKIDSS photometry into the
    training of XDQSO as an input to the NN-Combinator.  This led to a
    gain of $2-3$ high-$z$ quasars deg$^{-2}$ where UKIDSS data were
    available.  Overlap between adjacent imaging scans allowed improved
    photometry for objects observed more than once,
    (Sec.~\ref{sec:input_phot}), leading to a gain of $\sim0.3-0.5$
    quasars deg$^{-2}$ in Chunk 16. In Chunk 17, an optical-only trained
    version of the XDQSO (essentially what is used for CORE) was also used
    as an input to the NN-Combinator used for BONUS, with a gain of
    $\sim0.5$ quasars deg$^{-2}$.
    
    BOSS spectroscopic plates are designed by giving priority first to
    BOSS galaxy and quasar targets, followed by objects in various
    ancillary programs \citep[Section 2 of][]{Eisenstein11}.  If
    additional fibers are available, we assign them to previously known
    $1.8<z<2.15$ quasars; these are labeled as SUPPZ in
    Figure~\ref{fig:flowchart} and are flagged as QSO\_KNOWN\_SUPPZ in
    Table~\ref{tab:BOSS_TARGET_FLAGS}.  Reobserving these objects allows a
    measurement of the spectral structure from metal lines along the line
    of sight and spectral artifacts that may contaminate Ly$\alpha$
    structure measurements (\citealt{McDonald06}).

    \begin{table*}
      \begin{center}
        \begin{tabular}{l rr r rrr}
          \hline
          \hline
          Chunk  &  Observed            & Total      & CORE$^{a}$    & \# high-quality$^{b}$ & \# high-quality &  \# CORE high- \\
                      &  Area (deg$^2$)   & spectra  &  spectra          &      (zWarning=0)       &  $z>2.20$     & quality $z> 2.20$    \\
          \hline
          1 &    37.4 &      3811 ( 3174)   &   988 ( 849)   &  2313 (1909)    & 1211 ( 986)       &  411 ( 355)    \\
          2  &  117.6 &     9865 ( 9018)   &   2639 (2409)  & 7052 (6461)    & 2018 (1847)      &  880 ( 799)    \\
          3  &    33.1 &     2191 ( 2142)   &   630 ( 616)   & 1463 (1433)    &   521  ( 513)      &  268 ( 264)    \\
          4  &  168.3 &  11879 (11362)   &  3275 (3126) & 6603 (6302)    &  2527 (2417)     & 1320 (1269)  \\
          5  &  186.0 &  10344 (10154)   & 2924 (2875) & 7132 (7004)    &  3376 (3323)     & 1714 (1691)   \\
          6  &  121.7 &    8733 ( 8582)    & 2063 (2023) & 5091 (5003)    &  1914 (1878)     &   915 ( 896)    \\
          7  &  120.8 &    4615 ( 4506)    &   2647 (2581) & 3100 (3027)     & 1635 (1595)     & 1188 (1160)   \\
          8  &   67.0  &    2565 ( 2400)    &   1697 (1591) & 1891 (1762)     &   834 (  772)     &   657 ( 608)    \\
          9  &   26.2  &      906 (   900)    &   742 ( 738)   &  660  ( 655)      &   251 (   249)    &   226 ( 224)   \\
          \hline
          {\bf TOTAL}  & {\bf 878.14} &  {\bf 54909 (52238) } &   {\bf 17605 (16808)} &  {\bf 35305 (33556)}    &  {\bf 14287 (13580)}   &  {\bf 7579 (7266) } \\
          \hline
          \hline
        \end{tabular}  
      \end{center}
      \caption{Summary of the results from the first year of BOSS quasar
        observations, chunk by chunk.  
        Numbers in parentheses are for Unique objects. 
        $^{a}$CORE defined as target bit 10 for Chunks 1 and 2, bit 40 for
        Chunks 3--9 (Table~\ref{tab:BOSS_TARGET_FLAGS}).
        $^{b}$ High-quality redshifts are those for which the spectroscopic pipeline 
        zWarning flag is zero.} 
      \label{tab:raw_numbers}
    \end{table*}

    \subsection{The Sky Distribution of BOSS Quasar Targets}
    \label{sec:skydist}
\begin{figure*}
      \begin{center}   
        \includegraphics[height=8.0cm,width=16.0cm]
        {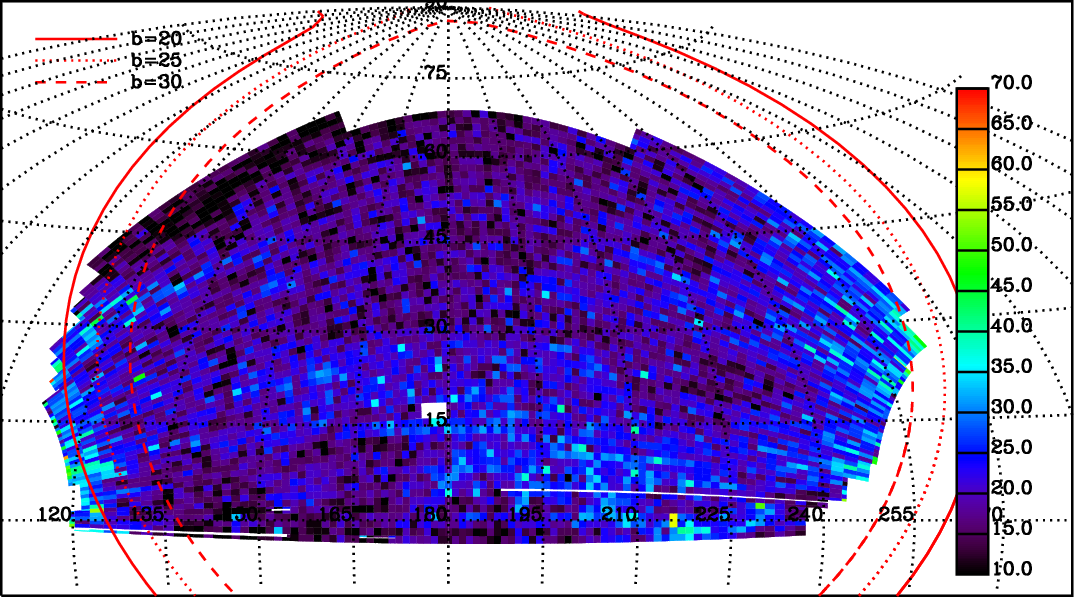}
        \caption{The quasar target density map in the NGC for the XDQSO
          CORE targets, 
          displayed in equatorial coordinates.
          The units are targets deg$^{-2}$. }
        \label{fig:ngc_core}
      \end{center}
    \end{figure*}

\begin{figure*}
      \begin{center}   
        \includegraphics[height=8.0cm,width=16.0cm]
        {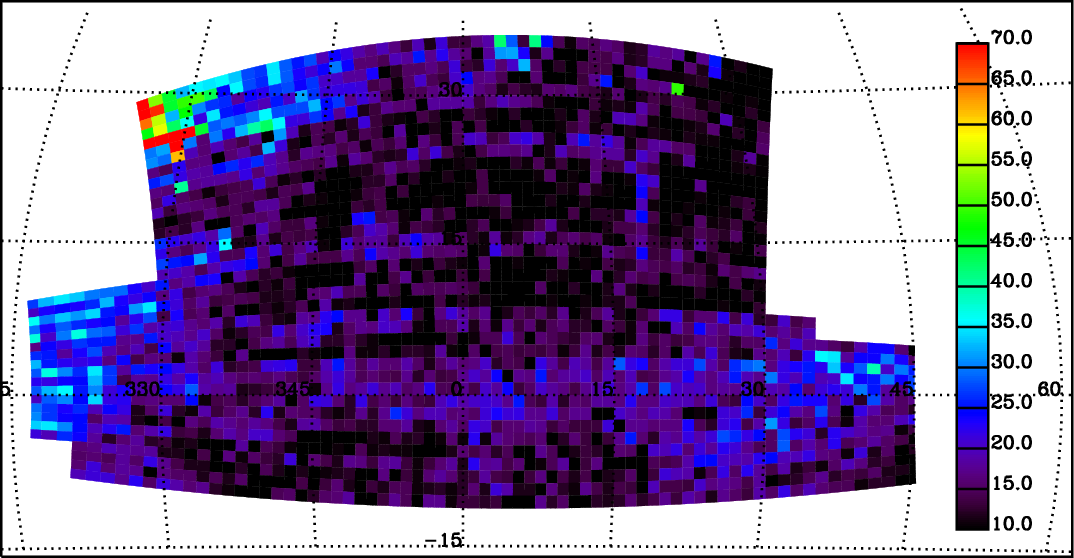}
        \caption[SGC ``HeatMap'' for CORE.]
        {The quasar target density map in the SGC for the XDQSO
          CORE targets, 
          displayed in equatorial coordinates.  The units are targets
          deg$^{-2}$.  }
        \label{fig:sgc_core}
      \end{center}   
    \end{figure*}

\begin{figure*}
      \begin{center}   
        \includegraphics[height=8.0cm,width=16.0cm]
        {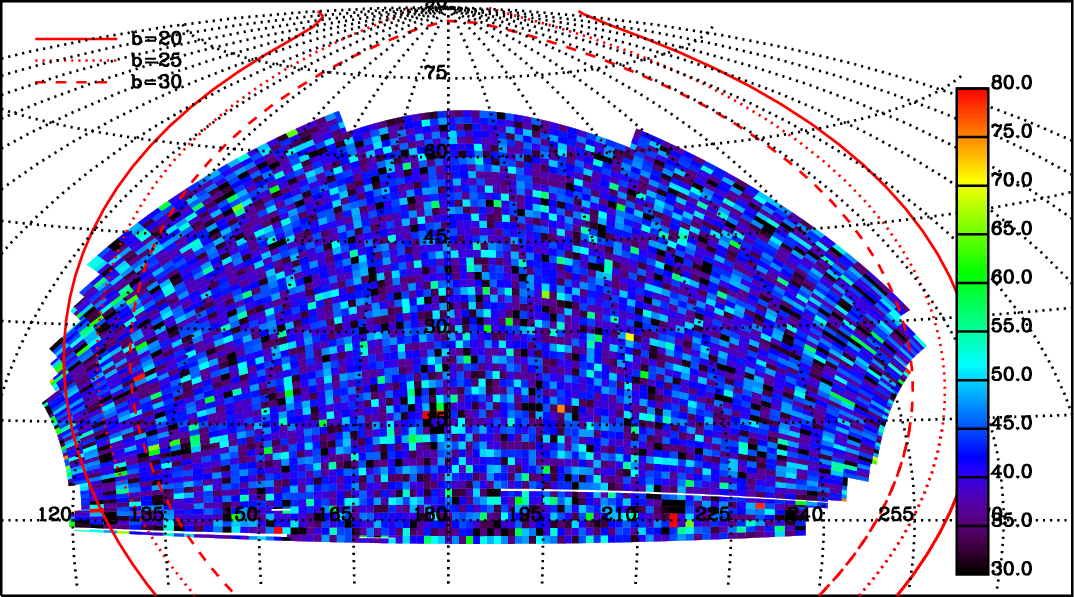}
        \caption{The quasar target density map in the NGC for all our targets,
          CORE+BONUS+KNOWN+FIRST,  displayed in equatorial coordinates. The units are targets deg$^{-2}$. }
        \label{fig:ngc_full}
      \end{center}   
    \end{figure*}

\begin{figure*}
      \begin{center}   
        \includegraphics[height=8.0cm,width=16.0cm]
        {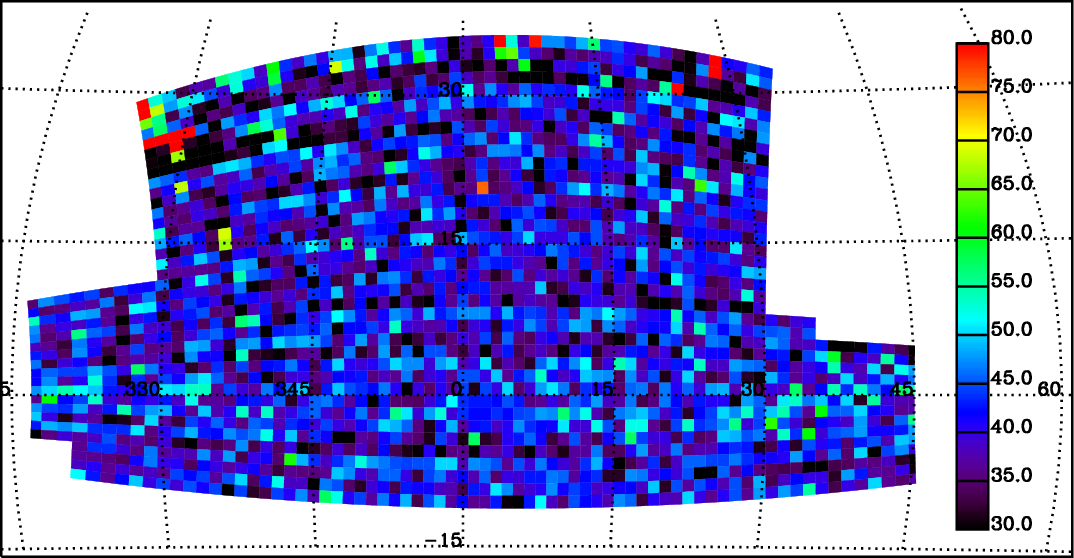}
        \caption[SGC ``HeatMap'' for all targets.]
        {The quasar target density map in the SGC for all our targets,
          CORE+BONUS+KNOWN+FIRST, displayed in equatorial coordinates. The units are targets deg$^{-2}$. }
        \label{fig:sgc_full}
      \end{center}  
    \end{figure*}

    The sky distribution of the BOSS quasar targets are shown in
    Figs.~\ref{fig:ngc_core}, \ref{fig:sgc_core}, \ref{fig:ngc_full} and
    \ref{fig:sgc_full}. In Figs.~\ref{fig:ngc_core} and
    \ref{fig:sgc_core}, we show the surface densities of BOSS quasar
    targets for the NGC and the SGC, respectively, as selected by the CORE
    method (XDQSO) for DR9. In Figs.~\ref{fig:ngc_full} and
    \ref{fig:sgc_full}, we show the surface densities of BOSS quasar
    targets for the NGC and the SGC, respectively, as selected by the CORE
    (XDQSO), BONUS (NN-Combinator) and FIRST methods, as well as the
    inclusion of all previously known $z > 2.2$ quasars.  The CORE sample
    is designed to produce a mean surface density of 20 targets
    deg$^{-2}$, and although it is reasonably uniform, the density of
    targets ranges from 10 to 30 targets deg$^{-2}$ over the footprint of
    the survey.  The largest variations are associated with Galactic
    structure, with excesses visible at low Galactic latitudes and in the
    Sagittarius stream.  The BONUS sample adds enough targets in each area
    of sky to give a much more uniform 40 targets deg$^{-2}$.
    
    
\begin{figure*}
  \centering
  \includegraphics[width=14.0cm]
  {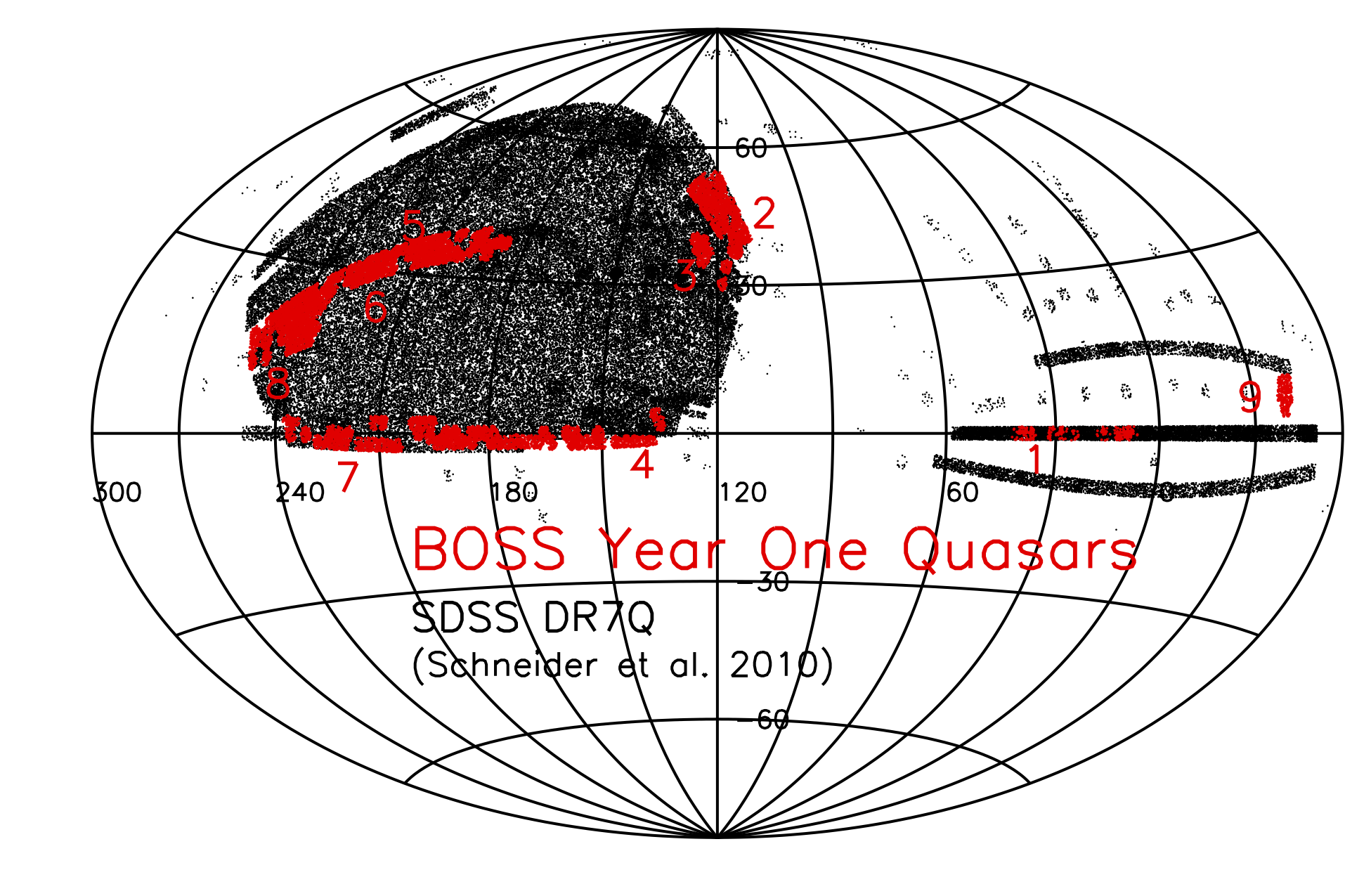}
  \caption[]
  {Sky distribution of the 14,287 quasars in the BOSS Year One 
    quasar survey (J2000 equatorial coordinates), in red. The nine chunks
    are labeled accordingly, and the dotted lines are drawn at Galactic
    latitudes  $b=\pm 25^{\circ}$. The spectroscopically confirmed
    SDSS-I/II DR7 quasar catalog
    \citep{Schneider10} is shown for comparison in black.}
  \label{fig:BOSS_quasar_RADec_YOD}
\end{figure*}

\begin{table*}
  \begin{center}
    \begin{tabular}{l p{0.5cm} rr p{0.5cm}  rr p{0.5cm} rr p{0.5cm}  rr}
      \hline
      \hline
      TARGET\_FLAG   &  & \multicolumn{2}{c}{No. of targets} & & \multicolumn{2}{c}{No. of targets with }  & & \multicolumn{2}{c}{zWarning=0 and} &  & \multicolumn{2}{c}{zWarning=0} \\
      &   & \multicolumn{2}{c}{(only)}             & & \multicolumn{2}{c}{zWarning=0 (only)}   & & \multicolumn{2}{c}{$z>2.20$ (only)}   &  & \multicolumn{2}{c}{and stars (only)} \\
      \hline
      CORE                & &  3627  &(1509)                                &          &   2693   &  (890)                            & &    1291  & (89)                                   & &   1007   & (619)  \\ 
      BONUS              & &  4071  &(2927)                                &          &     2631 & (1756)                          &  &     546  & (131)                                &    &    1558  & (1300) \\
      KNOWN\_ MIDZ & &  2975  &(529)                                  &        &    2831    & (490)                          &   &   2520  & (357)                                &   &      0      &    (0)   \\
      KNOWN\_LOWZ & &        0  &(0)                                     &     &       0  & (0)                                     &   &  0        &   (0)                                     &      &           0 &  (0) \\
      NN                   & & 17678  &(1111)                               &          &     13988  &  (791)                   &   &     8197 & (152)                                  &     &      3776 & (562) \\
      UKIDSS            & &      139  & (36)                                  &       &       119  & (33)                          &  &   80  &    (6)                                         & &       22 &   (21) \\
      KDE\_COADD  & &    2407  &(860)                                &         &     1517  & (309)                    &   &   890  &     (31)                                     &  &       324 & (107)\\ 
      LIKE                 & & 30534  &(2541)                               &         &      23022 & (1779)                &    &     11793 & (479)                                &    &       4712 &  (794) \\  
      FIRST               & &     986 & (530)                                &        &       791&  (400)                      &  &    403  &    (104)                                    &   &      35 &  (34) \\
      KDE                  & &  27145 & (0)                                  &       &     16068 & (0)                         &    &      7313 & (0)                                    &       &    5330 &   (0) \\
      CORE\_MAIN    & &  13978  &(0)                                 &   &    10652 & (0)                              &     &   6288 & (0)                                      &      &       2106 &  (0) \\
      BONUS\_MAIN  & &   40363 & (8)                               &   &    25218 & (2)                               &   &   10616&  (0)                                  &    &      7588  &  (2) \\
     \hline
     \hline
    \end{tabular}  
  \end{center}
  \caption{The total number of spectra of objects selected 
    with each target flag for Year One observations.  Objects can
    be counted more than once; the number of objects in only one
    category is also shown.  Also tabulated are the number of
    good (zWarning=0) redshifts, the number of $z>2.2$ quasars, and
    the number of stellar spectra obtained.}
 \label{tab:raw_flag_numbers}
\end{table*}
\section{Results}
\label{sec:results}
In this section, we present the results of spectroscopy carried out
during Year One after the completion of hardware commissioning, from
MJD 55176 (2009 December 11) through MJD 55383 (2010 July 06).  The
distribution of BOSS Year One quasars on the celestial sphere is shown
in Fig.~\ref{fig:BOSS_quasar_RADec_YOD}.

    \subsection{Global Properties and Efficiencies}
    Table~\ref{tab:raw_numbers} summarizes the results from the first
    year of BOSS quasar observations.  BOSS quasar targets are those which
    have one of the target bit flags listed in
    Table~\ref{tab:BOSS_TARGET_FLAGS} set.  There were 54,909 spectra of
    objects targeted as quasars, of which 52,238 were unique
    objects. These were observed over over a footprint of 878 deg$^{2}$,
    giving a mean surface density of 63.8 targets deg$^{-2}$.
    
    Of the \hbox{54,909} (52,238 unique) spectra, \hbox{35,305}
    (33,556) had high-quality redshifts, as designated by the ``zWarning''
    flag of the spectroscopic pipeline
    \citep{Adelman-McCarthy08,Aihara11}.  From visual inspection of the
    data, the zWarning flag is reliable at the 90-95\% level for the
    quasar target spectra; very few of the objects flagged as having
    high-quality redshifts (i.e., zWarning=0) are incorrect.  We present
    the performance of zWarning as a function of magnitude and S/N in
    Appendix~\ref{sec:zwarning}; most objects with $\rm zWarning \ne 0$
    are faint objects with low S/N spectra.  Given the faint magnitude
    limit of BOSS, it is not surprising that many of the targets that are
    not quasars lack the clearly identified spectral features required to
    assign a high-confidence redshift.  We will present a detailed
    examination of the performance of the reduction pipeline, the zWarning
    flag and the findings from the visual inspection of the data when we
    publish the BOSS Quasar DR9 Catalog in a separate paper.
    
\begin{figure*}
      \centering
      \includegraphics[height=8.5cm]
      {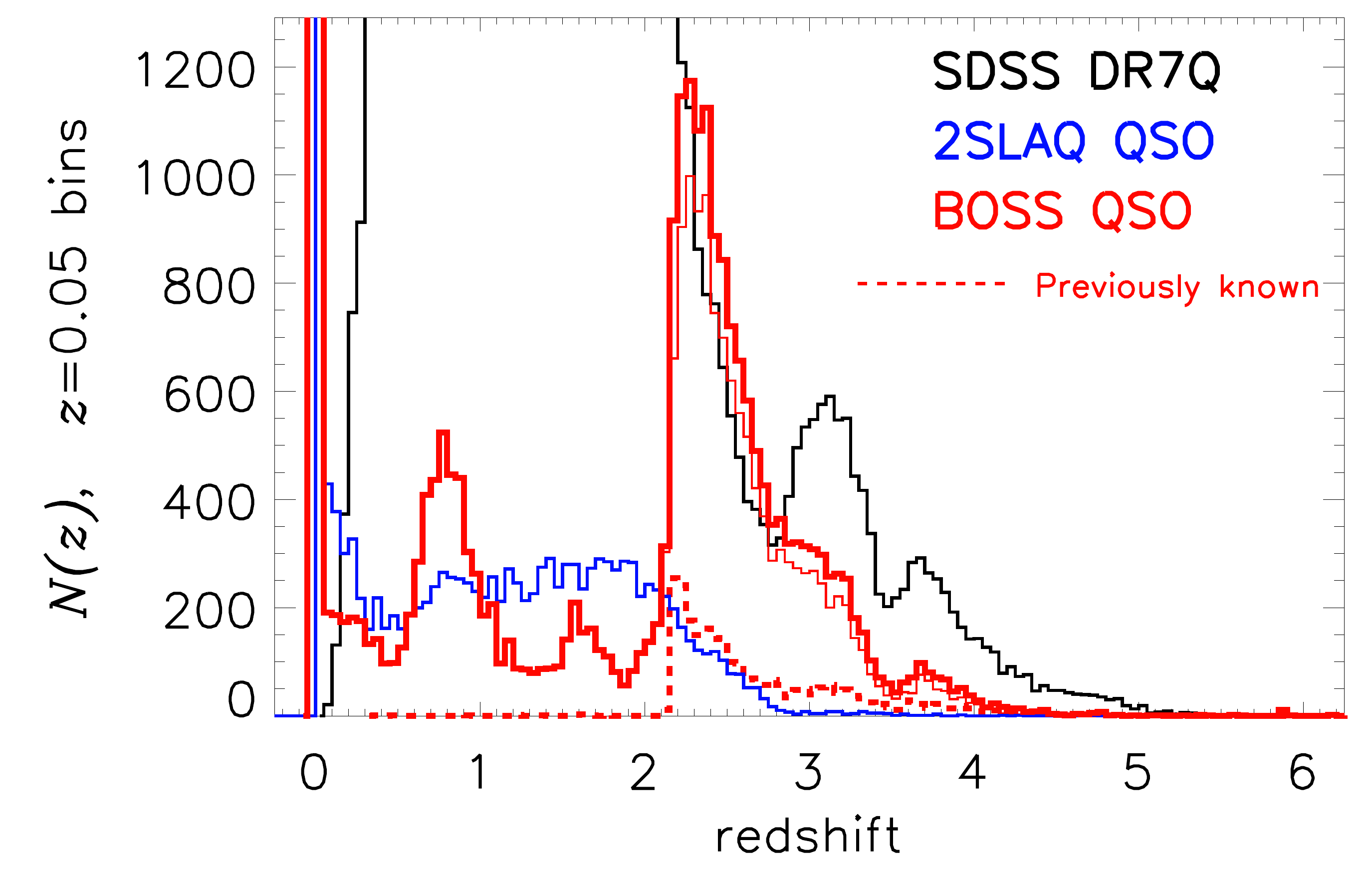}
      \caption[The BOSS Quasars $N(z)$.]
      {The redshift histogram of BOSS Year One quasars (solid red
        thick histogram). The dashed red line represents those objects known
        prior to BOSS observations, while the distribution of newly confirmed
        quasars is given by the thin red line. For comparison the SDSS DR7
        quasars from \citet{Schneider10} (selected over a much larger sky
        area) are shown by the black histogram, while the 2SLAQ quasar data
        \citep{Croom09a}, are in blue.}
      \label{fig:BOSS_Quasar_Nofz_hist}
    \end{figure*}

    Of the \hbox{33,556} unique objects with high-quality redshifts,
    \hbox{11,149} are stars, while \hbox{13,580} have $z>2.20$. The
    remaining 8,827 objects are mostly quasars at $z\sim0.8$ and
    $\sim1.6$, and low-$z$ compact galaxies; see
    Fig.~\ref{fig:BOSS_Quasar_Nofz_hist}.  Of the 13,580 high redshift
    objects, 2,317 had the QSO\_KNOWN\_MIDZ flag set; thus the first year
    of BOSS observations resulted in the spectroscopic confirmation of
    11,263 new $z>2.2$ quasars. A full breakdown of the number of objects
    associated with each target flag, the number of good (zWarning=0)
    redshifts and the number of $z > 2.2$ quasars obtained is given in
    Table~\ref{tab:raw_flag_numbers}.
    
    Figure~\ref{fig:BOSS_Quasar_Nofz_hist} shows the redshift
    distribution of BOSS quasars from the first year, and compares it
    with that from the SDSS DR7 quasar sample \citep{Schneider10} and
    the 2SLAQ survey \citep{Croom09a}.  This plot is very
    similar, but not identical, to that shown in the SDSS-III
    overview paper of \citet{Eisenstein11}.  
    Of course, the DR7 sample is selected over the full SDSS-II
    imaging area, approximately 9,380 deg$^2$, while the BOSS
    Year One data come from observations of 880 deg$^2$.  
    Already BOSS has slightly more quasars
    in the $z=2.2-2.8$ range, while at higher redshifts the
    DR7 sample remains larger.

    Degeneracies in the color-redshift relation of quasars lead to the
    selection of low-$z$ quasars in BOSS.  The quasars at $z\sim0.8$ have
    \mgii $\lambda$2800 \AA\ at the same wavelength as \lya\ at redshift
    $z\sim3.1$, giving these objects similar broad-band colors, while the
    large number of objects at $z\sim1.6$ is due to the confusion between
    $\lambda1549$ \civ and \lya\ at $z\approx2.3$. We shall come back to
    this feature when comparing the performance of the NN, KDE, and
    Likelihood methods in \S~\ref{sec:efficiencies}.  The tail of objects
    at $z\gtrsim3.5$ includes a significant contribution from
    re-observations of previously known quasars.  

    \begin{figure}
      \includegraphics[height=8.0cm,width=8.0cm]
      {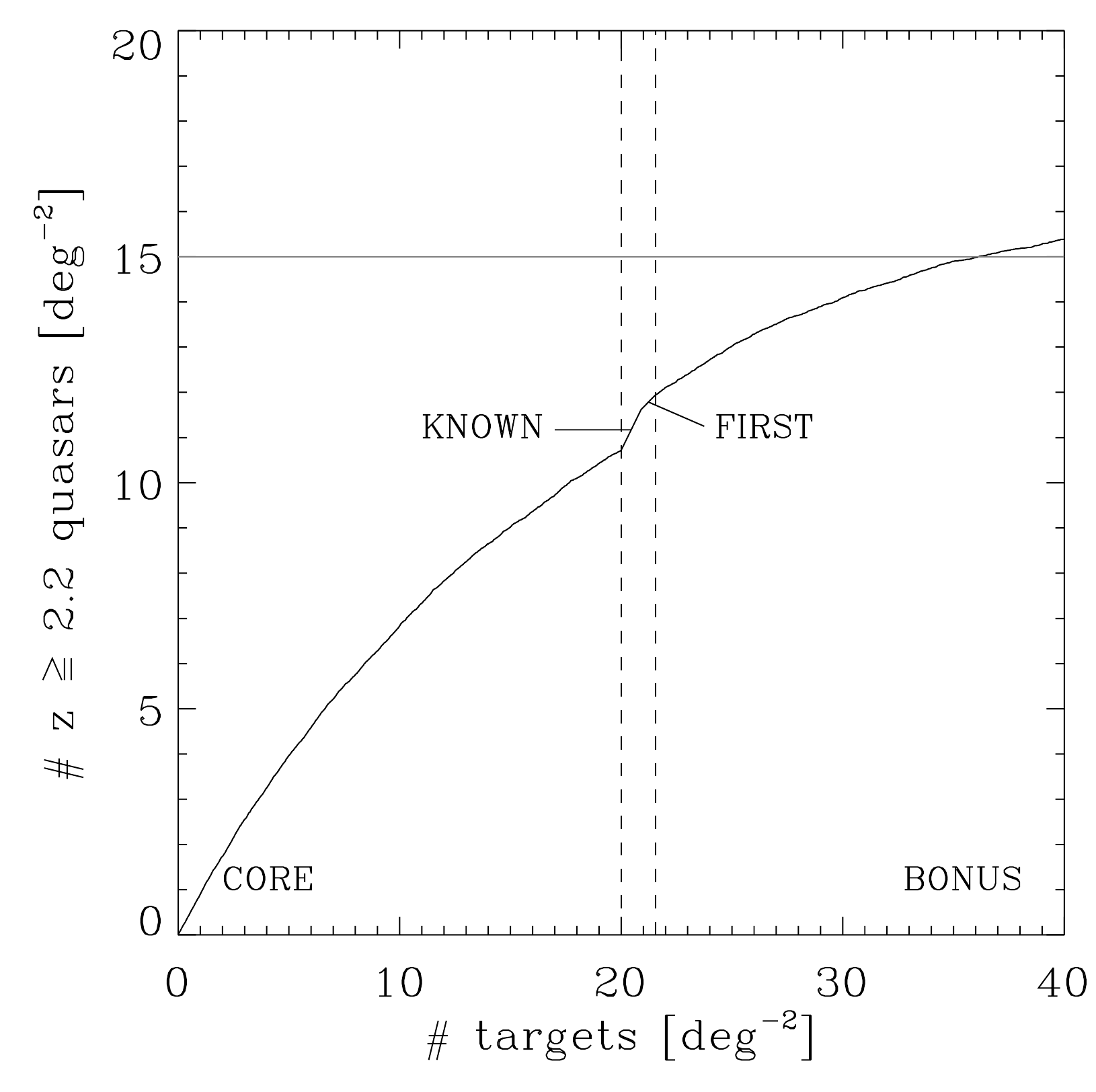}
      \caption[]
      {Cumulative number of quasars with $z>2.2$ as a function of the rank of
	the target for the Stripe 82 control sample with single-epoch
        photometry. At 20 fibers deg$^{-2}$, the XDQSO CORE algorithm
	selects 10.7 quasars deg$^{-2}$, while previously known and
	FIRST sources add an average of 1.5 quasars deg$^{-2}$.
	At 40 fibers deg$^{-2}$, the total surface density of
	$z > 2.2$ quasars selected by our current algorithms from
	single-epoch SDSS photometry is 15.4 deg$^{-2}$.
	Note that these numbers represent an average
	over a wide range of Galactic latitude, and therefore 
	stellar contamination.     }
      \label{fig:EfficiencyCurveSingleEpoch}
    \end{figure}
    \begin{figure}
        \includegraphics[height=8.0cm,width=8.0cm]
        {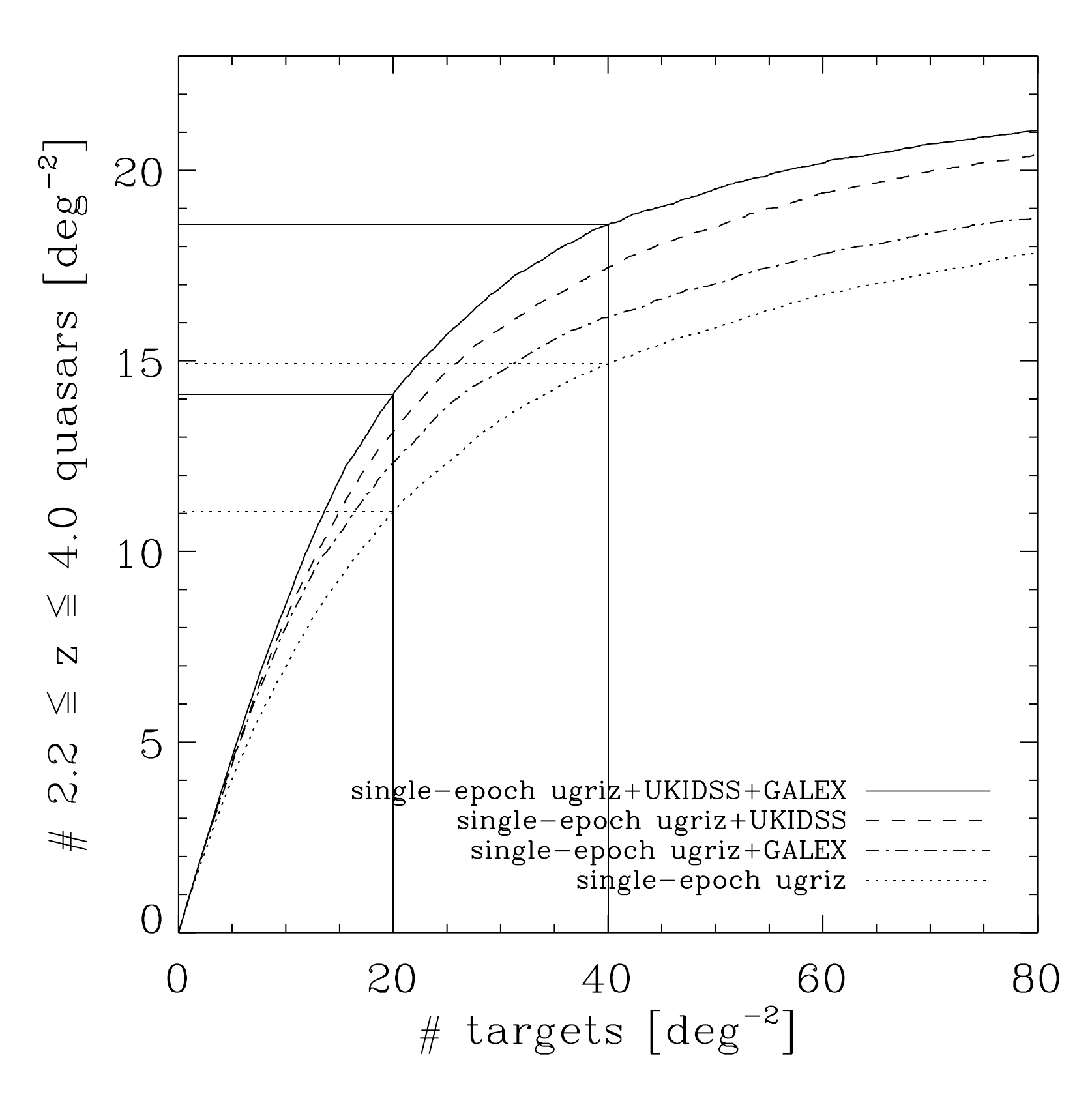}
        \caption[SDSS UKIDSS GALEX  comparisons]
        {Similar to Figure~\ref{fig:EfficiencyCurveSingleEpoch}, but
	  showing the impact of adding GALEX photometry, UKIDSS photometry, 
	  or both to SDSS single-epoch photometry.  This Figure is
	  based on Stripe 82 data and XDQSO selection for all targets.
          }
          \label{fig:AuxiliaryDataEfficiency}
      \end{figure}
  \begin{figure}
      \includegraphics[height=7.0cm,width=9.0cm]
      {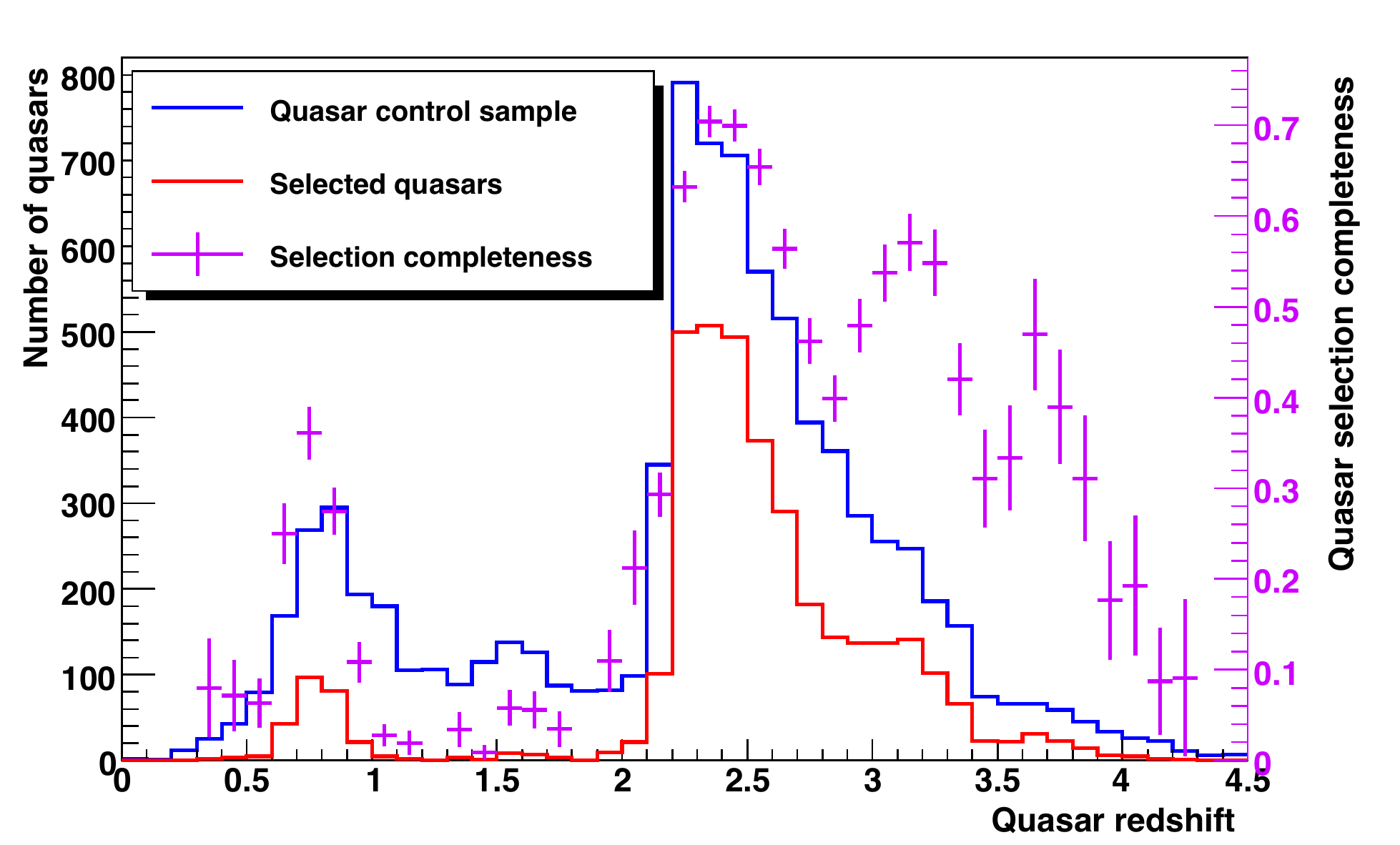}
      \caption[]
      {Completeness of BOSS single-epoch target selection vs. redshift,
       on Stripe 82.  The blue histogram shows the redshift 
       distribution of all spectroscopically confirmed quasars on Stripe 82.  
       The red histogram is for those quasars that pass the BOSS
       single-epoch target selection for a threshold tuned to produce 
       40 targets deg$^{-2}$.  
       Purple points with Poisson error bars show the ratio of the two, 
       i.e., the selection completeness (right-hand scale).}
      \label{fig:EfficiencyVsRedshift}
    \end{figure}

    Figures~\ref{fig:EfficiencyCurveSingleEpoch}
    and~\ref{fig:AuxiliaryDataEfficiency} present our key results, the
    efficiency of the current target selection algorithms.  For these
    tests, we have constructed a control sample of targets on Stripe 82,
    where our spectroscopy is more complete than anywhere else on the sky,
    albeit still not perfect.  Here we include data from Year Two from
    Chunk 11, where Stripe 82 was retargeted using a variability selection
    for quasars \citep{Palanque-Delabrouille11}.  Stripe 82 also has high
    completeness because quasars are selected from co-added photometry,
    with much smaller photometric errors.
    
    For Figure~\ref{fig:EfficiencyCurveSingleEpoch}, we select the quasar
    targets in our normal way from {\it single-epoch} data, with the first
    20 targets deg$^{-2}$ selected by the XDQSO CORE algorithm.  Targets
    are ranked in order of probability, and the plot shows the number of
    $z>2.2$ quasars deg$^{-2}$ vs. the number of targets deg$^{-2}$, with
    the slope of the curve indicating the efficiency of selection.  The
    CORE algorithm selects 10.7 $z>2.2$ quasars deg$^{-2}$ from its 20
    targets.  We then show the average contribution of KNOWN and FIRST
    quasars, totaling 1.6 high-$z$ quasars deg$^{-2}$.  This increment
    assumes a surface density of 0.9 known high-z quasars deg$^{-2}$ (and
    0.7 deg$^{-2}$ from FIRST), which is consistent with our Year One data
    (see Table~\ref{tab:raw_flag_numbers}) but lower than the surface
    density of known pre-BOSS high-z quasars on Stripe 82, which is
    unusually well studied.  Finally, we add the BONUS targets from the
    NN-combinator, again in rank order.  At 40 targets deg$^{-2}$, we are
    just above the minimum BOSS goal, with a mean density of 15.4 $z>2.2$
    quasars deg$^{-2}$.  Stripe 82 samples a wide range of Galactic
    latitude and thus stellar density; we therefore anticipate that this
    test should be representative of selection efficiency averaged over
    the full BOSS survey region. We also found from observations of early
    chunks, that adding additional fibers beyond the nominal 40
    deg$^{-2}$, led to only very minimal gains in yield.
    
    Figure~\ref{fig:AuxiliaryDataEfficiency} shows the impact of adding
    UKIDSS and GALEX data to single-epoch SDSS photometry.  For this test
    we use the XDQSO algorithm alone, since this is where these auxiliary
    data sets currently enter our selection procedures, and we extend the
    efficiency curves up to 80 targets deg$^{-2}$.  At 40 targets
    deg$^{-2}$, the efficiency for XDQSO with single-epoch SDSS imaging
    alone is 15.0 $z > 2.2$ quasars deg$^{-2}$.  Adding GALEX data
    improves the efficiency to 16.2 deg$^{-2}$, adding UKIDSS improves it
    to 17.3 deg$^{-2}$, and adding both improves it to 18.6 deg$^{-2}$.
    Thus, both of these data sets can significantly enhance the efficiency
    of BOSS quasar target selection in regions where they are available.
    Stripe 82 has medium-deep (``MIS'') GALEX data, and the improvement
    with shallower (``AIS'') coverage will be smaller, but our tests
    indicate that GALEX addition will still improve the selection.
    
    Fig.~\ref{fig:EfficiencyVsRedshift} shows the redshift
    distribution of all known quasars on Stripe 82 as a function of
    redshift, as well as those selected by the single-epoch SDSS
    algorithms illustrated in Fig.~\ref{fig:EfficiencyCurveSingleEpoch}
    above.  The ratio of the two measures the completeness of BOSS
    single-epoch quasar selection relative to known quasars in this well
    studied region, ranging from 40\% to 70\% over our critical redshift
    range $2.2 < z < 3.5$.  Of course, this remains a lower limit to the
    true completeness at the BOSS magnitude limit, though in the $2.2 < z
    < 3.5$ redshift range we anticipate that the BOSS Stripe 82 sample
    selected from co-added photometry and variability has high
    completeness \citep{Palanque-Delabrouille11}.

    \begin{figure}
      \includegraphics[height=12.0cm,width=8.0cm]
      {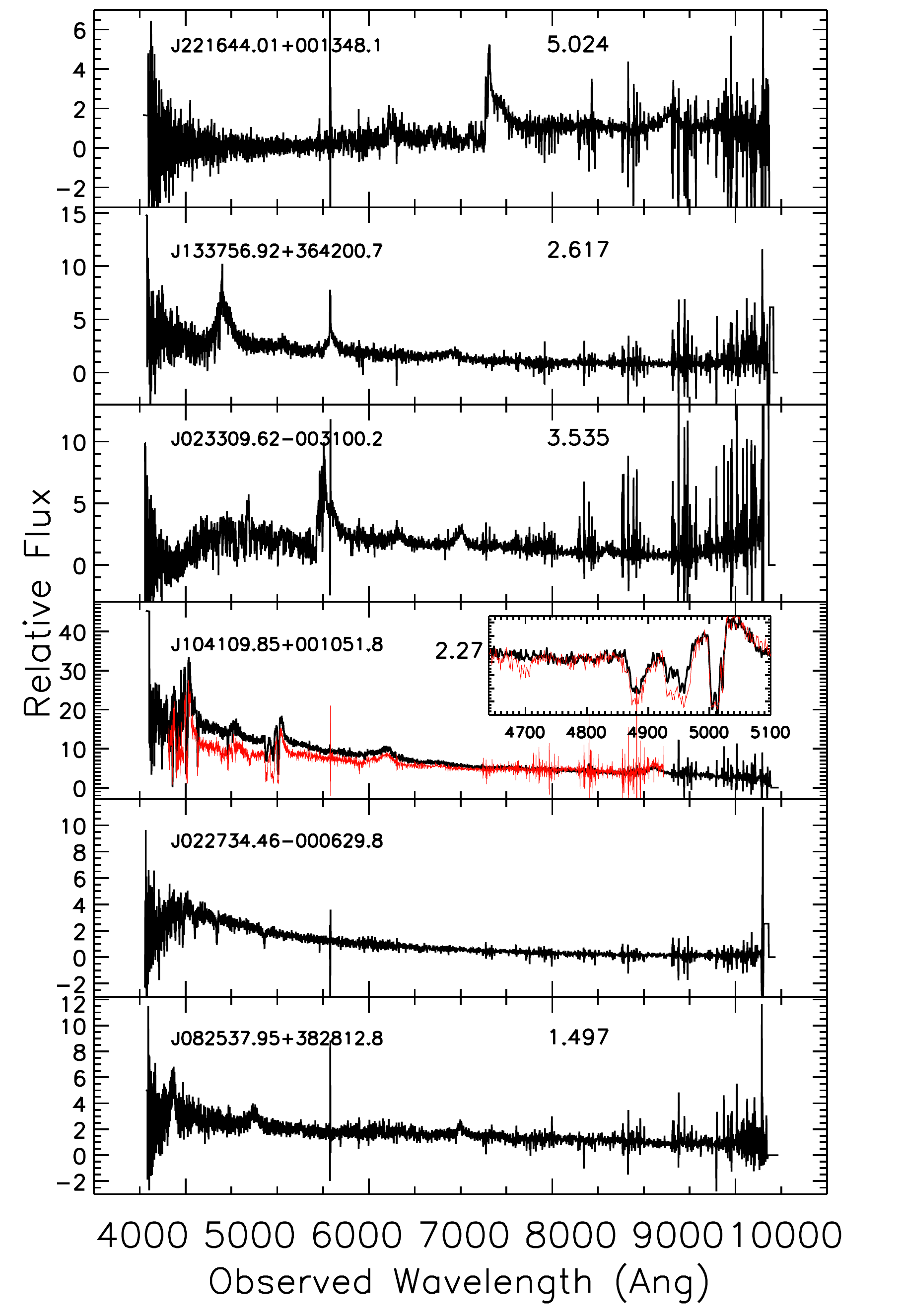}
      \caption[]
      {Examples of spectra of BOSS quasar targets.  The SDSS object
        name and pipeline redshift are given in each panel (except for
        the star). 
        From top to bottom: 
        a $z>5$ quasar found by the Likelihood method; 
        a newly discovered $z=2.6$ quasar at the typical S/N; 
        a $z=3.5$ quasar selected only by the KX method (\S~\ref{sec:UKIDSS}); 
        a re-observed BAL quasar showing spectroscopic variability (black line is
         the BOSS spectrum; red is from SDSS, a spectrum taken 3377
         days earlier); 
         a star with our typical S/N and 
        a $z=1.5$ quasar with our typical S/N.  The feature at 5577\AA\
	in all spectra is a residual from a sky line.}
      \label{fig:eg_spectra}
    \end{figure}

    Fig.~\ref{fig:eg_spectra} shows examples of BOSS spectra of quasar
    targets 
    from the Year One data. From top to bottom:
    a $z > 5$ quasar found by the Likelihood method (and not
    selected by any other method); 
    a newly discovered $z=2.6$ quasar at a typical S/N; 
    a $z = 3.5$ quasar selected only by the KX method; 
    a re-observed BAL quasar showing spectroscopic variability over
    3377 days in the
    observed frame; a star at our typical S/N; and a $z=1.5$ quasar 
    with our typical S/N.

    \begin{figure}
      \includegraphics[height=12.0cm,width=8.0cm]
      {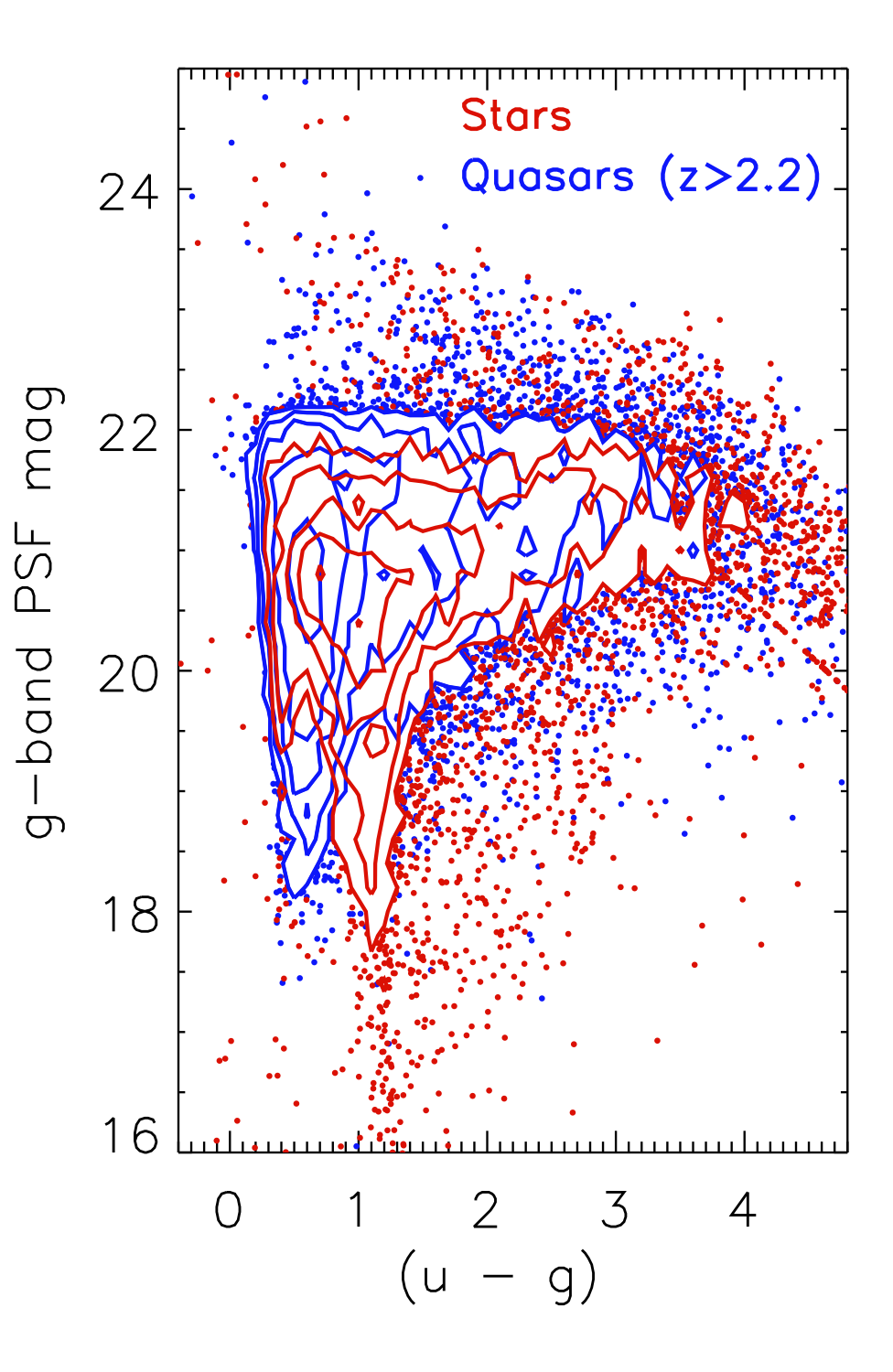}
      \caption[]
      {Color-magnitude diagram ($(u-g)$ vs.~$g$) for objects spectroscopically classified 
        as stars (red contours and points) and $z>2.2$ quasars
        (blue contours and points).  Only objects with zWarning=0 are
	shown.  The quasars are systematically bluer; there are very
	few quasars with $g < 18$.}
      \label{fig:BOSS_QSOs_g_vs_uminusg} 
    \end{figure}

    \begin{figure}
      \includegraphics[height=12.0cm,width=8.0cm]
      {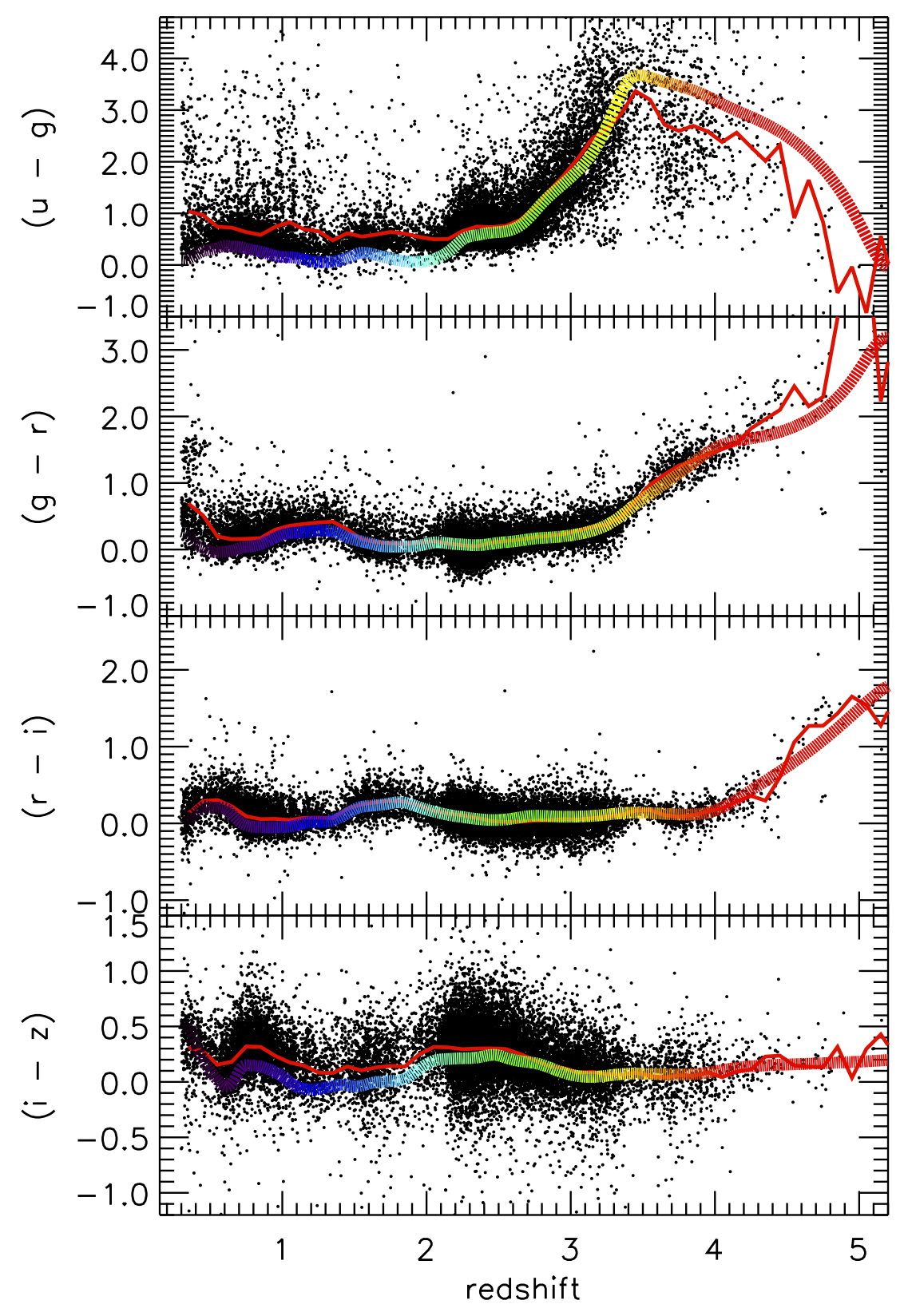}
      \caption[]
      {SDSS colors vs. redshift for quasars in the BOSS Year One data. The thin
	solid line is the mean color in bins of redshift, while the
	thick colorful line is from the model of Bovy et al.~(2011, in
      preparation).  The model is systematically bluer than the data
      at low redshift because BOSS systematically excludes UV-excess sources.}
      \label{fig:BOSS_QSOs_ugri_vs_redshift} 
    \end{figure}

  \begin{figure*}
      \includegraphics[height=16.0cm,width=16.0cm]
      {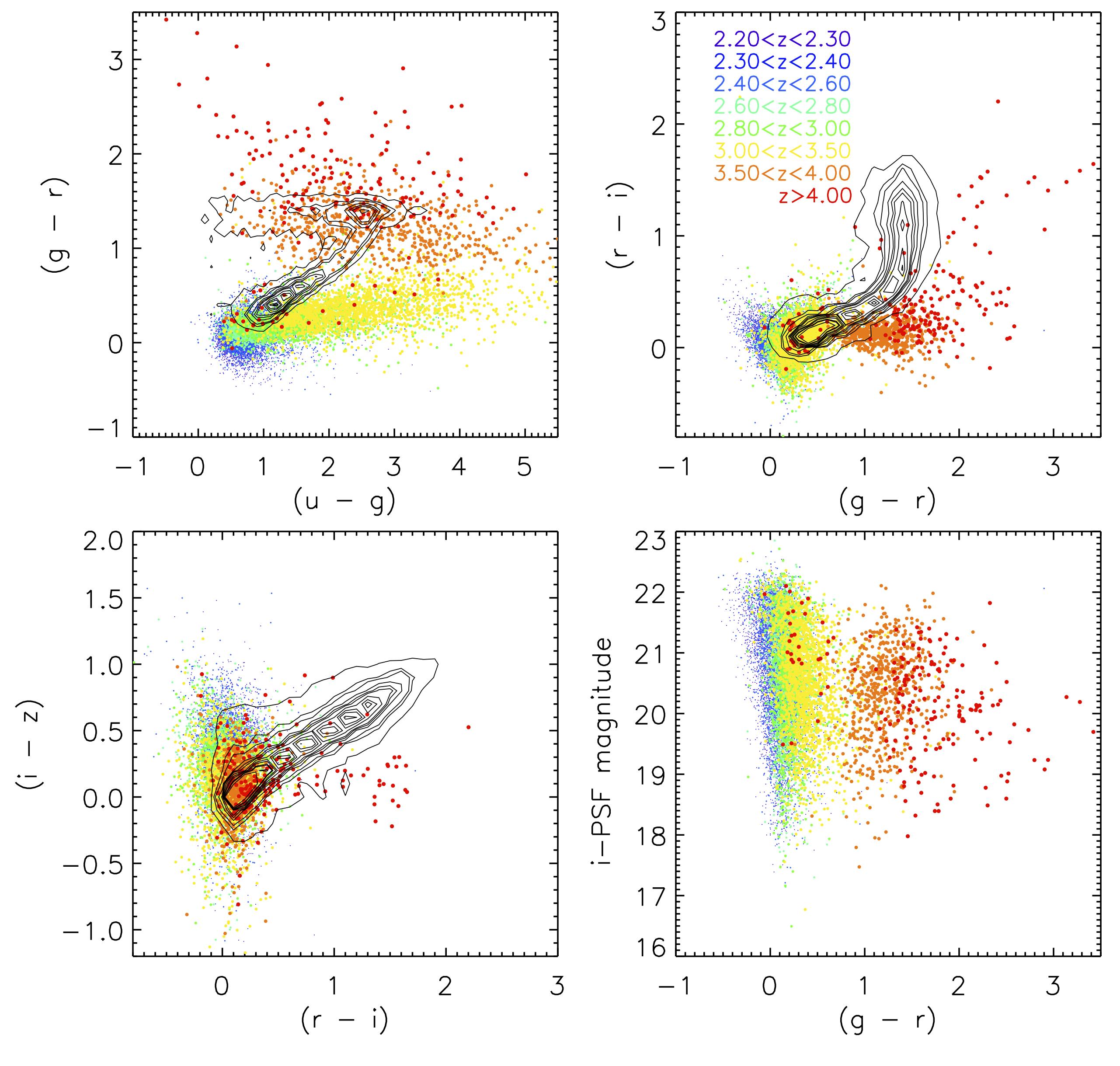}
     \caption[]
      {Color-color diagrams for the First Year data for all
	spectroscopically confirmed quasars with good
        (zWarning=0) redshifts above $z=2.2$.  The stellar locus is
	shown as contours.  
        {\it Top left, $ugr$; top right, $gri$, bottom left, $riz$.}
	The horizontal swath of both stars and quasars at $g-r \sim
	1.5$ in the $u-g,g-r$ color-color diagram is caused by the
	large $u$-band photometric errors in the reddest objects.  
        The colors of points encode their redshifts; the sizes of the
	points vary for clarity. The lower right panel shows the $i$
	magnitude as a function of the $g-r$ color.}
      \label{fig:color_color} 
    \end{figure*}

    \begin{figure*}
      \centering
      \includegraphics[height=8.0cm,width=14.0cm]
      {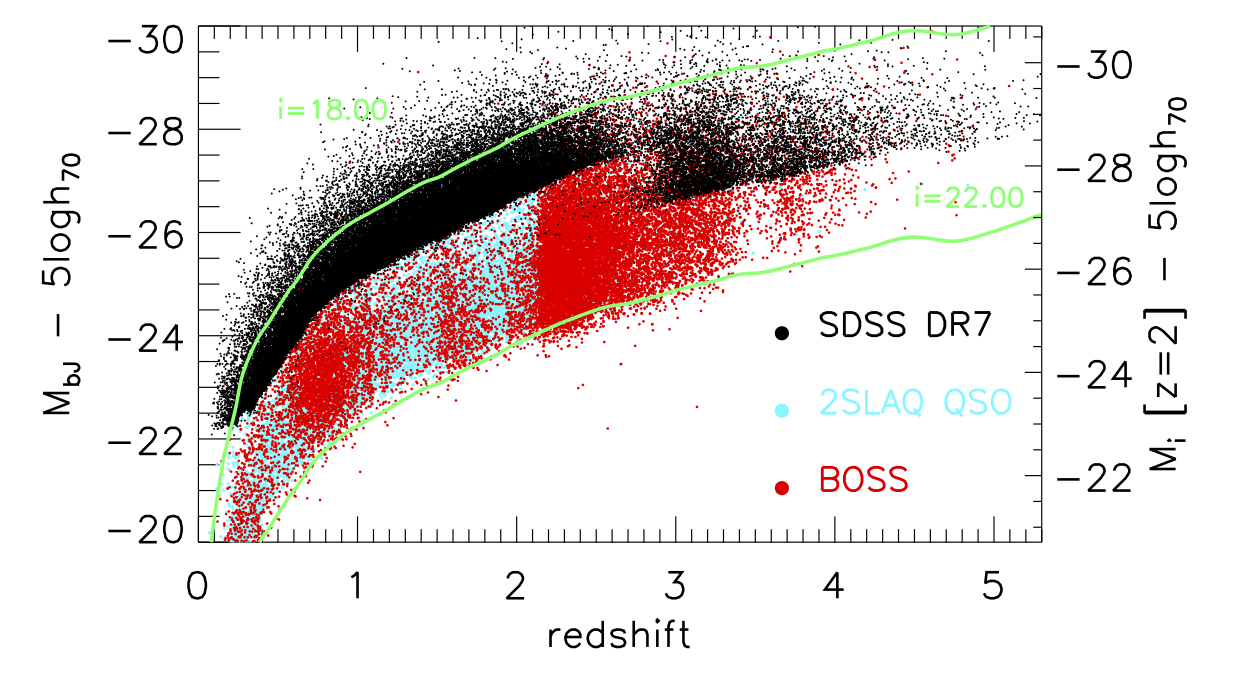}
      \caption[]
      {The $L-z$ plane for three recent quasar surveys: SDSS-I/II, 
        (black points), 2SLAQ (cyan) and BOSS (red).  The luminosity
	assumes $H_0 = 70\,\rm km\,s^{-1}\,Mpc^{-1}$.  There are
	$\approx 105,000$
        objects in the SDSS DR7 catalog and $\approx9,000$ 
        $g\leq21.85$ low-redshift quasars from the 2SLAQ Survey
        \citep{Croom09a}.  The three surveys together give a dynamic
        range in luminosity of $\approx4$ magnitudes at any given redshift up
        to $z\sim3.5$.  The luminosity corresponding to magnitude
	limits of $i=22$ on the faint end and $i=18$ on the bright end
	are shown.  The coverage here can be compared to Fig. 5 in \citet{Croton09}.}
      \label{fig:Croton_Lz}
    \end{figure*}
    \subsection{Magnitude, Color and the $L-z$ Plane}
    Fig.~\ref{fig:BOSS_QSOs_g_vs_uminusg} shows the distribution of
    quasar targets from the BOSS first-year data which are
    spectroscopically confirmed as either stars or $z>2.2$ quasars, in the
    {\it (u-g)} vs. {\it g} color-magnitude plane. The distribution of
    stars at the bright end, $g<18$, and the lack of bright $z > 2.2$
    quasars, led us to impose the bright $i=17.8$ limit. Objects fainter
    than $g = 22$ are brighter than our $r$ band limit of 21.85 mag.
    
    Fig.~\ref{fig:BOSS_QSOs_ugri_vs_redshift} shows the SDSS $(u-g)$,
    $(g-r)$, $(r-i)$, and $(i-z)$ colors as a function of redshift for the
    BOSS Year One data. Also shown are the mean color in redshift bins
    (thin solid line), and the model of Bovy et al. (2011b, in prep.;
    thick colored line).  This model is systematically bluer than the data
    at low redshift; BOSS target selection systematically excludes
    UV-excess quasars, and thus those low-redshift quasars that happen to
    enter the sample are redder than the average quasar.  The trends with
    redshift are due to various emission lines moving in and out of the
    SDSS broadband filters, and the onset of the Ly$\alpha$ forest and
    Lyman-limit systems (e.g., \citealt{Fan99}, \citealt{Richards02,
      Richards03}, \citealt{Hennawi10}, \citealt{Bovy11} and
    \citealt{Peth11}, but see also \citealt{Prochaska09} and
    \citealt{Worseck11}).  McGreer et al.~(2011, in preparation) will
    present a detailed analysis of this diagram, and its implications for
    our completeness.
    
    Fig.~\ref{fig:color_color} shows the SDSS color-color diagrams for
    the first year BOSS quasars, for all quasars with good (zWarning=0)
    redshifts above $z=2.2$.  This figure illustrates the redshift
    dependence of quasar colors as the Ly$\alpha$ emission line moves from
    the $g$ band to the $r$-band at $z\approx3.5$.  Quasars with
    $2.2<z<3.5$ lie in the range $-0.3<(g-r)<0.6$, while objects with
    $z>3.5$ generally have $(g-r)>0.8$.
    
    Fig.~\ref{fig:Croton_Lz} shows the distribution of objects in the
    redshift-luminosity (``$L-z$'') plane for three recent large quasar
    surveys: SDSS (black points), 2SLAQ (cyan) and BOSS (red).  There are
    $\approx 105,000$ objects in the SDSS DR7 catalog, and $\approx9,000$
    $g\leq21.85$ low-redshift quasars from the 2SLAQ Survey
    \citep{Croom09a}. We calculate the absolute {\it i}-band magnitudes,
    $M_{i}$, using the observed $i$-band PSF magnitudes and the
    $k$-corrections given in Table 4 of \citet{Richards06}.  The three
    surveys together cover the $L-z$ plane well, with a dynamic range in
    luminosity of $\approx4$ magnitudes at any given redshift up to
    $z\sim3.5$. This coverage will be vital for calculating the evolution
    of the faint end of the quasar luminosity function, and placing strong
    constraints on the luminosity dependence of quasar clustering.

   \subsection{Comments on Several Chunks}
   Because of the BOSS hardware commissioning in Fall 2009, only 37.4
   deg$^{2}$ (out of a possible 220 deg$^{2}$) were observed in Chunk 1
   under survey-quality conditions after MJD 55169. Thus Stripe 82 was
   re-targeted, re-tiled and re-observed for Year Two as Chunk 11
   \citep{Palanque-Delabrouille11}. However, the non-survey quality data
   from prior to MJD 55169 were visually inspected during the very early
   part of the survey, and used to inform subsequent QTS decisions.
   
   In Chunks 1-6, the quasar target selection algorithm was generous,
   allocating 60-80 targets deg$^{-2}$.  Chunk 7 was the first time we
   ran the BOSS QTS at the nominal 40 targets deg$^{-2}$.  Of the 4,506
   unique targets in this chunk, 1,595 (35\%) are classified as $z>2.20$
   objects with zWarning=0 (Table~\ref{tab:raw_numbers}).  Although this
   does not reach the BOSS efficiency goal of 50\%, there are several
   reasons that this number can be considered a lower bound.  First,
   Chunk 7 is in the region of sky known to have a high density of faint
   stellar sources, due to the presence of the tidal stream of the
   Sagittarius dwarf spheroidal galaxy \citep[see ][ and our
   Fig.~\ref{fig:NGC_heatmap}]{Ibata95, Ibata97, Belokurov06}.  Second,
   visual inspections of the spectra identified 0.5-1 more high-$z$
   quasars per square degree than the pipeline, and while not all of
   these might be suitable for Ly$\alpha$F analyses (e.g., due to BALs
   which cause the pipeline to fail), there should be a net gain upon
   production of the final BOSS quasar catalogs. Finally, and potentially
   most importantly, we know that our target selection methods and
   algorithms have continued to improve, with the incorporation of XDQSO
   and ancillary data such as UKIDSS and GALEX (see also the discussion
   on a variability based QTS in \S~\ref{sec:conclusions}).
   
   In this context, the performance of QTS in Chunks 8 and 9, with
   only 11.5 and 9.5 $z > 2.2$ quasars deg$^{-2}$ respectively, was
   disappointing.  Chunk 8 lies at relatively low Galactic latitudes, and
   is affected by stellar contamination.  Chunk 9 is in a region of sky
   where there was neither previously known quasars nor FIRST radio
   coverage.  We continue to observe the rest of Chunks 8 and 9 in Year
   Two.

    \begin{figure}
      \includegraphics[height=6.0cm,width=8.0cm]
      {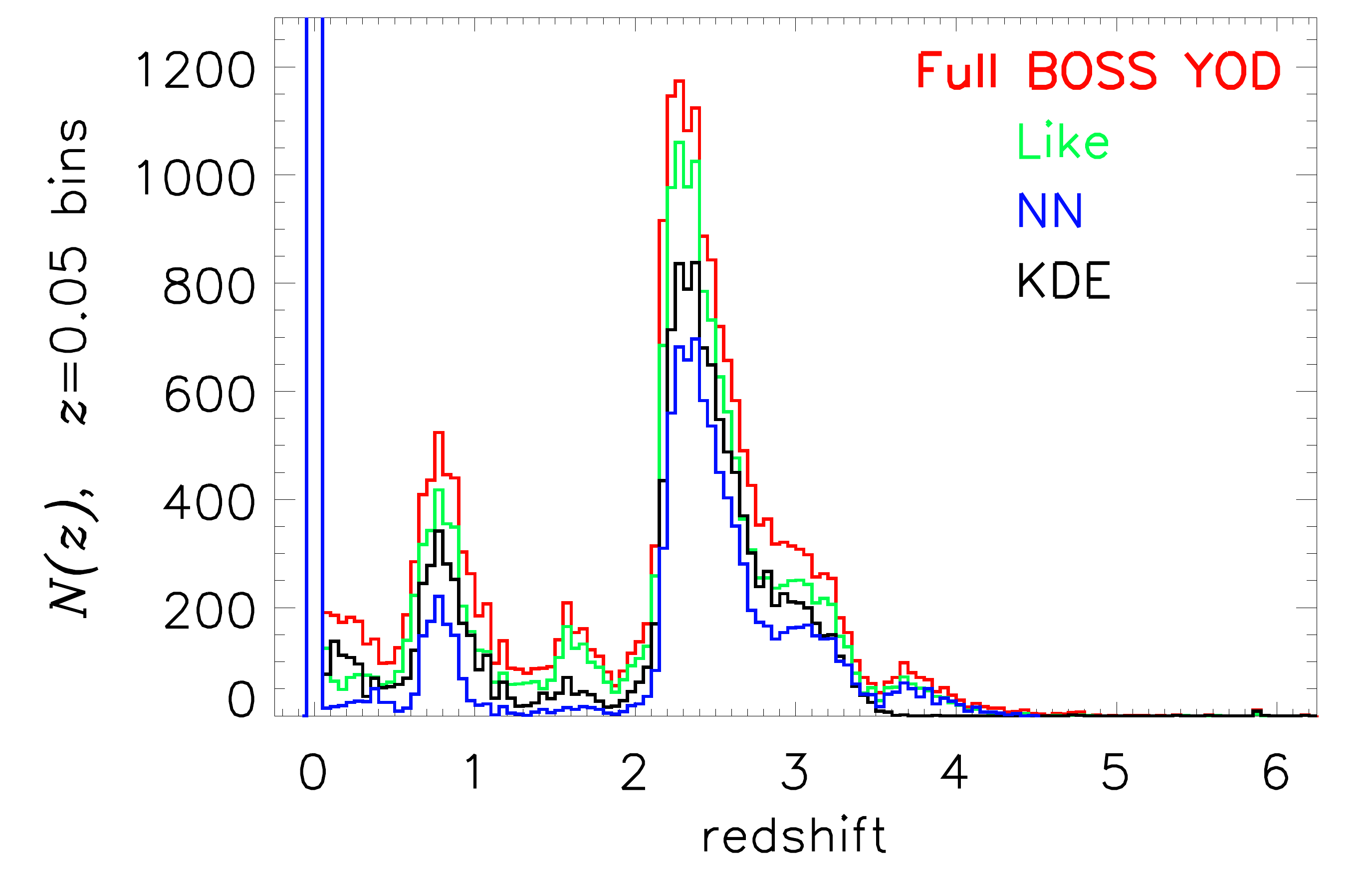}
      \caption[]
      {The BOSS quasar redshift distribution for objects with reliable 
        redshifts (zWarning=0), selected by our three main methods from
	Year One. The green, blue
        and black histograms give the redshift distributions for the
        Likelihood, NN and KDE methods, respectively.  The red histogram
        is the full sample from
	Figure~\ref{fig:BOSS_Quasar_Nofz_hist}.  These methods were
	not applied uniformly through Year One, so this plot is shown
	for {\it qualitative and informative} purposes only, and 
        should not be used as a direct comparison between the methods.
	The KDE, NN and Likelihood algorithms 
        are {\it not} mutually exclusive, with many objects selected
	by more than one method.}
      \label{fig:BOSS_Quasar_Nofz_flags_global} 
    \end{figure}
    \begin{table}
      \begin{center}
        \begin{tabular}{l ll ll}
          \hline
          \hline
          Selection & \# Quasar  & \# with    & and with        & or are        \\ 
                          & targets  &  zWarning=0  & $z>2.20$ &stars \\ 
         \hline 
          Totals & 52,238            & 33,556            & 13,580       & 11,149 \\
          KDE   & 34,503 (4794) & 20,993 (2693) &   9,050 (229) &   7,607 (1,856) \\
          NN  & 16,747 ( 975)  & 13,267 ( 710)  &   7,743 (135) &   3,604 ( 504) \\
         Likelihood  & 29,150 (2325) & 21,975 (1647) & 11,244 (447) &   4,483 ( 724) \\
         \hline
         \hline
       \end{tabular}
      \end{center}
         \caption[The number of targets, from the Year One data, broken down by the three key selection methods. ]
         {The number of unique quasar targets from the first year of
	   BOSS spectroscopy, broken down by the
            three key selection methods.  Numbers in parentheses
	    indicate the number of objects selected by the indicated
	    method {\it only}.  Because these methods were
	    applied non-uniformly, this table is
           provided as an {\it informational guide}, and not as a direct
           comparison between methods (see text for further explanation). }
         \label{tab:raw_numbers_3methods}     
    \end{table}

    \subsection{Comparison of Algorithms}\label{sec:efficiencies}
    The original motivation for the implementation of multiple target
    selection algorithms was the lack of evidence prior to BOSS
    observations that a single method could select $z>2.2$ quasars down to
    $g\approx22$ with our required efficiency. With the Year One data now
    in hand, we can compare the effectiveness of our different
    methods. However, due to the continually changing nature of the BOSS
    QTS over this year, where different methods were used as CORE and
    BONUS, these comparison will be generally qualitative in nature. The
    interested reader is referred to the discussions in \citet{Bovy11} for
    further comparisons.
    
    As an aid for our discussions, we give a condensed version of
    Table~\ref{tab:raw_flag_numbers} in
    Table~\ref{tab:raw_numbers_3methods}, where we list the number of
    targets from this first year, broken down by the three key selection
    methods.  Again, given the non-uniform selection over this year, this
    table is provided as an {\it informational guide} only; it should not
    be used as a direct comparison between methods.
    
    The redshift distributions for objects with reliable redshifts
    selected by our three main methods (NN, KDE, and Likelihood) are given
    in Fig.~\ref{fig:BOSS_Quasar_Nofz_flags_global}.  Again, because of
    the non-uniform manner in which these methods were applied during Year
    One, this plot should not be interpreted as a quantitative comparison
    between the methods.  There is substantial overlap between the
    methods; many objects are selected by more than one technique.  The
    three histograms have similar shapes over the range $2.2<z<3.5$.
    While NN avoids being confused by $z\sim1.5$ objects, and KDE avoids
    objects at $z>3.5$, all three methods select a substantial number of
    objects at $z \sim 0.8$.
    
\begin{figure}
      \includegraphics[height=8.0cm,width=8.0cm]
      {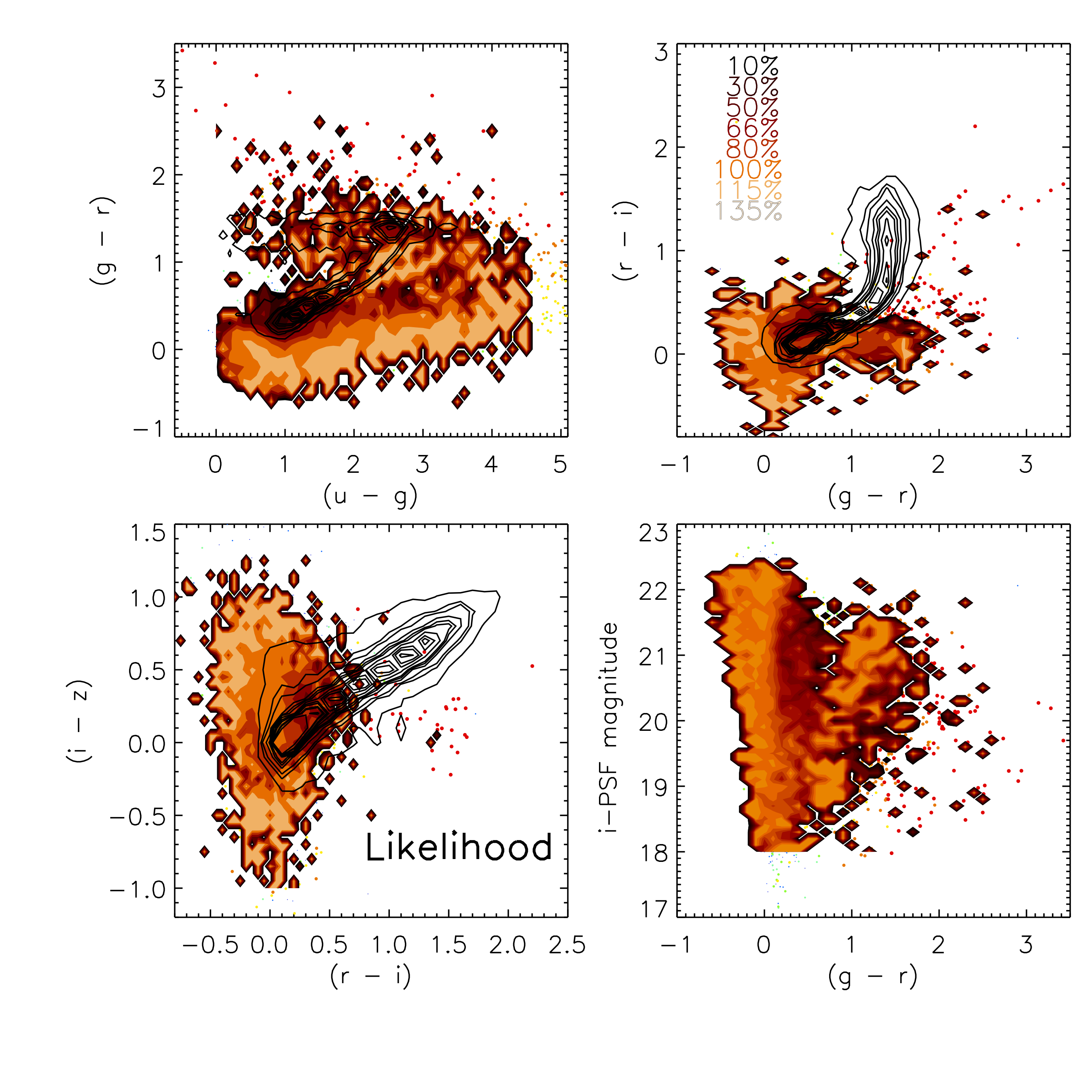}
      \caption[]
      {Distributions in color-color and color-magnitude space for $z >
	2.2$ quasars selected by the Likelihood 
	method in Year One.  The black contours give the location of
	the stellar locus, while the orange contours give the ratio,
	at each point of color space, of $z > 2.2$ quasars selected by
	Likelihood to all Year One BOSS quasars, normalized to the
	global ratio of the two.  Quasar numbers were smoothed with a
	tophat of width 0.10 mag in $u-g$ and $g-r$, and 0.05 mag in
	$r-i$ and $i-z$, before taking ratios.}
      \label{fig:color_color_Like} 
    \end{figure}

\begin{figure}
      \includegraphics[height=8.0cm,width=8.0cm]
      {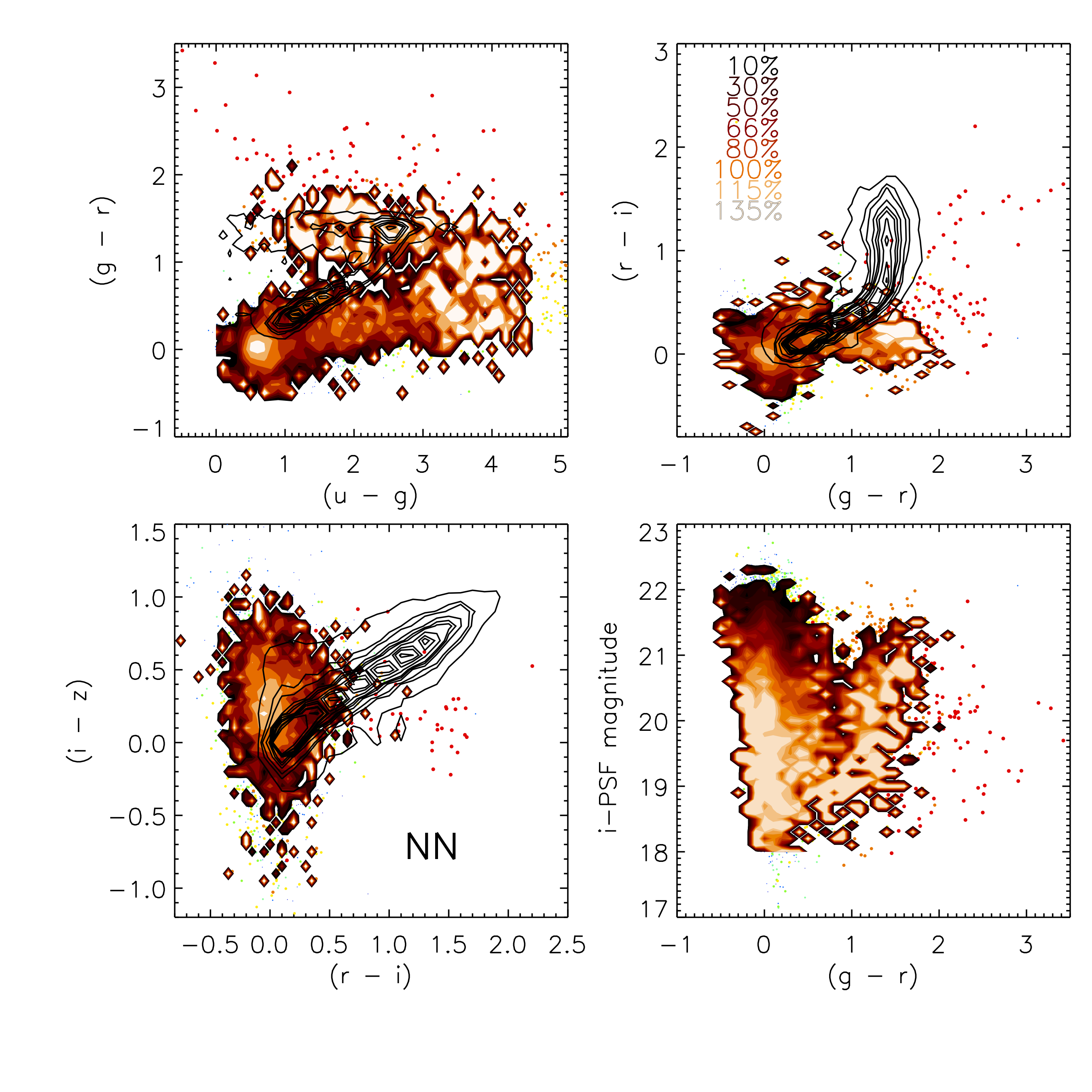}
     \caption[]
      {As in Figure~\ref{fig:color_color_Like}, for the NN method. }
      \label{fig:color_color_NN} 
    \end{figure}

\begin{figure}
      \includegraphics[height=8.0cm,width=8.0cm]
      {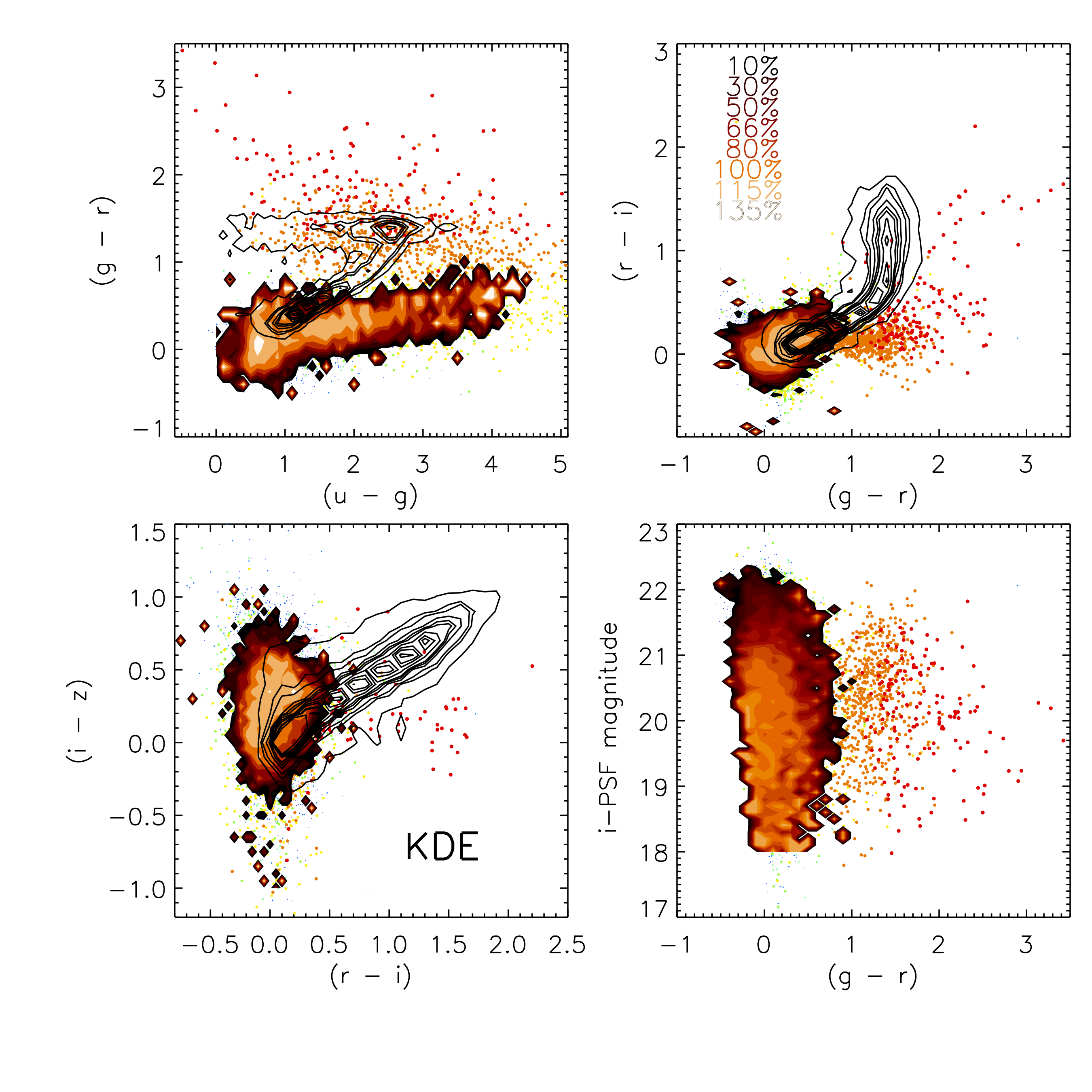}
     \caption[]
      {As in Figure~\ref{fig:color_color_Like}, for the KDE method. }
      \label{fig:color_color_KDE} 
    \end{figure}

    Figs.~\ref{fig:color_color_Like}, \ref{fig:color_color_NN}
    and~\ref{fig:color_color_KDE} show the color-color and the
    color-magnitude distributions of $z > 2.2$ quasars selected by the
    Likelihood, NN and KDE methods, respectively.  The figures show in
    orange and black the ratio of numbers of objects selected by each
    method to the total number of Year One quasars, at each point in color
    space.  This ratio is normalized to the global ratio of targets from
    Column 4 of Table~\ref{tab:raw_numbers_3methods}; thus a point in
    color space with a value $>100$\% is one where the method in question
    outperforms the total selection on average.  The difference between
    the three methods is clear in the $(u-g)$ vs.\ $(g-r)$ diagrams.  The
    contours for the Likelihood method are fairly flat away from the
    stellar locus.  NN performs well at $(u-g)\sim0.6$, $(g-r)\sim0$ and
    in those regions of color-color space corresponding to higher-redshift
    quasars, but does more poorly elsewhere.  KDE selects objects only
    over a very narrow range in $(g-r)$. From the $(g-r)$ vs.\ $i$-band
    color-magnitude diagram (bottom right panels of the figures), we see
    that the Likelihood method was more efficient at selecting fainter,
    $i\gtrsim21.0$ quasars, while the NN tends to select the brighter
    $i\lesssim20.0$ objects at all $(g-r)$ colors.
    
    These trends can be understood given the methodology of these
    algorithms.  The Likelihood method downweights objects close to the
    stellar locus as the denominator of
    equation~(\ref{eq:likelihood_ratio}) gets large, which is why
    Likelihood selects few objects there.  Otherwise, the Likelihood
    method traces the overall BOSS Year One sample in color-color and
    color-magnitude space.  The Likelihood method did not place any cuts
    on photometric redshift, and hence samples the high redshift
    distribution of the BOSS data well, especially at $(g-r)\gtrsim1$
    (corresponding to redshift $z>3.5$). We refer the interested reader to
    \citet{Kirkpatrick11} for full details of the Likelihood performance.
    
    At the crux of an artificial neural network is the sample of
    objects used to train it (see \citealt{Yeche10} and references
    therein, and Section~\ref{sec:nn}).  The training set for the NN we
    have used was based on the SDSS quasar catalog and the 2SLAQ surveys,
    and did {\it not} use data from the MMT pilot survey
    (Appendix~\ref{sec:MMT}) or the AUS survey. Thus, this training set
    was geared towards brighter quasars ($i<20.2$), giving rise to the
    tendency for NN to select the brighter quasars.
    
    The KDE training set included only $2.2<z<3.5$ quasars, and thus
    the redshift histogram drops to zero at $z = 3.5$
    (Fig~\ref{fig:BOSS_Quasar_Nofz_flags_global}).  This is related to the
    fact that KDE quasars inhabit a much narrower range of the $(g-r)\
    vs.\ (r-i)$ color-color plane than the other two methods.  In summary,
    Figures~\ref{fig:color_color_Like}-\ref{fig:color_color_KDE} reflect
    the relative strengths and trainings of these methods; ultimately, the
    three methods complemented each other well.
    
    \begin{table*}
      \begin{center}
        \begin{tabular}{l cc cc cc}
          \hline
          \hline
          Method & Threshold    & Threshold     & $N_{\rm quasar}$ (deg$^{-2}$) & $N_{\rm quasar}$ (deg$^{-2}$)  & Score (deg$^{-2}$) & Score (deg$^{-2}$)\\
          & @ 20 deg$^{-2}$ & @ 40 deg$^{-2}$ & @ 20 deg$^{-2}$      & @ 40 deg$^{-2}$             & @ 20 deg$^{-2}$   & @ 40 deg$^{-2}$ \\
          \hline
          KDE             & 0.904           & 0.599           & 9.45                          & 11.35                          & 4.79 & 5.71\\ 
          Likelihood   & 0.543           & 0.234           & 8.70                        & 12.23 & 4.39 & 5.89 \\
          Weighted Likelihood & 0.262          &  0.108          & 8.89            &  12.33 &  4.58 & 5.98 \\
          NN              & 0.852          & 0.563           &  7.62       & 10.84 & 4.00  & 5.51 \\
          NN Combinator        & 0.853         & 0.573 & 9.37 & 12.81 & 4.69 & 6.26 \\
          Color Box                &  n/a            & n/a   & 6.45 &   & 3.41 &  \\
          \hline
          \hline
        \end{tabular}  
        \caption{The surface density of
          spectroscopically confirmed $2.2 < z < 3.5$ quasars from
          early (Chunk 1, 2 and 3) BOSS spectroscopic data that would be
          recovered by various methods, and the thresholds of the
          key parameters (Table~\ref{tab:key_params}) required to
          yield a surface density of 20 or 40 deg$^{-2}$ in the 
          blind survey region (\S~\ref{sec:blind_test}).  The Weighted Likelihood
          incorporated a weighting function which optimizes the
          S/N of the Ly$\alpha$ forest clustering signal.  The
          redshift and flux distribution of the resulting
          quasar sample determines this signal, as quantified by
          the score in the last two columns. }
        \label{tab:blind_test} 
      \end{center}
    \end{table*}
    
    \subsection{The Blind Test Area}
    \label{sec:blind_test}
    After spectroscopy from the first few chunks had been
    analyzed, it became clear that the survey would have to decide on a
    single method for the CORE, and that we would have to restrict
    ourselves to the nominal target density of 40 targets
    deg$^{-2}$. Thus, we designed a test to decide which combinations of
    methods gave the best yields for the CORE and BONUS selections.
    
    The ``Blind Test Area'' is a region of sky of $\sim 1000$
    deg$^2$ in the NGC at high declination ($\delta>+40^\circ$) and high
    Galactic latitudes, shown by the thin white line in
    Fig.~\ref{fig:NGC_heatmap}. This area is used for tuning the threshold
    of each method to a particular target density.  The resulting
    thresholds were then applied to existing data to determine the
    selection efficiency.
    
    Table~\ref{tab:blind_test} summarizes these tests.  This table
    gives the surface density of $2.2 < z < 3.5$ quasars from early (Chunk
    1, 2 and 3) BOSS spectroscopic data that would be recovered by various
    methods at various thresholds of their key parameters when they are
    tuned to yield a surface density of 20 or 40 deg$^{-2}$ in the blind
    survey region.  The effectiveness of each quasar spectrum for
    Ly$\alpha$ forest studies depends on its redshift (and thus the
    spectral coverage of the forest) and its brightness (and thus the S/N
    of the spectrum).  This ``value'' is quantified by a score of each
    quasar, motivated by the checks performed in \citet{McDonald07};
    summing this over the expected quasars per square degree gives the
    numbers in Table~\ref{tab:blind_test}.  These scores do not include
    contributions from quasars outside the redshift range $2.2 < z < 3.5$.
    ``Weighted Likelihood'' was an adaption of the Likelihood method to
    maximize this score, as discussed in detail by \citet{Kirkpatrick11}.
    
    We also tried selecting quasars using a simple color region
    isolating the region where $z \sim 2.7$ quasars are found, akin to the
    mid-$z$ box used by \citet{Richards02}, but this did not deliver an
    efficiency close to our requirements.
    
    Although Table~\ref{tab:blind_test} shows that the KDE method
    returns the most $z>2.2$ quasars (9.45 deg$^{-2}$) at the CORE target
    density of 20 deg$^{-2}$, after much deliberation, we selected the
    Likelihood method as CORE for the latter stages of Year One, since it
    is a simpler algorithm to understand and explain, it has a more
    uniform spatial selection, and is easier to reproduce.  Further tests
    showed that using the Neural Network in its ``Combinator'' mode for
    BONUS would yield the highest number of high-$z$ quasars overall.  The
    difference when weighting by the Ly$\alpha$ forest score was too small
    to motivate us to include it; see the discussion in \citet{McQuinn11}.
    
    However, tests of the Year One data with the XDQSO method
    \citep{Bovy11} showed it selected about 1 $z > 2.2$ quasar deg$^{-2}$
    more than Likelihood.  Thus in Chunks 12 and 13
    (Section~\ref{sec:chunks12and13}) the union of Likelihood and XDQSO
    was treated as CORE, allowing us to test them directly against one
    another \citep{Bovy11}.  In Chunks 12 and 13, 2426 out of 4710 XDQSO
    targets had spectra with zWarning=0 and $2.2 < z < 3.5$, for an
    efficiency of 52\%, while Likelihood obtained 2296 quasars from 5086
    targets, for a 45\% efficiency.  This result is our motivation for
    declaring XDQSO to be CORE for the rest of the BOSS quasar survey.


\section{The Completeness of CORE in Year One}
\label{sec:discussion}
Studies of clustering in the \lyaf\ are not biased by the distribution
of background quasars used to illuminate \lyaf\ absorption. Thus the
Year One BOSS quasar sample can be used for these studies.  Indeed,
\citet{Slosar11} have performed a first clustering analysis of
Ly$\alpha$ forest flux from the BOSS Year One data.

However, given the changes in QTS throughout the first year, the
quasar sample described in this paper is far from sufficiently uniform
to be used directly for studies of the statistics of the quasars
themselves, such as measurements of their luminosity function or
clustering.  The goals of the CORE sample is to have such a
uniformly-selected sample of quasars, but as the definition of CORE
changed several times during commissioning, CORE objects in the first
year do not represent a statistical sample.

\begin{table}
  \begin{center}
    \begin{tabular}{rrrr}
      \hline
      \hline
                    &                            &Effective \\
                    &  Area (deg$^2$)  & Area (deg$^2$)     &  Mean \\
      Chunk    &  $C \ge 0.75$   &       $C \ge 0.75$   & $C$  \\
      \hline 
      11 &     70.6  &     58.2 &   0.654\\
      2 &    130.1  &    120.4 &   0.905\\
      3 &     85.9  &     79.4 &   0.830\\
      4 &    246.1  &    230.4 &   0.861\\
      5 &    243.0  &    232.0 &   0.952\\
      6 &    182.6  &    171.2 &   0.933\\
      7 &    205.0  &    185.8 &   0.836\\
   8 &     75.5  &     65.7 &   0.814\\
   9 &     84.1  &     71.6 &   0.822\\
   10 &     71.7  &     60.7 &   0.813\\
   \hline 
      \hline
    \end{tabular}  
    \caption{Fraction $C$ of objects that would have been targeted by the
      {\em a posteriori} XDQSO CORE algorithm, which were actually targeted,
      for each Year One chunk.  Chunk 11 has greater area coverage than
      Chunk 1, thus we list it instead.  The second column gives the solid
      angle (in deg$^2$) of the region of each chunk in which the
      completeness is greater than 0.75, the third column lists the same
      value but for {\em effective area} (i.e area $\times$ completeness)
      and the fourth column tabulates the mean completeness over the
      chunk. See also Fig.~\ref{fig:Blue_yellow_compl}.}
    \label{tab:Blue_yellow_compl}
  \end{center}
\end{table}

\begin{figure*}
  \begin{center}
    \includegraphics[height=10.0cm,width=14.5cm]
    {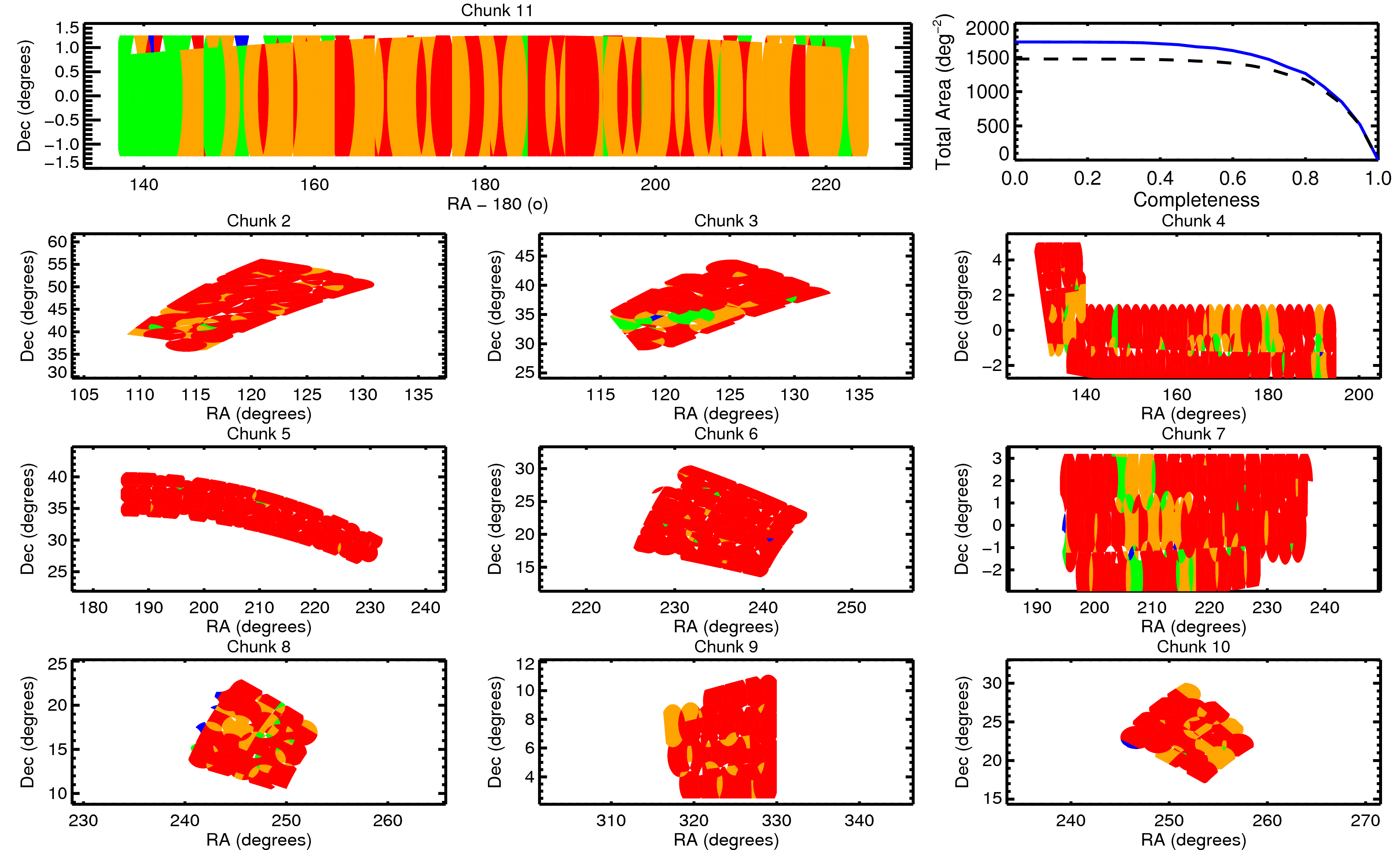}
    \caption[Year One Completeness]
    {The fraction of the objects would be targeted using the final version
      of the XDQSO CORE quasar target selection, that were actually targeted
      in Year One.  Each panel shows the area covered by a Chunk (2-11) from
      Year One. We use Chunk 11 on Stripe 82 in place of Chunk 1 (top-left
      panel) as Chunk 11 has superior areal coverage. Note that in some
      chunks, the scales on the RA and Dec axes are quite different.  Color
      coding shows the spectroscopic completeness of the {\em a posteriori}
      XDQSO CORE sample for each area.  Those areas in red have a targeting
      completeness above 0.75, orange have a completeness of 0.5--0.75,
      green have a completeness of 0.25--0.5 and the few areas in blue have
      a completeness below 0.25. The top right panel shows the cumulative
      area (blue solid line) and effective area (area $\times$ completeness;
      black dashed line) above a given level of targeting completeness for
      the XDQSO CORE sample.}
    \label{fig:Blue_yellow_compl}
  \end{center}
\end{figure*}

The project settled on the XDQSO algorithm (\S~\ref{sec:exd};
\citealt{Bovy11}) for the CORE method at the end of Year One, and will
use it for the rest of the survey.  It is therefore useful to apply
this algorithm to the photometry used in the Year One spectroscopy,
and determine the completeness of the Year One targeted chunks.
Table~\ref{tab:Blue_yellow_compl} and Fig.~\ref{fig:Blue_yellow_compl}
give the results of this test. Given the placement and overlap of the
spectroscopic plates, each chunk can be uniquely divided into sectors
covered by a unique combination of plates.  The completeness of the
targeting: i.e., the fraction of the XDQSO CORE sources that were
actually targeted in Year One, is measured for each sector
separately. Encouragingly, these targeting completeness values are
generally 80\% or higher, which indicates that statistical analyses of
the final CORE sample should be able to incorporate Year One data by
introducing moderate weighting factors.  The lower targeting
completeness (65\%) on Chunk 11 highlights a subtle point: the
completeness for CORE-selected {\it quasars} should be higher than the
completeness for CORE targets as a whole, because the true quasars are
the most likely to also be selected by one of our other algorithms.
In the case of Chunk 11, the deeper Stripe 82 photometry eliminates
many noisy stellar contaminants in the single-epoch XDQSO target list,
but it probably selects nearly all of the true quasars selected by
CORE.
   
{\it For Year Two and the remainder of the BOSS quasar Survey, the core
sample is defined by boss\_target1 flag QSO\_CORE\_MAIN (bit 40) and
QSO\_CORE\_ED for Chunks 12 and 13, and QSO\_CORE\_MAIN (bit 40) only
for later chunks (Table~\ref{tab:core_bits}).}

For calculations of the quasar luminosity function, one must also
account for the incompleteness of the XDQSO CORE sample relative to
the full population of quasars. This can be quantified, for example,
using the extensive targeting on Stripe 82
\citep{Palanque-Delabrouille11}.  Similarly, to determine completeness
as a function of position on the sky for quasar clustering work it is
necessary to determine the fraction of quasars hiding among the
unclassifiable spectra (see Appendix~\ref{sec:zwarning}).  Ongoing
visual inspections of these spectra will address this question to some
extent.
   
\begin{table}
  \begin{center}
    \begin{tabular}{ll}
      \hline
      \hline
      Chunk                 &  Bits to Select \\
      \hline 
      12, 13                 & 40 AND 42 \\
      14 and onwards  & 40 \\
      \hline 
      \hline
    \end{tabular}  
    \caption{The {BOSS\_TARGET1} flag values that need to be
      set in able to select a CORE sample from Year Two observations
      onward.}
    \label{tab:core_bits}
  \end{center}
\end{table}

\section{Conclusions and Future Prospects}
\label{sec:conclusions}
This paper describes the BOSS quasar target selection algorithms
during the first two years of BOSS observations.  BOSS aims to obtain
spectra of a sample of $\sim$150,000 $z>2.2$ quasars, in order to
probe structure in the \lyaf\ to provide a percent-level measurement
of the expansion history of the Universe, by measuring baryon
oscillations in the \lyaf\ clustering.  This first year was a
commissioning period for quasar target selection, and the algorithms
for identifying quasar candidates varied significantly over the year.

Our key results are:
\begin{itemize}
\item{We have performed quasar target selection (QTS) over
    10,200 deg$^{2}$ of the SDSS-III imaging footprint, producing a list of
    488,000 targets. These objects are selected to be at redshift $z>2.2$,
    motivated by the need to observe the Ly$\alpha$ forest in the BOSS
    wavelength coverage}.
\item{After a year of testing and evolution of the BOSS QTS, we
    settled on the Extreme Deconvolution method as our
    uniformly-selected subsample (CORE) and a neural network
    Combinator for the BONUS sample. }
  \item{Having the BONUS  selection allows us to implement
    improvements throughout the survey, 
      e.g., through auxiliary photometric data. This has already been 
      achieved with the inclusion of NIR $YJHK$ photometry from 
      the UKIDSS and UV data from GALEX, increasing our $z>2.2$
      quasar yields by 
      $\sim2-3$ deg$^{-2}$.}
  \item{We obtained spectra of 54,909 objects selected by the quasar
    target selection algorithms 
      over a footprint of 878 deg$^{2}$ during the first year of
      observations, the mean target density is 63.8 targets
      deg$^{-2}$.}
  \item{Of these 54,909 spectra, \hbox{33,556} were unique objects and
      had high quality spectra. \hbox{11,149} had redshifts $z<0.02$, and
      \hbox{13,580} had redshifts of $z>2.20$ (of which 11,263 were
      not previously known).}
  \item{Our mean $z > 2.2$ quasar surface density was 15.46 $z>
      2.20$ quasar deg$^{-2}$, with a global efficiency of 26.0\%.}
  \item{The $z > 2.2$ objects selected by the three main methods used
    during Year One 
      are found in different regions in color-color and color-magnitude
      space, reflecting in part the fact that the methods were trained
      for different redshift ranges.  The three methods complemented
      each other well, and together select 60-70\% of {\em all}
      quasars in our magnitude range with $2.2 < z < 3.5$.}
  \item{Working with single-epoch SDSS data, our current target selection
      algorithms slightly exceed the BOSS technical goal of selecting
      15 $z > 2.2$ quasars deg$^{-2}$ from 40 targets deg$^{-2}$
      \citep{Eisenstein11}.  The tests on Stripe 82 indicate an 
      efficiency of 15.4 quasars deg$^{-2}$, of which 11.2 deg$^{-2}$
      come from known quasars plus the CORE selection at 20 targets 
      deg$^{-2}$ (Fig.~\ref{fig:EfficiencyCurveSingleEpoch}).
      We anticipate that use of auxiliary imaging data, including 
      GALEX, UKIDSS, and additional SDSS epochs in overlap regions,
      will boost our efficiency by $1-4$ quasars deg$^{-2}$,
      significantly increasing the statistical power of BOSS
      Ly$\alpha$ forest clustering measurements.}
  \item{All BOSS spectra from the first two years of observations,
      August 2009 through to July 2011, will be made publicly available in the
      next SDSS data release, DR9. }
\end{itemize}

We continue to investigate ways to improve quasar target selection.
We have already described the incorporation of data from ultraviolet
(GALEX) and near-IR (UKIDSS).  Data from the Wide-field Infrared
Survey Explorer (WISE; \citealt{WISE}) will provide photometry at
mid-infrared wavelengths for our targets; it is deep enough to detect
at least the brighter quasars in the BOSS sample.  Variability as measured from repeat
scans is an important method, independent of colors, to separate
quasars from stars.  
Building on the SDSS Stripe 82 study by \citet[][]{Sesar07}, recent
investigations by \citet{Palanque-Delabrouille11}, \citet{Butler11},
\citet{MacLeod11}, \citet[][]{Richards11}, 
\citet{Kozlowski11} and \citet{Sarajedini11}, have re-invigorated the
field of AGN identification through variability selection.

In addition to Stripe 82, roughly 50\% of the SDSS imaging footprint
has been imaged more than once \citep[][]{Aihara11}, primarily in overlaps
between adjacent stripes.  However, most of this area is observed only
a few times, over timescales of days, rather than the desired
month or year baselines that lead to efficient AGN selection. 

In this regard, the Palomar Transient Factory
\citep[PTF;][]{Law09}\footnote{http://www.astro.caltech.edu/ptf/}
could be a natural dataset to use for this purpose. The PTF is an automated, wide-field
imaging survey aimed at the exploration of the optical transient sky.
PTF uses the 1.2m Schmidt telescope at Palomar Observatory
with a 8 deg$^{2}$ field-of-view to perform large area transient
searches. An area of several hundred deg$^{2}$ can be imaged in one
night, typically in the Mould {\it R}-band but also in the SDSS {\it
g}-band. We are actively investigating the inclusion of PTF imaging
data into BOSS QTS.

PTF could also potentially
aid BOSS QTS by improving star/galaxy separation at the faint end.
Potentially any of the PTF
variability methods could work with other transient/variability based
surveys as well, e.g. the Pan-STARRS survey, \citep{Kaiser02}.

\acknowledgments
We kindly thank Nurten Filiz Ak for providing us with the variable BAL
information and spectra shown in Fig.~\ref{fig:eg_spectra}.  This work
was partially supported by the National Science Foundation grant
AST-0607634 (N.P.R. and D.P.S.). A.D.M. was partially supported by
NASA (grant NNX08AJ28G) and is a research fellow of the Alexander von
Humboldt Foundation.  J.B. was partially supported by NASA (grant
NNX08AJ48G) and the NSF (grant AST-0908357). M.A.S. acknowledges the
support of NSF grant AST-0707266. N.P.R. thanks Gabor Worseck, Nick
Mostek and Anna Rosen for helpful discussions.

The observations reported here were obtained in part at the MMT
Observatory, a facility operated jointly by the Smithsonian
Institution and the University of Arizona. Some MMT telescope time was
granted by NOAO (program 2008B-0282), through the Telescope System
Instrumentation Program (TSIP). TSIP is funded by NSF.

Funding for SDSS-III has been provided by the Alfred P. Sloan
Foundation, the Participating Institutions, the National Science
Foundation, and the U.S. Department of Energy. The SDSS-III web site
is \href{http://www.sdss3.org/}{http://www.sdss3.org/}.  SDSS-III is
managed by the Astrophysical Research Consortium for the Participating
Institutions of the SDSS-III Collaboration including the University of
Arizona, the Brazilian Participation Group, Brookhaven National
Laboratory, University of Cambridge, University of Florida, the French
Participation Group, the German Participation Group, the Instituto de
Astrofisica de Canarias, the Michigan State/Notre Dame/JINA
Participation Group, Johns Hopkins University, Lawrence Berkeley
National Laboratory, Max Planck Institute for Astrophysics, New Mexico
State University, New York University, Ohio State University,
Pennsylvania State University, University of Portsmouth, Princeton
University, the Spanish Participation Group, University of Tokyo,
University of Utah, Vanderbilt University, University of Virginia,
University of Washington, and Yale University.  \\ {\it Facilities:
SDSS, MMT}

\appendix
   
\section{Appendix A: Quasar Targeting Logic Cuts}
\label{sec:appflags}
\newcommand{\fb}{{\em bitmask }}
\newcommand{\fbpunc}{{\em bitmask}}

This Appendix describes the various quality cuts that objects from the
SDSS photometric pipeline must satisfy to be considered for
selection using the algorithms described in \S~\ref{sec:methods}.
Target selection is restricted to sources that are unresolved in SDSS
imaging, as determined by the difference between the model and PSF
magnitudes \citep{Stoughton02}; such objects are flagged with {\tt
  OBJC\_TYPE $=$ 6} in the outputs of the SDSS photometric pipeline
\citep{Lupton01}. 

  To reduce processing time, we precalculate a number of
  combinations of flags from the 
  photometric pipeline and the photometric calibration
  \citep{Padmanabhan08a}.  These flags are used in different ways for
  different target selection algorithms, as summarized in
  Table~\ref{tab:PhotoFlags}: for example, we are not as
  stringent for objects selected as FIRST radio sources
  (\S~\ref{sec:Radio_Sec2}) as we are for those objects which are selected
  by their colors.  In the main text, we refer to various combinations
of the six flag combinations described in this Appendix. 

   \subsection{Is the Photometry Clean?}
    \label{sec:FlagLogic}
  The photometric pipeline sets a series of flag bits for each
  detected object which identify problems with the processing of the
  SDSS photometry, ranging from the presence of bad columns to issues
  with deblending \citep{Stoughton02}.  These are particularly useful
  in recognizing when the photometry might be poor, and therefore
  color selection of targets unreliable.  The detailed meaning of the
  specific flag bits in what follows is described in
  \citet{Stoughton02} and the SDSS-III web page\footnote{\tt
    http://www.sdss3.org/dr8/algorithms/photo\_flags.php}; the
  logic behind these flag combinations is given in
  \citet{Richards02}.  

 Note that unlike the latter paper, we
  did not calculate and apply the flag checks on each
  band separately, and just use the flags associated with the union of
  the detections in the five SDSS bands.  While this could cause us to
  reject some genuine quasars, checks on Stripe 82 (where the flag
  checking on the coadded data was significantly less strict; see
  below) showed only a statistically
  insignificant 1\% difference in the number of quasars identified.  

We first define a combination of flag bits that denotes whether the
source in question was adversely affected by interpolation across bad
pixels, bad columns, or bleed trails: \\
    
    \noindent
    {\tt INTERP\_PROBLEMS} = ({\tt PSF\_FLUX\_INTERP} \&\& ($ {\rm gerr}>0.2 \parallel {\rm rerr} >0.2 \parallel {\rm ierr}>0.2$)) $\parallel$ {\tt BAD\_COUNTS\_ERROR} $\parallel$ ({\tt INTERP\_CENTER} \&\& {\tt CR}), \\

\noindent a combination that identifies objects in which the deblending of
  overlapping images may be questionable: \\

    \noindent
    {\tt DEBLEND\_PROBLEMS} = {\tt PEAKCENTER} $\parallel$ {\tt NOTCHECKED} $\parallel$ ({\tt DEBLEND\_NOPEAK} \&\& (${\rm gerr}>0.2 \parallel {\rm rerr}>0.2 \parallel {\rm ierr}>0.2$))\\
    
\noindent and a combination which identifies objects with detectable proper
motion between the exposures in the different SDSS filters
(asteroids): \\

\noindent  {\tt MOVED} =  {\tt DEBLENDED\_AS\_MOVING} \&\&
    (rowv/rowverr)$^2$ + (colv/colverr)$^2 > 3^2$.\\

    \noindent Here, the symbols (\&\&, $\parallel$, !) have their
    standard meanings from 
    Boolean logic.  The quantities rerr, gerr, and ierr are the quoted uncertainties
    in the PSF photometry in $g$, $r$, and $i$ respectively, rowv and colv are the measured proper
    motion along the rows and columns of the CCD, and rowverr and
    colverr are their errors. 

A source is considered to have clean
    photometry if it satisfies the following:\\
    
    \noindent
    {\tt GOOD} = {\tt BINNED1} \&\& {\tt !BRIGHT} \&\& {\tt
      !SATURATED} \&\& {\tt !EDGE} \&\& {\tt !BLENDED} \&\& {\tt
      !NODEBLEND} \&\& {\tt !NOPROFILE} \&\& {\tt !INTERP\_PROBLEMS}
    \&\& {\tt !DEBLEND\_PROBLEMS} \&\& {\tt !MOVED} . \\
    
    \begin{table*}
      \begin{center}
        \begin{tabular}{lcl ccc}
          \hline
          \hline
          Flag Name  &  Bitmask &  Description & CORE/BONUS & FIRST & KNOWN  \\
          \hline
          {\tt GOOD}               & 11 & Target has clean SDSS photometry                     & $\surd$ & $\times$  & $\times$ \\
          {\tt GMAG\_BITMASK}               & 11 & Target meets the magnitude limits                      & $\surd$ & $\times$  & $\times$ \\
          {\tt GMAG\_BITMASK\_NOB}     & 12 & Used when no bright cut is required                    &  $\times$ & $\surd$   & $\times$\\
          {\tt RESOLVE\_BITMASK}          & 13  & Target is a primary target in SDSS photometry   &$\surd$  & $\surd$ &$\surd$\\
          {\tt BOUNDS\_BITMASK}           & 16 & Target lies within the SDSS target footprint           & $\times$  & $\surd$ & $\surd$\\
          {\tt FIRST\_COLOR\_BITMASK}  & 17 & Color cut for objects that match a radio source &$\times$  &$\surd$  &$\times$\\
          \hline
          \hline
        \end{tabular}  
        \caption{Flags used by BOSS Quasar Target Selection. }
        \label{tab:PhotoFlags} 
      \end{center}
    \end{table*}

    \subsection{Magnitude Limits}
    \label{sec:MagLogic}
    The {\tt GMAG\_BITMASK} records whether a target satisfies  the 
    magnitude limits required to be targeted as a quasar. Magnitude cuts
    are made on PSF magnitudes measured by the SDSS, corrected for \citet{SFD}
    Galactic extinction.  These limits are encoded in a flag: \\

\noindent {\tt GMAG\_BITMASK} = $(g \le 22 \parallel r \le
    21.85)~\&\&~i \ge 17.8$.\\

This includes a cut at the bright end, reflecting the fact
that bright $z > 2.2$ quasars are extraordinarily rare.  We also
define a variant of this flag: \\

\noindent {\tt GMAG\_BITMASK\_NOB}  = $(g \le 22 \parallel r \le
    21.85)$,\\

\noindent to be used when no bright
    cut is required---such as when retargeting known quasars or FIRST
    objects.

    \subsection{Resolving Image Overlaps}
   The DR8 paper \citep{Aihara11} describes the algorithm used to
   define the primary detection of a given object, if it lies in the
   $\sim 50$\% of the SDSS footprint covered by more than one scan.  
    The {\tt RESOLVE\_BITMASK} records whether a source is a primary
    target in the SDSS photometry. 
   
   \subsection{Boundary Logic}
    \label{sec:BoundsLogic}   
    The {\tt BOUNDS\_BITMASK} records whether a source is within the
    footprint of the SDSS imaging, which is useful for keeping track
    of data from the ancillary surveys (FIRST, UKIDSS, GALEX) used in
    the target selection. 
    
    \subsection{FIRST Color Logic}
    \label{sec:FIRSTLogic}
    We saw in \S~\ref{sec:Radio_Sec2} that we could limit the number of $z
    < 2.2$ sources targeted by FIRST with $u-g$ color cut.  Thus we
    define:\\

\noindent {\tt FIRST\_COLOR\_BITMASK} = ($u-g > 0.4$).  \\

    \subsection{Conditions for Generation of Stripe 82 Coadded Photometry}
    \label{sec:VarCats}
    The single-epoch photometry used for coaddition on Stripe 82 is
    first vetted by a series of quality cuts.  All fluxes used in the
    coaddition are limited by the following conditions: 
\begin{itemize} 
\item They must be primary, i.e., {\tt RESOLVE\_BITMASK} must be true;
\item They must be observed under photometric conditions (an important
  issue from Stripe 82, as it was repeatedly observed under
  non-photometric conditions as part of the SDSS Supernova Survey; see
  \citealt{Frieman08}); 
\item They must have a positive estimated inverse flux variance (zero
  values are indicative of problems with the data);
\item They must pass various flag cuts: \\
\noindent (!{\tt DEBLEND\_TOO\_MANY\_PEAKS} \&\& !{\tt  SATUR} \&\& !{\tt  BADSKY} \&\& !{\tt  SATUR\_CENTER} \&\& !{\tt 
    INTERP\_CENTER} \&\& !{\tt  DEBLEND\_NOPEAK} \&\& !{\tt  PSF\_FLUX\_INTERP}). 
\end{itemize}

\section{Appendix B: Flowchart for Year One QTS and Target Selection Versions}
\label{sec:flowchart_YearOne}
\begin{figure*}
  \begin{center}
    \includegraphics[height=18.7cm,width=14.4cm]
    {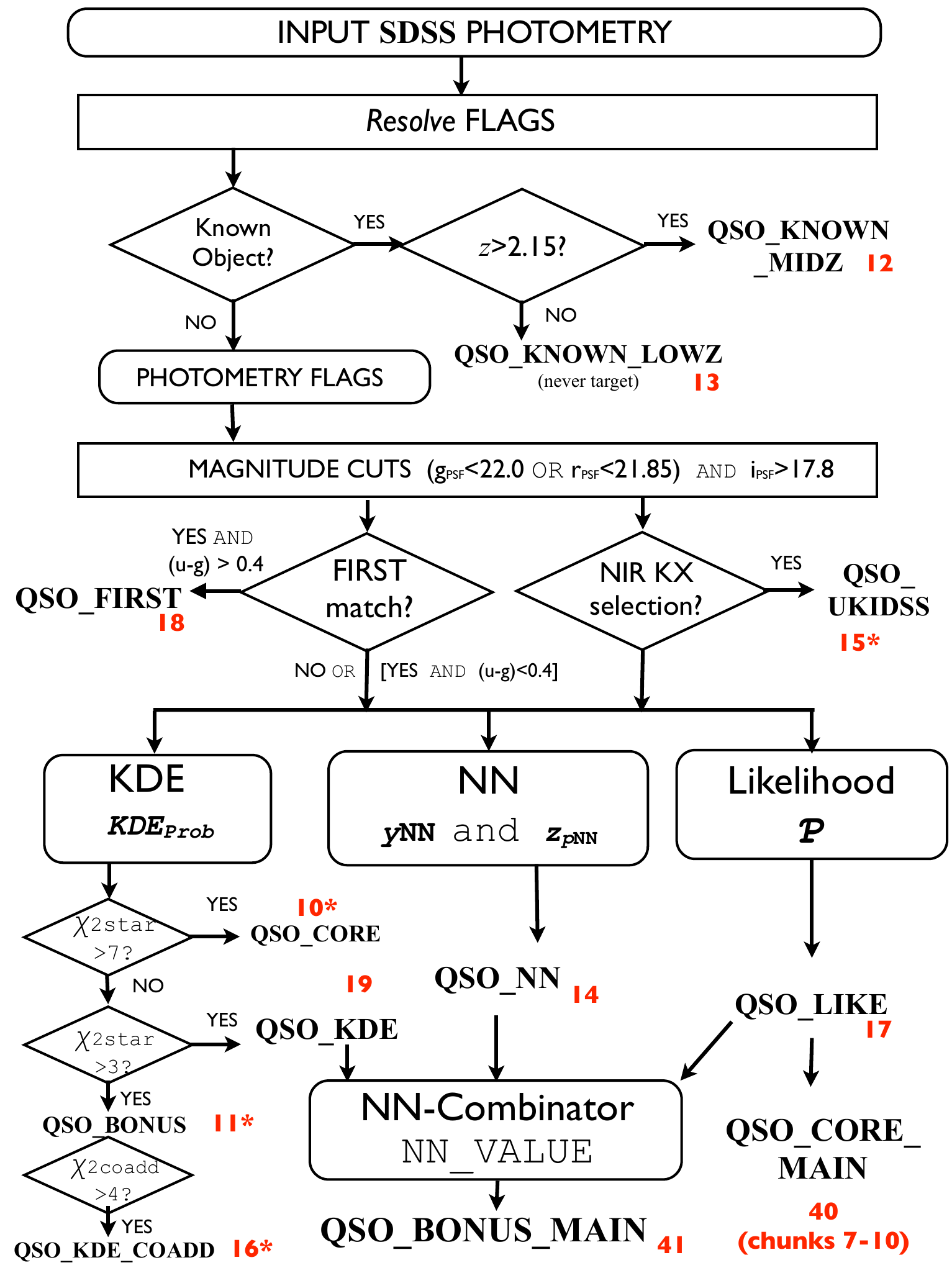}
    \caption{Schematic flowchart for the BOSS quasar target selection during
      the first year of observations, to be compared with the Year Two
      version in Figure~\ref{fig:flowchart}. The red numbers give the bitwise 
      value for the {\tt boss\_target1} flag (see Table~\ref{tab:BOSS_TARGET_FLAGS}). The red numbers with
      asterisks have target flags that were obsolete after the first year
      of target selection. The input SDSS photometry is described
      in Section~\ref{sec:input_phot} and the algorithm to resolve
      overlapping images is explained in \citet{Aihara11}.  Previously known
      objects are described in Section~\ref{sec:prev_known} and the FIRST
      radio selection is given in Section~\ref{sec:Radio_Sec2}. The
      photometry flags are discussed in Section~\ref{sec:phot_flags} and in
      Appendix A. The three target selection methods (KDE, NN, and
      Likelihood) are described in \citet[][ and references
      therein]{Richards09}, \citet{Yeche10} and \citet{Kirkpatrick11},
      respectively, and are outlined in Section~\ref{sec:methods}.  Each of
      these methods produces one or more continuous parameters that quantify
      the confidence that the object in question is a high-redshift quasar:
      $\mathcal{P}$ for the Likelihood method; 
      KDE$_{Prob}$ and $\chi^{2}_{\rm star}$ for the KDE method; 
      $y_{NN}$and $z_{p, {\rm NN}}$ for the first Neural Network and 
      {\tt NN\_VALUE} for Neural Network Combinator.  The target selection flag bits
      are also shown, with
      descriptions in Table~\ref{tab:BOSS_TARGET_FLAGS}.  Objects
      with $i<17.8$ with FIRST counterparts are selected for spectroscopy.}
    \label{fig:flowchart_YearOne}.
  \end{center}
\end{figure*}

Figure~\ref{fig:flowchart} is a flowchart which describes quasar
target selection as it was carried out in Year Two and beyond.
Fig.~\ref{fig:flowchart_YearOne} gives the equivalent for Year One. The red numbers give the bitwise
value for the {\tt boss\_target1} flag. Those values with asterisks
have target flags that were obsolete after the first year of target
selection.

Table~\ref{tab:chunk_targets2} gives the BOSS quasar target selection 
version code label for each chunk. Sheldon et al. (2011, in prep.) will
describe in detail the differences between these versions.  

Because of the $62''$ diameter of the cladding around each optical
fiber, two objects with separation smaller than that angle cannot both
be observed on a given
spectroscopic plate, which means that an algorithm to decide
which of two objects in such a collision should take precedence is needed.
Our thinking on this evolved throughout Year One; the rules for each
chunk are given in Table~\ref{tab:chunk_targets2}.  By Chunk 14, we
settled on giving 
KNOWN, CORE, and FIRST quasar targets higher priority than galaxy
targets, with BONUS at lower priority.  

\begin{table}
      \begin{center}
        \begin{tabular}{lcc}
          \hline
          \hline
          Area Name & Target Selection Version label  & Tiling Priority \\
          \hline
          Chunk 1       &       comm                             &  all quasar targets before galaxy targets\\
          Chunk 2       &        comm2                           &  $''$ \\
          Chunks 3, 4  &    main002                            &  all galaxy targets before quasar targets \\
          Chunks 5, 6     &  main005                            &  $''$ \\
          Chunks 7, 8    &   main006                             & $''$ \\ 
          Chunk 9       &      main006-masksgc1             & $''$ \\
          Chunk 10    &  main006-collate-maskngc          &  KNOWN before galaxies; galaxies before CORE, BONUS, FIRST.\\          
          Chunk 11    & vcat-2010-07-02                          & $''$ \\
          Chunks 12, 13 & main008-sgc40                        & $''$ \\
          Chunk 14  & main008-edcore-maskngc40          & KNOWN, CORE, FIRST over galaxies; galaxies before BONUS. \\
          Chunk 15  & main008-edfinal-maskngc40          & $''$ \\
          Chunk 16  & main010-maskngc40                       & $''$ \\
          Chunk 17  &  main011-maskngc40                      & $''$  \\
          Chunk 18  &   main012-nosuppz-maskngc40      &  $''$ \\
          \hline
          \hline
        \end{tabular}  
      \end{center}
\caption{This table lists the internal label of the version of target
  selection code used in each chunk, and also explains the relative
  priority of different classes of target in the case of fiber collisions.}
     \label{tab:chunk_targets2}
    \end{table}

\section{Appendix C: Quasars from the MMT Pilot Program}
\label{sec:MMT}
Prior to the commencement of BOSS spectroscopy, we carried out
spectroscopy of quasar candidates selected from coadded photometry in
SDSS Stripe 82 to increase the number of faint quasars available in
the BOSS redshift range for testing and training of BOSS targeting
algorithms. Candidate quasars for these observations were selected in
two ways: first, using very inclusive cuts in the ($\chi^{2}_{\rm
  phot}$, $\chi^{2}_{\rm star}$) plane, where these $\chi^{2}$
statistics are as defined in \citet{Hennawi10}, and second, using the
methods outlined in \citet{Richards09, Richards09b}.  These
observations were intended to include as large a sample of $z > 2.2$
quasars as possible, but do not represent a statistically well-defined
sample, so we do not describe their selection in greater detail.

Observations of these candidates were carried out in queue mode
between 2008 September and 2009 January using the Hectospec
multi-fiber spectrograph \citep{Fabricant05} on the 6.5m Multiple
Mirror Telescope (MMT). The data were reduced using Juan Cabanela's
ESPECROAD\footnote{\url{http://iparrizar.mnstate.edu/~juan/research/ESPECROAD/index.php}}
pipeline, an external version of the SAO SPECROAD pipeline
\citep{Mink07}. Quasars were identified by eye, and redshifts were
measured using IRAF.

The MMT program was conducted before the release of the SDSS DR7
quasar catalog. In addition, BOSS targets all confirmed quasars from
the MMT program for re-observation (\S~\ref{sec:prev_known}). Thus,
most of the MMT observations have been superseded by subsequent SDSS
DR7 or BOSS spectroscopy at better resolution, wavelength coverage and
signal-to-noise ratio. In
Tables~\ref{tab:MMTquasars}--\ref{tab:MMTquasarsnonprim}, we provide
positions, PSF photometry (as observed, uncorrected for Galactic extinction), and redshifts for 
confirmed quasars from the MMT survey. Objects that are not flagged
Primary in the CAS are listed separately. Over 99\% of quasars that {\it
were} observed a second time have redshifts in agreement (to $\Delta z
< 0.05$) between the MMT survey and the SDSS/BOSS pipelines.

\begin{table*}
   \begin{center}
      \begin{tabular}{rrlllllllllll}
         \hline \hline
           \colhead{RA} & \colhead{DEC} & \colhead{$\mathrm{u}$} &
	   \colhead{$\mathrm{u_{err}}$} & \colhead{$\mathrm{g}$} &
	   \colhead{$\mathrm{g_{err}}$} & \colhead{$\mathrm{r}$} &
	   \colhead{$\mathrm{r_{err}}$} & \colhead{$\mathrm{i}$} &
	   \colhead{$\mathrm{i_{err}}$} & \colhead{$\mathrm{z}$} &
	   \colhead{$\mathrm{z_{err}}$} &  \colhead{redshift $z$} \\
         \hline
          00 46 00.48 & +00 05 43.7 & 23.488 &  0.384 & 22.241 &  0.070 & 21.775 &  0.062 & 21.365 &  0.060 & 20.455 &  0.119 & 2.460 \\
          00 46 31.22 & -00 11 46.2 & 22.802 &  0.381 & 21.535 &  0.048 & 20.541 &  0.032 & 20.217 &  0.041 & 19.445 &  0.062 & 2.451 \\
          00 46 42.32 & -00 07 53.7 & 20.947 &  0.074 & 20.282 &  0.028 & 20.031 &  0.027 & 19.882 &  0.039 & 19.719 &  0.072 & 2.234 \\
          00 46 47.93 & -00 06 17.4 & 20.005 &  0.044 & 19.296 &  0.023 & 19.140 &  0.021 & 19.070 &  0.035 & 19.002 &  0.047 & 2.850 \\
          00 47 20.78 & +00 18 06.7 & 21.252 &  0.091 & 20.969 &  0.034 & 20.747 &  0.036 & 20.544 &  0.039 & 20.919 &  0.198 & 1.610 \\
          00 47 21.06 & +00 09 32.3 & 25.176 &  0.529 & 21.628 &  0.045 & 20.690 &  0.030 & 20.393 &  0.033 & 20.272 &  0.104 & 3.573 \\
          00 47 32.61 & -00 16 35.7 & 23.181 &  0.300 & 22.038 &  0.062 & 21.825 &  0.067 & 21.606 &  0.076 & 20.872 &  0.160 & 2.610 \\
          00 47 43.04 & -00 23 32.0 & 23.872 &  0.475 & 22.251 &  0.074 & 22.077 &  0.084 & 21.928 &  0.100 & 21.560 &  0.281 & 2.837 \\
          00 47 51.18 & -00 15 44.9 & 21.865 &  0.111 & 21.065 &  0.031 & 20.686 &  0.038 & 20.524 &  0.038 & 20.174 &  0.089 & 2.477 \\
          00 47 55.49 & +00 14 42.3 & 23.320 &  0.546 & 21.762 &  0.053 & 21.431 &  0.056 & 21.214 &  0.064 & 20.483 &  0.138 & 0.822 \\
          \hline 
          \hline
      \end{tabular}
      \caption{Quasars discovered in the MMT survey. Many of these objects were
               subsequently confirmed in the SDSS DR7 quasar catalog or in the BOSS.
               Imaging information is taken from the SDSS DR8 Catalog Archive Server. 
               The first 10 objects are given to show the format of the table. 
               This table is available in its entirety in machine-readable and Virtual Observatory
               (VO) forms in the online journal.}
      \label{tab:MMTquasars}
  \end{center}
\end{table*}

\begin{table*}
   \begin{center}
      \begin{tabular}{rrlllllllllll}
         \hline \hline
           \colhead{RA} & \colhead{DEC} & \colhead{$\mathrm{u}$} &
	   \colhead{$\mathrm{u_{err}}$} & \colhead{$\mathrm{g}$} &
	   \colhead{$\mathrm{g_{err}}$} & \colhead{$\mathrm{r}$} &
	   \colhead{$\mathrm{r_{err}}$} & \colhead{$\mathrm{i}$} &
	   \colhead{$\mathrm{i_{err}}$} & \colhead{$\mathrm{z}$} &
	   \colhead{$\mathrm{z_{err}}$} &  \colhead{redshift $z$} \\
         \hline
          00 46 39.91 & -00 05 03.7 & 21.784 &  0.184 & 21.572 &  0.075 & 21.498 &  0.096 & 21.396 &  0.114 & 20.750 &  0.264 & 2.235 \\
          00 57 16.14 & +00 21 04.7 & 21.896 &  0.153 & 21.700 &  0.090 & 21.649 &  0.226 & 21.116 &  0.316 & 22.734 &  0.385 & 2.110 \\
          02 33 23.79 & -00 02 11.1 & 23.946 &  0.659 & 21.480 &  0.054 & 22.083 &  0.138 & 21.755 &  0.151 & 22.929 &  0.861 & 2.402 \\
          03 37 10.37 & +00 23 55.1 & 20.082 &  0.074 & 19.279 &  0.123 & 18.978 &  0.150 & 18.922 &  0.166 & 18.938 &  0.129 & 2.920 \\
          03 37 33.89 & -00 03 04.7 & 21.458 &  0.120 & 20.264 &  0.259 & 19.667 &  0.021 & 19.282 &  0.024 & 19.127 &  0.050 & 0.671 \\
          22 58 58.68 & -00 20 38.0 & 21.924 &  0.317 & 21.373 &  0.078 & 21.077 &  0.085 & 20.967 &  0.105 & 20.497 &  0.294 & 2.421 \\
          23 07 33.34 & -00 17 58.9 & 21.981 &  0.186 & 21.884 &  0.088 & 22.491 &  0.209 & 21.815 &  0.163 & 21.323 &  0.334 & 2.765 \\
         \hline \hline
      \end{tabular}
      \caption{Quasars discovered in the MMT survey that are non-primary in SDSS
               DR8 imaging}
      \label{tab:MMTquasarsnonprim}
   \end{center}
\end{table*}

\section{Appendix D: Performance of zWarning} 
\label{sec:zwarning}
Here we present the fraction of spectroscopically observed quasar
targets which are flagged with zWarning${}\ne 0$ by the spectroscopic pipeline.  As
described in \citet{Adelman-McCarthy08}, this is an indication that
the automatically derived redshift and classification are not
reliable.  

\begin{table*}
  \begin{center}
    \begin{tabular}{lclr}
     \hline
      \hline
      {\tt zWarning} flag          &   bit  &  Description  & No. of
      objects in Year One (unique) \\
      \hline
      No flag set                    &   -  & Spectrum has no known problems. & 35,305 (33,556) \\
      SMALL\_DELTA\_CHI2  &  2   & $\chi^{2}$ best fit is too close to
      that of second best ($< 0.01$ in reduced $\chi^{2}$) & 16,765    (15,982) \\ 
      NEGATIVE\_EMISSION   & 6    & a quasar line exhibits negative emission. & 620 (597) \\
     \hline
     \hline
    \end{tabular}  
 \end{center}
  \caption[The zWarning flags]
  {zWarning flag bits and Year One Quasar Spectroscopy}
    \label{tab:BOSS_zWarning_FLAGS}
\end{table*}
Table~\ref{tab:BOSS_zWarning_FLAGS} gives the three most common of the
zWarning flag bits for quasar targets, a short description of each, and 
    the number of objects with these bits set.  1851 objects have both
    bits 2 and 6 set. All other zWarning bits are set in 200 or
    fewer objects, representing less than 1\% of the sample.

\begin{figure}
  \begin{center}
    \includegraphics[height=8.0cm,width=8.0cm]{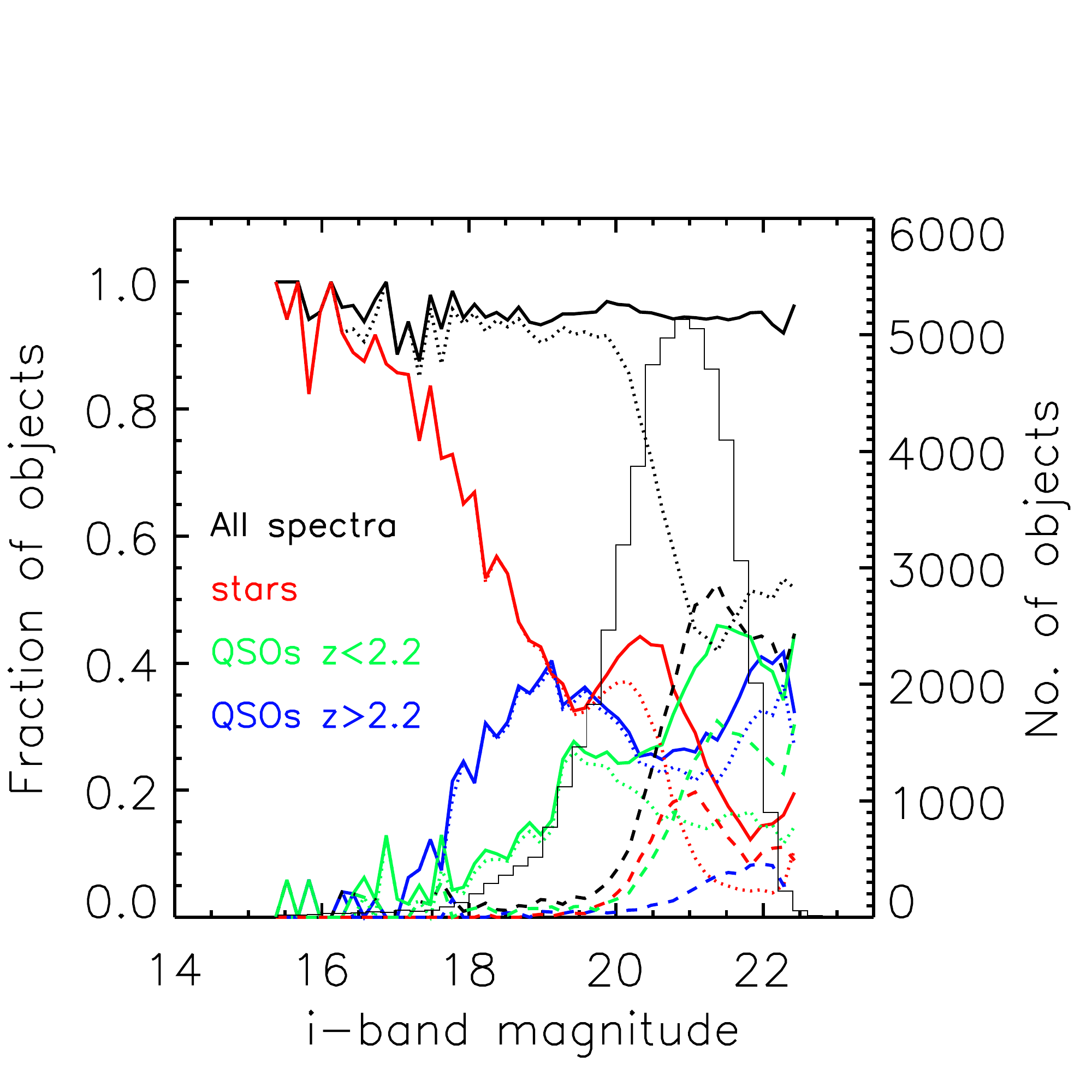}
    \includegraphics[height=8.0cm,width=8.0cm]{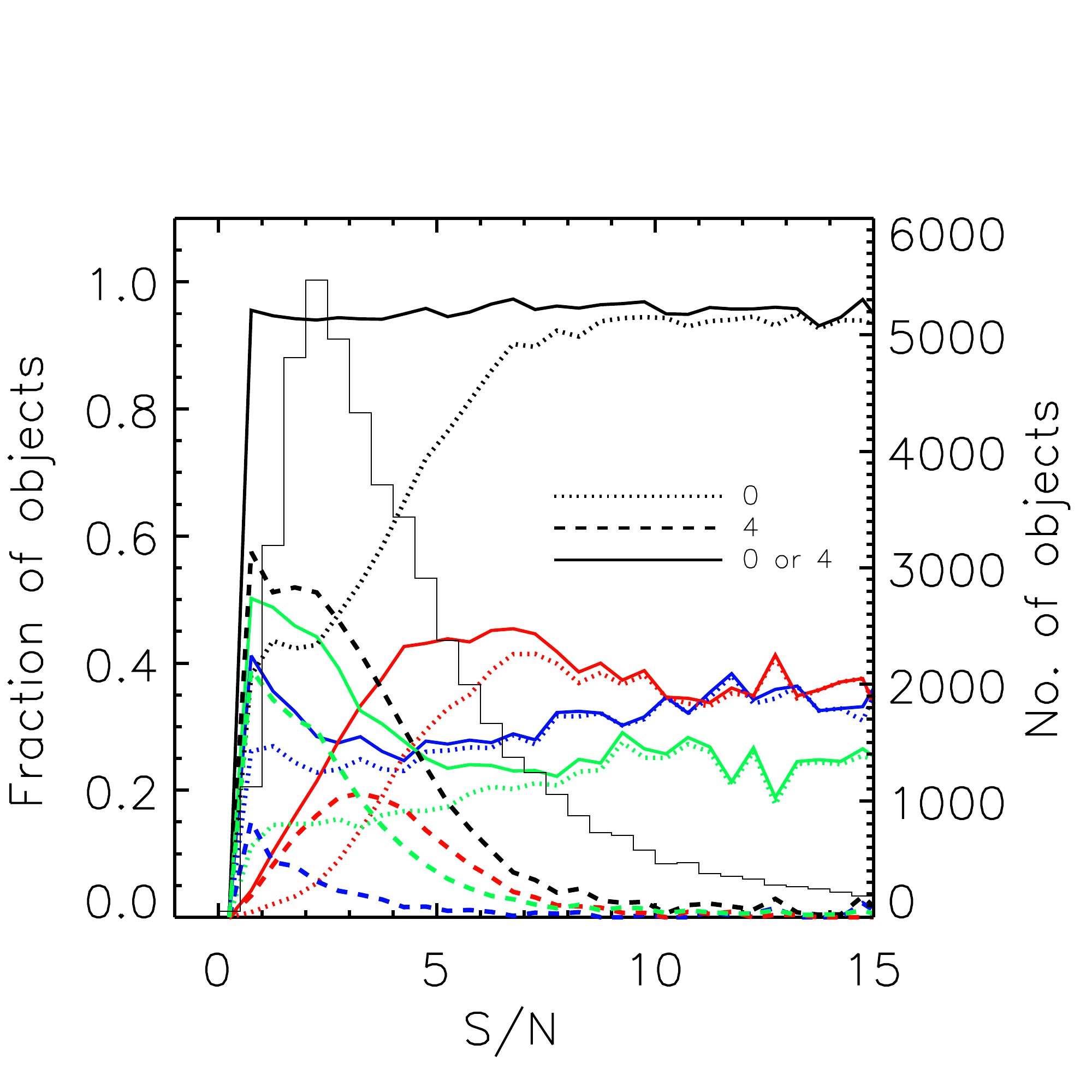}
    \caption{The fraction of Year One quasar targets with good
      redshifts (zWarning=0) as
      a function of $i$-band magnitude ({\it left}) and median spectroscopic S/N
      ({\it left}). Objects with zWarning=0 are given by the dotted lines, 
      objects with zWarning=4 are given by the dashed lines, and 
      objects with zWarning=0 $\parallel$ zWarning=4 --representing 95\% of
      our sample-- are given by the solid lines (i.e. dashed+dotted = solid). 
      Also shown separately are objects classified spectroscopically as stars (red) and
      high (blue) and low (green)-redshift quasars, as indicated. 
      Their sum is given by the black lines.  Also
    shown as the histogram and the right-hand y-axis is the
    distribution function of objects.}
    \label{fig:BOSS_zWarning}
  \end{center}
\end{figure}

Fig.~\ref{fig:BOSS_zWarning} gives the fraction
of objects with good, zWarning=0, spectra as a function of $i$-band
magnitude and spectroscopic S/N per pixel (median over the spectrum).  
The (black) histogram shows the distribution of all objects to give a sense
of where the majority of the signal arises from.  The most common
flag is {\tt SMALL\_DELTA\_CHI2}, indicating that there is more than
one template that fits the spectrum.  This is most commonly seen in
low S/N spectra.  We hope that planned visual
inspections of those objects with zWarning${}\ne 0$ will allow 
positive identification of many of these objects, boosting the number of
confirmed high-redshift quasars. 

\bibliographystyle{mn2e} 
\bibliography{Quasar_TS_paper}

\end{document}